\documentclass[twocolumn]{aastex62}

\newcommand{\logne}{log($N_\mathrm{e}$\,[cm$^{-3}$])~}

\received{\today}
\revised{}
\accepted{}
\submitjournal{ApJ}

\shorttitle{Electron Densities in the Solar Corona Measured in EUV and NIR}
\shortauthors{Dud\'{i}k et al.}

\begin{document}

\title{Electron Densities in the Solar Corona \\ Measured Simultaneously in the Extreme-Ultraviolet and Infra-Red}

\correspondingauthor{Jaroslav Dud\'{i}k}
\email{dudik@asu.cas.cz}

\author[0000-0003-1308-7427]{Jaroslav Dud\'{i}k}
\affil{Astronomical Institute of the Czech Academy of Sciences, Fri\v{c}ova 298, 251 65 Ond\v{r}ejov, Czech Republic}

\author[0000-0002-4125-0204]{Giulio Del Zanna}
\affil{Department of Applied Mathematics and Theoretical Physics, CMS, University of Cambridge, Wilberforce Road, Cambridge CB3 0WA, United Kingdom}

\author[0000-0003-3128-8396]{J\'an Ryb\'ak}
\affil{Astronomical Institute, Slovak Academy of Sciences, Tatransk\'a Lomnica, Slovakia}

\author[0000-0002-9690-8456]{Juraj L\"{o}rin\v{c}\'{i}k}
\affil{Astronomical Institute of the Czech Academy of Sciences, Fri\v{c}ova 298, 251 65 Ond\v{r}ejov, Czech Republic}
\affil{Institute of Astronomy, Charles University, V Hole\v{s}ovi\v{c}k\'{a}ch 2, CZ-18000 Prague 8, Czech Republic}

\author[0000-0003-2629-6201]{Elena Dzif\v{c}\'akov\'a}
\affil{Astronomical Institute of the Czech Academy of Sciences, Fri\v{c}ova 298, 251 65 Ond\v{r}ejov, Czech Republic}

\author[0000-0002-6418-7914]{Helen E. Mason}
\affil{Department of Applied Mathematics and Theoretical Physics, CMS, University of Cambridge, Wilberforce Road, Cambridge CB3 0WA, United Kingdom}

\author[0000-0001-7399-3013]{Steven Tomczyk}
\affil{High Altitude Observatory, National Center for Atmospheric Research, P.O. Box 3000, Boulder CO 80307-3000, USA}

\author[0000-0002-5872-7531]{Michael Galloy}
\affil{High Altitude Observatory, National Center for Atmospheric Research, P.O. Box 3000, Boulder CO 80307-3000, USA}

\begin{abstract}
Accurate measurements of electron density are critical for determination of the plasma properties in the solar corona. We compare the electron densities diagnosed from \ion{Fe}{13} lines observed by the Extreme-Ultraviolet Imaging Spectrometer (EIS) onboard the \textit{Hinode} mission with the near infra-red (NIR) measurements provided by the ground-based \textit{Coronal Multichannel Polarimeter (COMP)}. To do that, the emissivity-ratio method based on all available observed lines of \ion{Fe}{13} is used for both EIS and \textit{CoMP}. The EIS diagnostics is further supplemented by the results from \ion{Fe}{12} lines. We find excellent agreement, within 10\%, between the electron densities measured from both EUV and NIR lines. In the five regions selected for detailed analysis, we obtain electron densities of \logne\,=\,8.2--8.6.  Where available, the background subtraction has a significant impact on the diagnostics, especially on the NIR lines, where the loop contributes less than a  quarter of the intensity measured along the line of sight. For the NIR lines, we find that the line center intensities are not affected by stray light within the instrument, and recommend using these for density diagnostics. The measurements of the \ion{Fe}{13} NIR lines represent a viable method for density diagnostics using ground-based instrumentation.

\end{abstract}



%
\section{Introduction}
\label{Sect:1}

The solar corona is an expansive environment composed of relatively tenuous and hot plasma. As it is not yet possible to derive its properties directly from in-situ measurements, we are forced to rely on analyses of remote-sensing observations. The optically thin nature of the corona allows for derivation of the basic plasma properties, such as temperature and electron density, using analyses of multiple emission line intensities \citep[e.g.,][]{Phillips08,DelZanna18} that are produced by the coronal plasma.

Although the temperature structure of the solar corona can be relatively precisely determined from imaging observations as well as from spectroscopy \citep[e.g.,][]{DelZanna03,Ugarte-Urra09,Brooks09,Schmelz09,Warren11,Warren20,Winebarger11,Hannah12,Hannah13,DelZanna13b,Cheung15,Parenti17,Su18,Bak-Steslicka19,Goddard20}, the electron density requires analyses of ratios of emission line intensities obtained from spectroscopic observations \citep[e.g.,][]{Gabriel82,Young94,Young09,Ugarte-Urra05,Tripathi08,Watanabe09,Winebarger11,Reale14,Polito16,DelZanna18,Lorincik20}.

The electron density is an important plasma parameter. For example, the total radiative losses from the corona depend on the square of electron density, having important implications for the coronal energetics and the heating input required for corona to exist \citep[e.g.,][]{RTV78,Warren06,Klimchuk08,Colgan08,Dudik11,Lionello13,Xia14}. Moreover, comparisons of the measured electron densities with those predicted from static loop models suggest that a significant portion of the loops are neither static nor uniformly heated \citep[e.g.,][and references therein]{Lenz99,Warren03,Gupta15,Schmelz15,Brooks19}.

Furthermore, measurements of propagating waves observed from the ground with the \textit{Coronal Multichannel Polarimeter} \citep[\textit{CoMP},][]{Tomczyk08} instrument were recently used by \cite{Yang20} to provide for the first time a global map of the coronal magnetic field. These measurements would not be possible without the observational determination of the Alfvén phase speed 
and the electron density. The densities from \textit{CoMP} were obtained by using the \ion{Fe}{13} near-infrared (NIR) line ratio (see Table~1), assuming that the emission was originating only from plasma in the plane of the sky. 

Various issues can affect the accuracy of such measurements. The observations only provide emissivity-weighted values along the line of sight, and the actual distribution of densities could be more variable than the values obtained with the plane-of-sky assumption \citep[as also pointed out in][]{Yang20}. Furthermore, different line ratios can have different sensitivity to the density distribution along the line of sight. For example, the allowed transitions occurring in the extreme-ultraviolet (EUV) or UV parts of the spectrum, which decay to different excited states, could be produced by different plasma regions \citep[see, e.g.][]{doschek:1984}. The forbidden transitions in the visible or NIR, such as the \textit{CoMP} \ion{Fe}{13} lines have a different sensitivity than the allowed transitions, as they are strongly affected by photo-excitation from the solar disk \citep[see, e.g.][ and references therein]{chevalier_lambert:69,Pineau73,Young94,DelZanna18}. 
Therefore, they are naturally produced by different plasma regions in the corona.

These issues are particularly relevant for future observations, since routine measurements of the \ion{Fe}{13} NIR lines will be obtained with the new improved version of
\textit{CoMP} \citep{ucomp2019} and the 4-m \textit{Daniel K. Inouye Solar Telescope (DKIST)}  
Cryogenic Near Infrared Spectropolarimeter \citep[Cryo-NIRSP,][]{Fehlmann16}, which will provide unprecedented high-resolution data. 

The above-mentioned issues have hardly been explored in the literature. There are several density measurements of the solar corona using the \ion{Fe}{13} NIR lines (following \citet{chevalier_lambert:69}; see also, for example, \citet{Penn94}; \citet{Singh02}). There is also ample literature on densities obtained from EUV/UV \ion{Fe}{13} and \ion{Fe}{12} ratios, see e.g. \cite{dere_etal:1979} and references in the review by \cite{DelZanna18}. However, as far as we are aware, EUV and NIR simultaneous measurements of densities, as presented here, have not been published. Here, we compare density measurements using space-borne EUV observations from \textit{Hinode}/EIS with those obtained by the ground-based \textit{CoMP} instruments. 
We also discuss several instrumental issues and uncertainties associated with such measurements.

%
\begin{figure*}
   \centering
   \includegraphics[width=6.83cm,viewport= 0 0 330 410,clip]{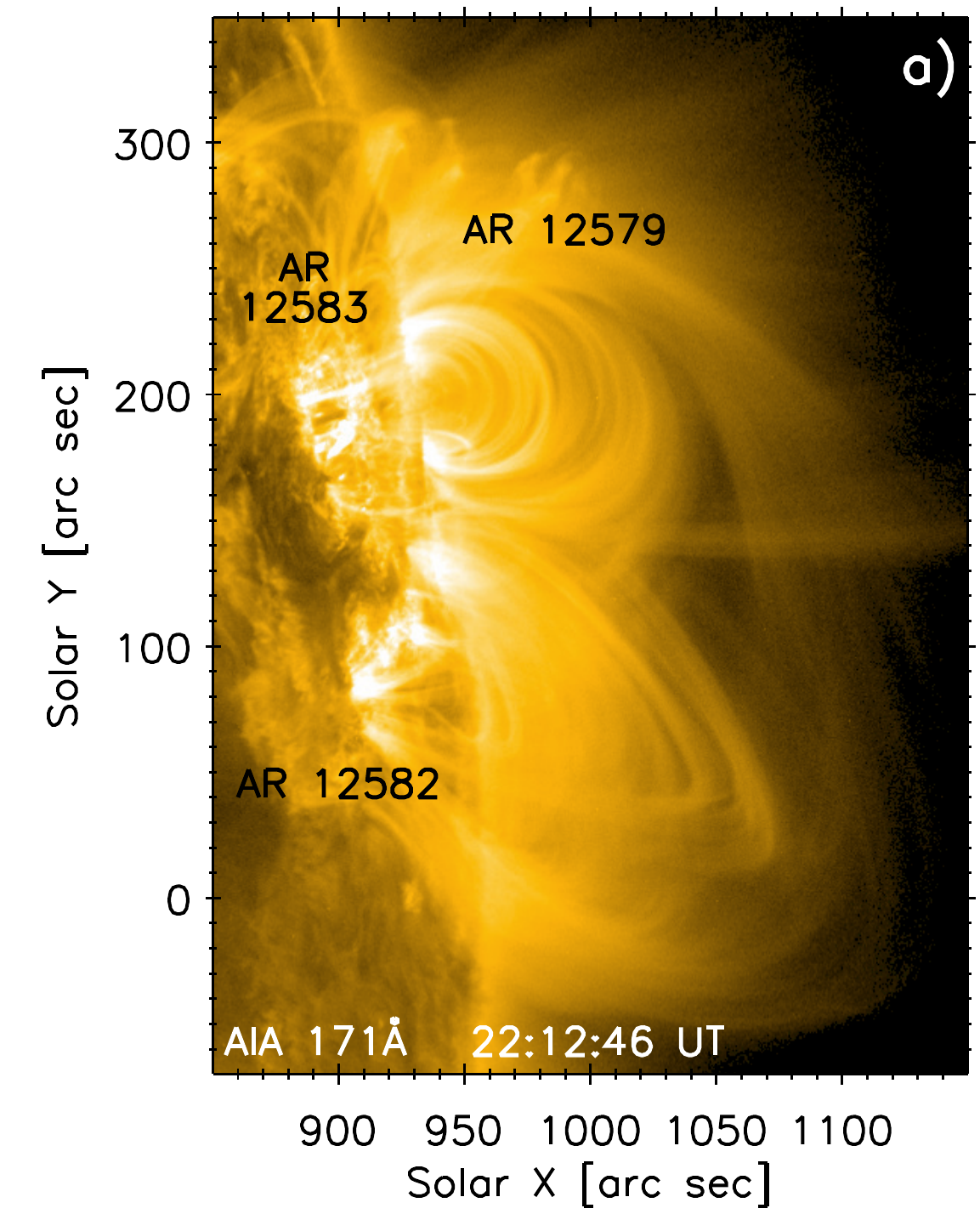}
   \includegraphics[width=5.38cm,viewport=70 0 330 410,clip]{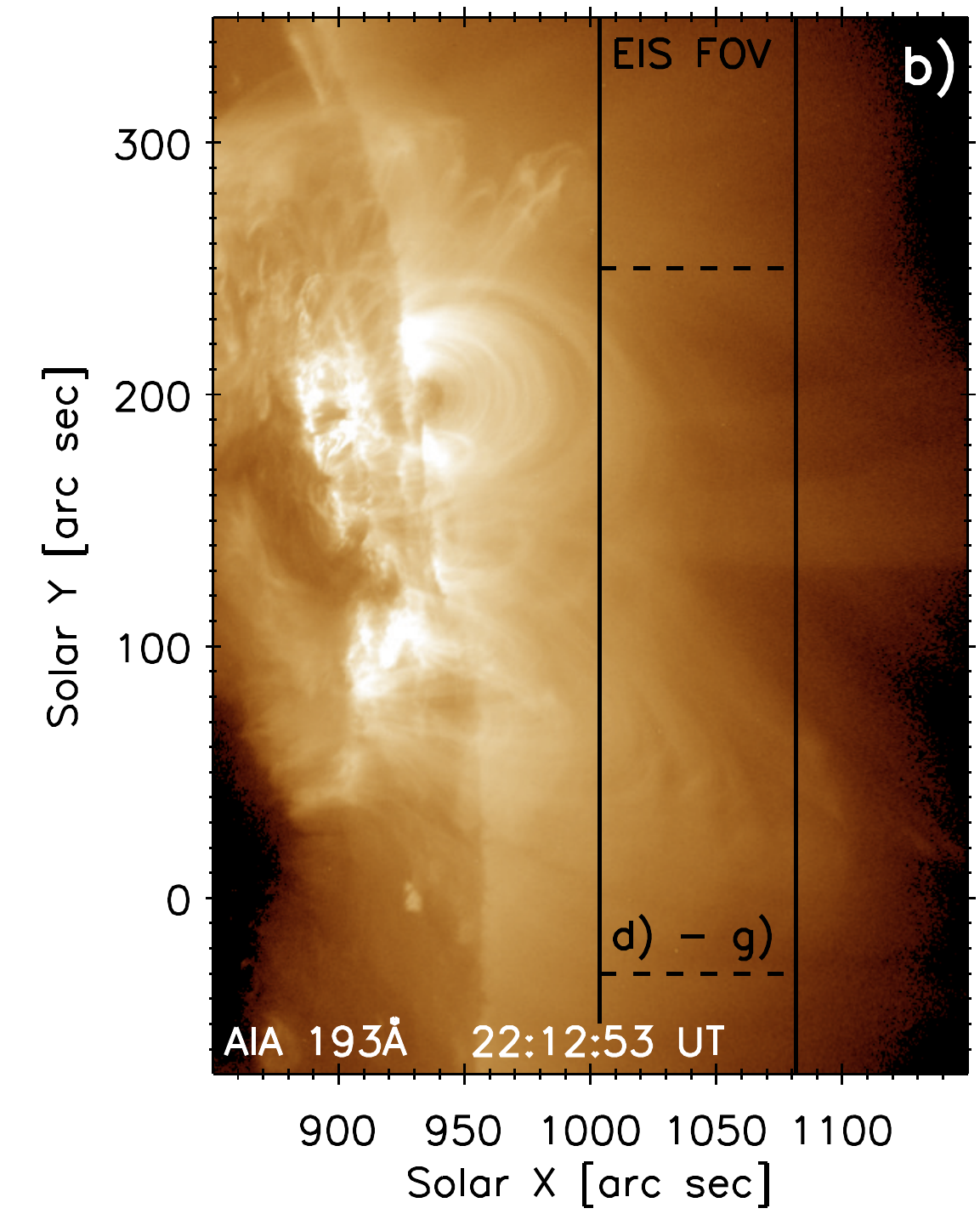}
   \includegraphics[width=5.38cm,viewport=70 0 330 410,clip]{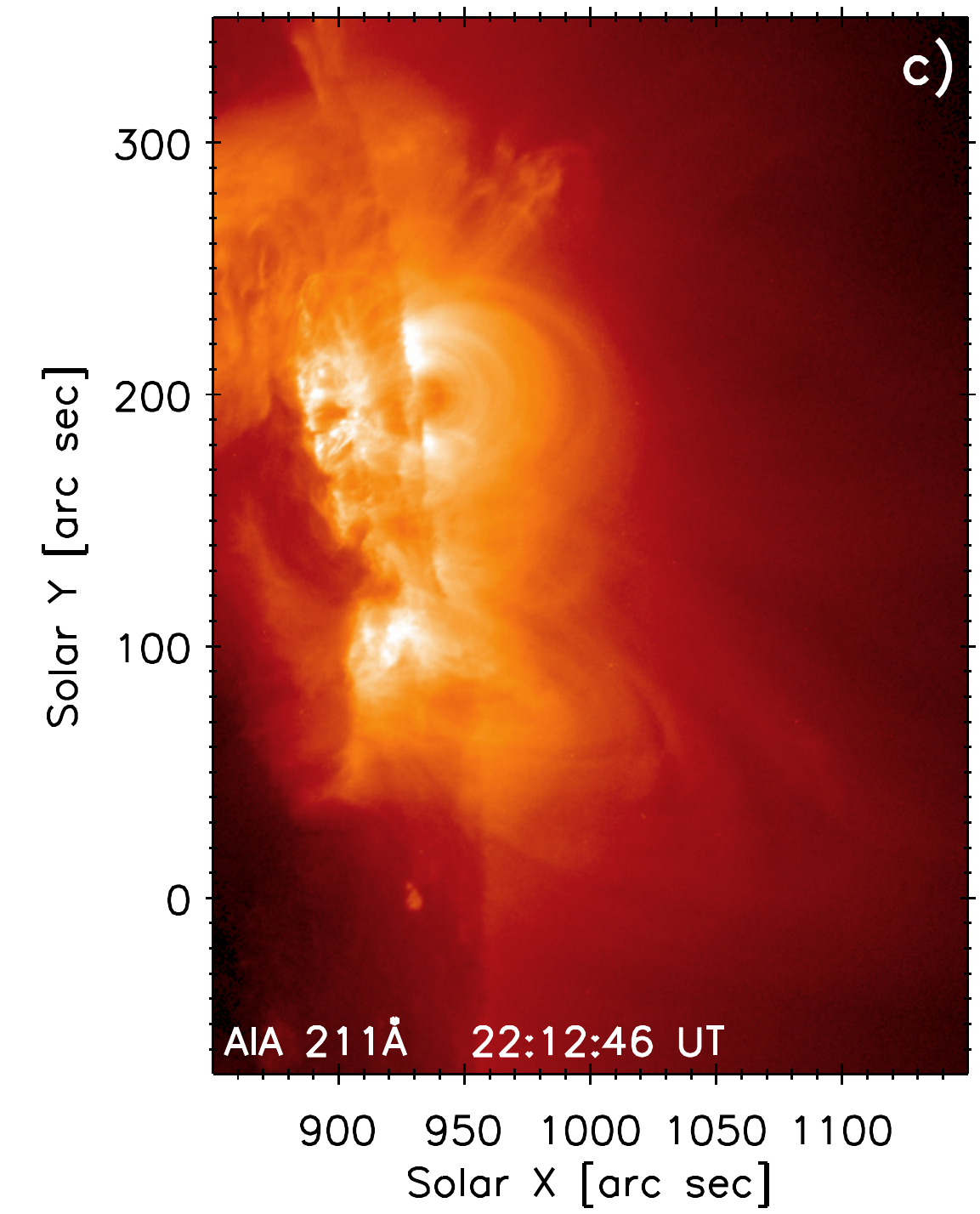}
   \includegraphics[width=5.50cm,viewport= 0 0 225 575,clip]{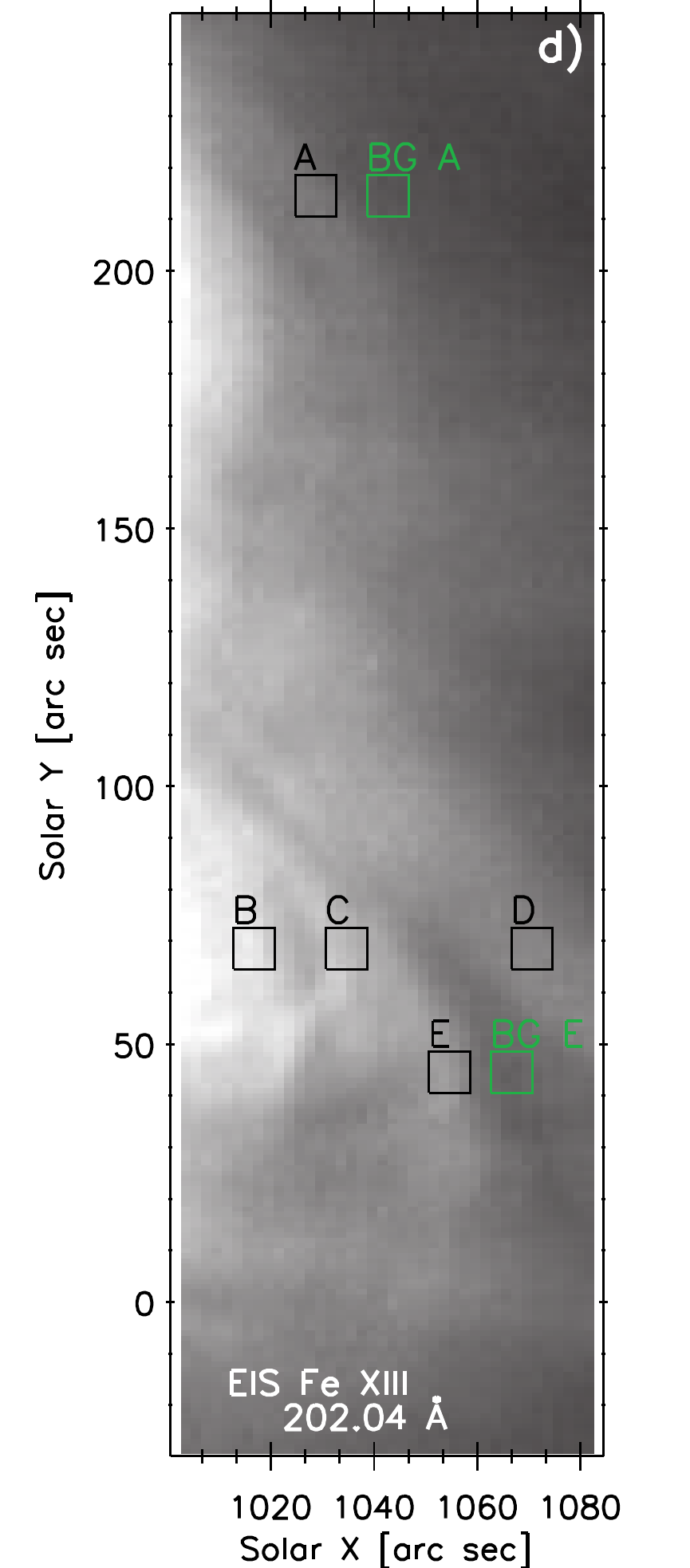}
   \includegraphics[width=4.03cm,viewport=60 0 225 575,clip]{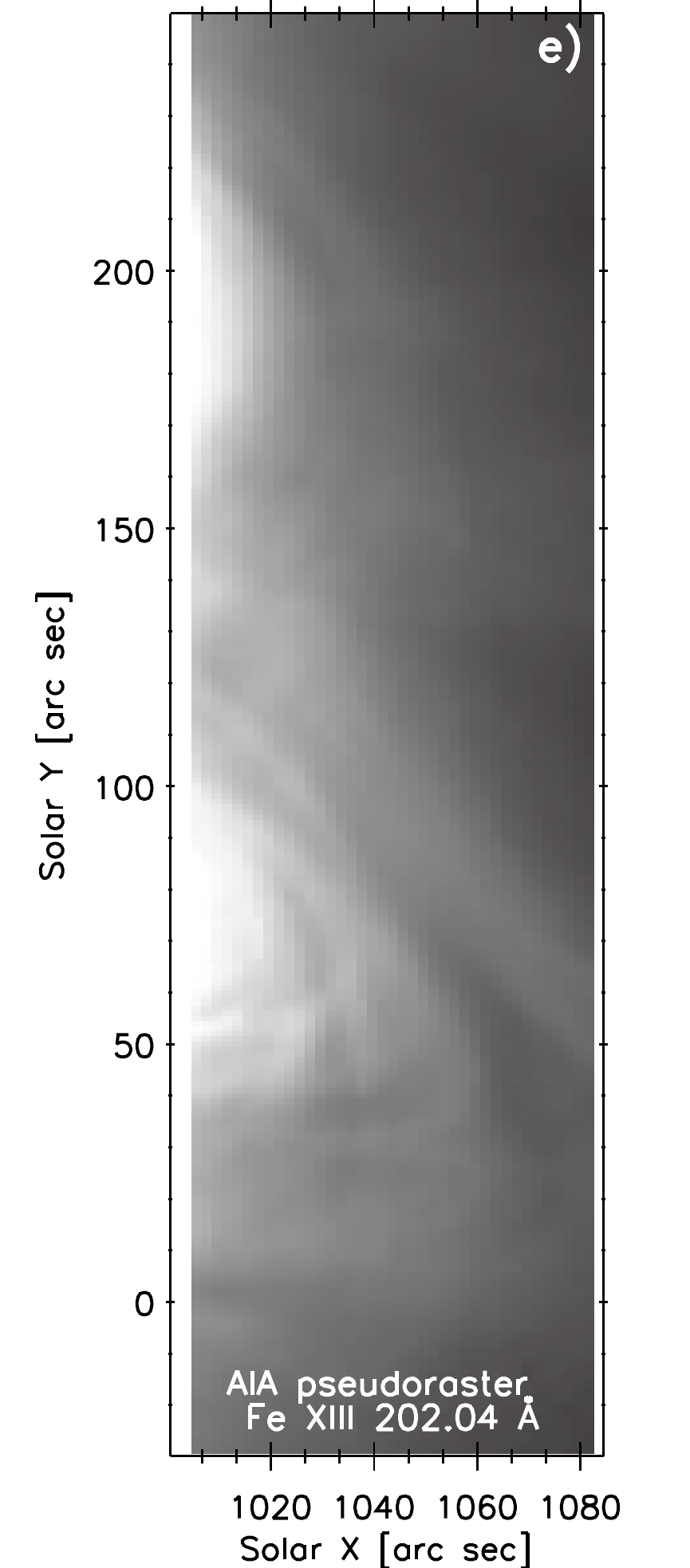}
   \includegraphics[width=4.03cm,viewport=60 0 225 575,clip]{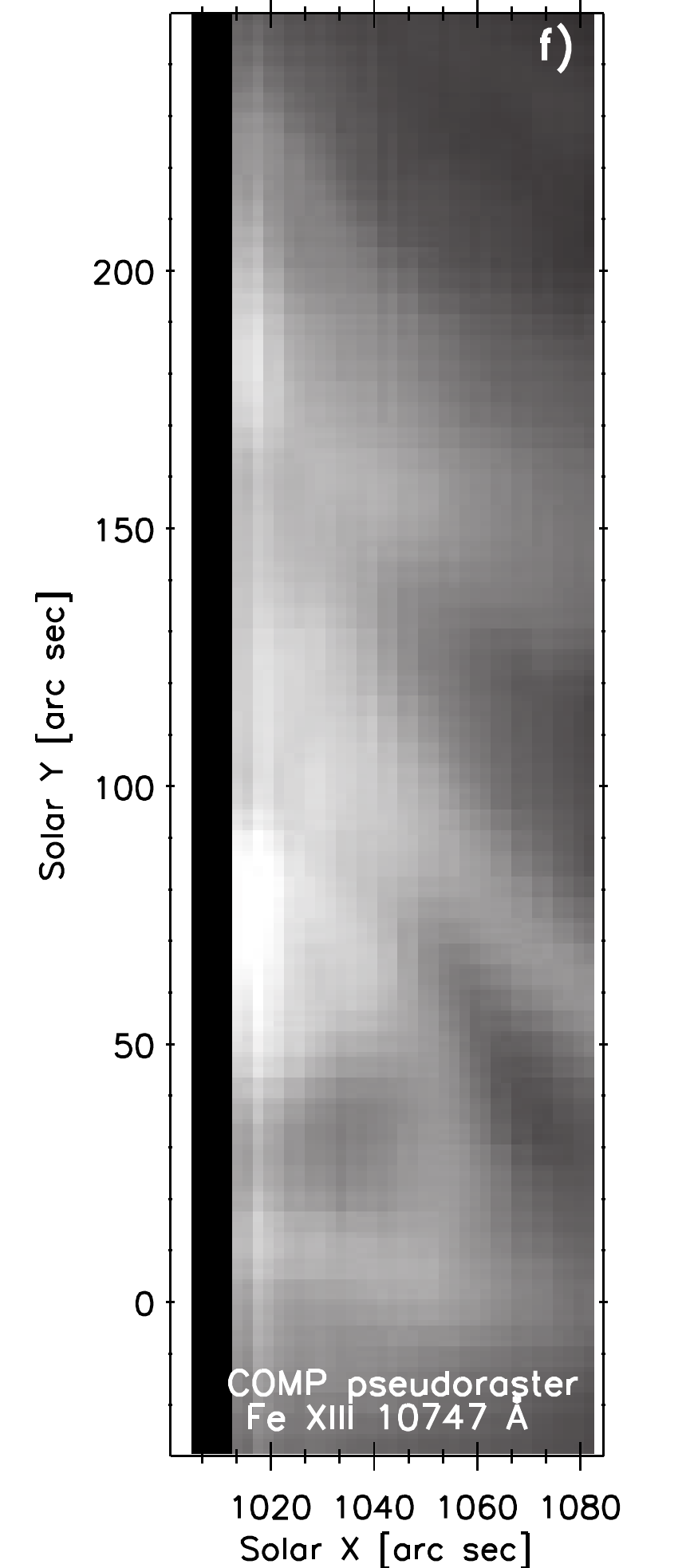}
   \includegraphics[width=4.03cm,viewport=60 0 225 575,clip]{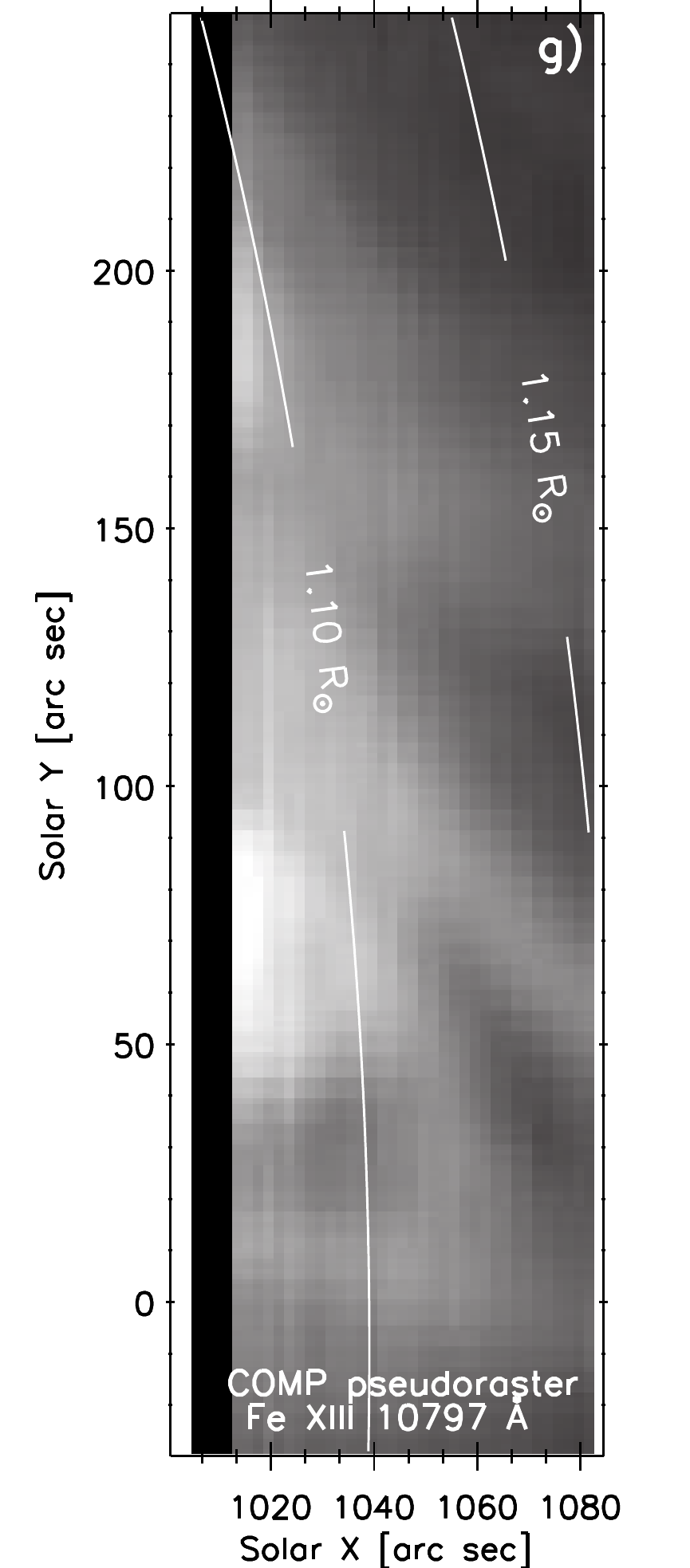}
   \caption{Observations of the AR complex off limb on 2016 August 30. Panels (a)--(c) show the context observations from \textit{SDO}/AIA in the 171, 193, and 211\,\AA~channels. Panel (d) shows a portion of the \textit{Hinode}/EIS raster (indicated by the box in panel (b)) in the \ion{Fe}{13} 202.04\,\AA~line, while panels (e)--(g) show the pseudo-rasters created from (e) the synthetic \ion{Fe}{13} 202.04\,\AA~from AIA DEM reconstruction, (f) \textit{CoMP} 10747\,\AA, and (g) \textit{CoMP} 10797\,\AA~line center intensities. The regions selected for further analysis are shown in panel (d), with background regions in green. The height above the solar limb is denoted in panel (g) in units of solar radius $R_\odot$.}
   \label{Fig:Context}
\end{figure*}
%
\begin{table*}
\begin{center}
\caption{The strong \ion{Fe}{13} lines used in the present analysis including their self-blends. The wavelengths of the NIR lines are given in air. Adapted from Table B5 of \citet{Dudik14b}.}
\label{Table:fe13}
\begin{tabular}{llcll}
\tableline
\tableline
$\lambda$~[\AA] & levels & level description				&  selfblends	 	& notes  \\
\tableline
196.53    & $4 - 26$	&   $3s^2 3p^2 ~{}^1D_2 - 3s^2 3p 3d ~{}^1F_3$	& 			&			\\
202.04    & $1 - 20$	&   $3s^2 3p^2 ~{}^3P_0 - 3s^2 3p 3d ~{}^3P_1$	& 			&			\\
203.83    & $3 - 25$	&   $3s^2 3p^2 ~{}^3P_2 - 3s^2 3p 3d ~{}^3D_3$	& $203.77 ~(7-60) $	& bl \ion{Fe}{12}	\\
	  &		&						& $203.80 ~(3-24) $	&			\\
	  &		&						& $203.84 ~(8-60) $	&			\\
\tableline
10747     & $1 - 2 $	&   $3s^2 3p^2 ~{}^3P_0 - 3s^2 3p^2 ~{}^3P_1 $	&           		&			\\
10797     & $2 - 3 $	&   $3s^2 3p^2 ~{}^3P_1 - 3s^2 3p^2 ~{}^3P_2 $	&           		&			\\
\tableline
\tableline
\end{tabular}
\end{center}
\end{table*}
%
%
%
%
\begin{table*}
\begin{center}
\caption{The strong \ion{Fe}{12} lines used in the present analysis including their self-blends. Adapted from Table B4 of \citet{Dudik14b}.}
\label{Table:fe12}
\begin{tabular}{llcll}
\tableline
\tableline
$\lambda$~[\AA] & levels & level description					&  selfblends	 	& notes  \\
\tableline
186.89    & $3 - 39$	&   $3s^2 3p^3 ~{}^2D_{5/2} - 3s^2 3p^2 3d ~{}^2F_{7/2}$ & $186.85 ~(2-36)  $	&	\\
	      &		        &							                             & $186.93 ~(22-135)$	&	\\
192.39    & $1 - 30$	&   $3s^2 3p^3 ~{}^4S_{3/2} - 3s^2 3p^2 3d ~{}^4P_{1/2}$ &          	    	&	\\
193.51    & $1 - 29$	&   $3s^2 3p^3 ~{}^4S_{3/2} - 3s^2 3p^2 3d ~{}^4P_{3/2}$ & $193.50 ~(7-80)  $	&	\\
195.12    & $1 - 27$	&   $3s^2 3p^3 ~{}^4S_{3/2} - 3s^2 3p^2 3d ~{}^4P_{5/2}$ & $195.08 ~(31-151)$	& sbl \\
	      &		        &							                             & $195.18 ~(22-135)$	&	\\
	      &		        &							                             & $195.22 ~(13-115)$	&	\\
196.64    & $3 - 34$	&   $3s^2 3p^3 ~{}^2D_{5/2} - 3s^2 3p^2 3d ~{}^2D_{5/2}$ & $196.65 ~(23-131)$	&	\\
203.73    & $3 - 32$	&   $3s^2 3p^3 ~{}^2D_{5/2} - 3s^2 3p^2 3d ~{}^2D_{5/2}$ &         	        	& bl \ion{Fe}{13}	\\
\tableline
\tableline
\end{tabular}
\end{center}
\end{table*}
%
%
%
%
%
\section{Observations}
\label{Sect:2}

The observations analyzed here were performed as part of the joint \textit{IRIS}--\textit{Hinode} Operation Plan (IHOP) 316 that also involved coordinated observations with the \textit{CoMP} instrument. The campaign ran over a period of several weeks. Here, we focus on the off-limb observations obtained on 2016 August 30 during about 22:00--00:00\,UT. The observations analyzed were performed by the Extreme-Ultraviolet Imager \citep[EIS][]{Culhane07} onboard on the \textit{Hinode} mission, as well as \textit{CoMP}. The EIS field of view (FOV) was chosen to be sufficiently far off-limb so that it is above the \textit{CoMP} occulter, thus ensuring overlap between EIS and \textit{CoMP} (Figure \ref{Fig:Context}). The \textit{IRIS} instrument also participated in the campaign, observing the forbidden \ion{Fe}{12} 1349.4\,\AA~line \citep{Testa16}. However, upon review of the data we found that in the present observations this line is weak and detected only close to the limb, i.e., outside of the field of view of EIS and \textit{CoMP}. For this reason, we do not analyze the \textit{IRIS} data here.

\subsection{Context observations from \textit{SDO}/AIA}
\label{Sect:2.1}

The coronal context for the EIS and \textit{CoMP} was examined using the observations provided by the Atmospheric Imaging Assembly \citep[AIA,][]{Lemen12,Boerner14} onboard the \textit{Solar Dynamics Observatory} \citep[\textit{SDO},][]{Pesnell12}. AIA provides high-resolution (1.5$\arcsec$), high-cadence (12\,s) imaging observations of the solar corona in 7 EUV wavelength channels. The warm corona at 1--2\,MK is observed in 171\,\AA, 193\,\AA~and 211\,\AA~passbands (Figure \ref{Fig:Context}a--c) dominated respectively by \ion{Fe}{9}, \ion{Fe}{12}, and \ion{Fe}{14}, although contributions from other ions formed at different temperatures are also present \citep[][]{ODwyer10,DelZanna13b}. We also reviewed the other EUV channels of AIA, those at 94\,\AA, 131\,\AA, and 335\,\AA, to examine the emission originating at higher temperatures, as well as to derive the differential emission measures in turn used in coalignment with EIS and \textit{CoMP} (see Sect. \ref{Sect:2.3.4}). The 304\,\AA~observations were also checked to exclude the presence of optically thick plasma at transition-region temperatures, such as filaments.

The AIA observations of the off-limb region reveal that it consisted of a series of coronal loops within the complex of active regions (AR) NOAA 12579, 12582, and 12583. The AR 12579 was located westward of the two, with AR 12583 being located about 10$^\circ$ to the north of the AR 12582. The AIA images reveal the presence of many coronal loops, with some being visible as series of two arches located approximately in the plane of sky (Figure \ref{Fig:Context}a--c). Examining the time-series for the AIA 171\,\AA, 193\,\AA, and 211\,\AA~observations for the duration of the observations, we found that the morphology of the observed corona did not show any significant changes in the period analyzed. We note however that a small $\approx$C3 flare occurred in the low corona of AR 12582 during the observing period, but is outside of the EIS and \textit{CoMP} fields of view.

\subsection{\textit{Hinode}/EIS}
\label{Sect:2.2}

\textit{Hinode}/EIS \citep{Culhane07} is an extreme-ultraviolet (EUV) spectrometer observing in two spectral windows, 171--210\,\AA~and 250--290\,\AA\ with a spectral resolution of 22\,m\AA~and spatial resolution of about 3$\arcsec$ (with pixel size along the slit of 1$\arcsec$). For the present study, the EIS employed the \texttt{CompS\_NonMax} observing sequence and rastered a 80$\arcsec$\,$\times$\,512$\arcsec$ off-limb region using 40 exposures with the 2$\arcsec$ slit. The exposure duration for each slit position was chosen to be 60 and 120\,s, performing one raster each. In the present study, we focus on the 60\,s exposures, which were taken first (22:14 -- 22:54\,UT), i.e., earlier in the day for \textit{CoMP}, to take advantage of best seeing conditions. Furthermore, using shorter exposures helps avoid possible saturation in stronger EIS lines \citep[see][]{DelZanna19}. The EIS dataset contains three \ion{Fe}{13} lines at 196.5, 202.0, and 203.8\,\AA~\citep[see, e.g.,][]{Watanabe09,Young09} which have strong signal off-limb, alongside a number of other lines of \ion{Fe}{9}--\ion{Fe}{16} as well as \ion{Si}{7} included for purposes of emission-measure loci \textbf{and non-Maxwellian} analyses, and finally the \ion{He}{2} for co-registration with other instruments. 

The EIS observations were processed using custom-written codes which closely follow most of the EIS standard processing. The spectra were cleaned of cosmic rays and hot/warm pixels. In addition, the missing pixels were replaced by smoothing procedures. The processed spectra were then fitted in the units of DN by using Gaussian line profiles (dominated by instrumental width) and a linear background. Finally, radiometric calibration was applied to the line intensities. The choice of radiometric calibration is a difficult issue \citep[see][]{DelZanna13a,Warren14,DelZanna19}. The earlier in-flight calibrations provided an improvement to data up to 2012. While these should clearly be not valid for the present dataset (2016), we opted for using the ground calibration \citep{Lang06} and lines close in wavelength in the short-wavelength (SW) channel only, as large discrepancies are present with the lines in the long-wavelength (LW) channel. As shown below, with the ground calibration we obtain consistent electron densities from \ion{Fe}{13} and \ion{Fe}{12}. We note that the EIS instrument underwent its first bakeout in February 2016, which is possibly the reason why the instrument relative calibration,  for the wavelengths considered here, is close to the ground one. The EIS bakeout also resulted in significant decrease in the number of warm pixels \citep[see Figure 3 in][]{Kennedy20}. \footnote{\url{ https://hesperia.gsfc.nasa.gov/ssw/hinode/eis/doc/eis_notes/06_HOT_WARM_PIXELS/eis_swnote_06.pdf}} 

The strong \ion{Fe}{13} lines observed at 196.5\,\AA, 202.0\,\AA~(Table \ref{Table:fe13}), are unblended, while the 203.8\,\AA~line is a self-blend of four transitions \citep[as identified by][]{DelZanna11} and is further blended in its blue wing with the \ion{Fe}{12} transition at 203.73\,\AA~\citep{Young09}. The \ion{Fe}{12} contribution is estimated from the other \ion{Fe}{12} lines, and subtracted from the total intensity of the \ion{Fe}{13} 203.8\,\AA~line. Other lines used here include multiple \ion{Fe}{12} lines that are listed in Table \ref{Table:fe12}. \ion{Fe}{12} is used because it produces many strong emission lines observable in the same SW channel of EIS as \ion{Fe}{13}, and because it is formed close in temperature. We use these lines in the density diagnostics as well as \ion{Fe}{13} lines, and provide a discussion on the results in Sect. \ref{Sect:4.2}. The \ion{Fe}{12} lines are strong and do not require deblending. However, as  described by \citet{DelZanna19}, opacity effects in the strongest \ion{Fe}{12} 193.5 and 195.1\,\AA~lines are very common in active region observations, 
while the weaker 192.4\,\AA~line is least affected. The effects in the present dataset (which is quite far off limb) are however small. 

\subsection{Coronal Multichannel Polarimeter}
\label{Sect:2.3}

The \textit{CoMP} instrument was a ground-based coronograph located at Mauna Loa Solar Observatory\footnote{\url{https://www2.hao.ucar.edu/mlso/mlso-home-page}}, Hawaii, USA. It had a 20\,cm $f/11$ objective measuring the complete polarization states of the \ion{Fe}{13} NIR doublet as well as the \ion{He}{1} line at 10830\,\AA~\citep{Tomczyk08}. To do that, a tunable four-stage calcite birefringent filter was used together with a Wollaston polarizing beamsplitter. The width of the passband is 1.3\,\AA~FWHM. Higher orders were blocked with a 17\,\AA~interference filters.  

During IHOP 316, the \textit{CoMP} instrument was observing the polarization state of the off-limb corona in both \ion{Fe}{13} lines. Each line was scanned using 5 points per profile spaced 1.2\,\AA~apart. The wavelengths sampled were 10743.8, 10745.0, 10746.2, 10747.4, 10748.6\,\AA~ for the \ion{Fe}{13} 10747\,\AA~line, and 10795.4, 10796.6, 10797.8, 10799.0, and 10780.2\,\AA~for the \ion{Fe}{13} 10797\,\AA~line. Additionally, the local continuum located 5.85\,\AA~away from the outer points of line profiles and on either side were recorded. The observations lasted from 22:12\,UT to about 23:00\,UT, alternating between both lines. The typical cadence for each line is 72\,s, with a gap between 22:17--22:25\,UT caused by tracking difficulties. Finally, the \textit{CoMP} instrument has a pixel size of 4.35$\arcsec$ and its resolution is thus coarser than that of \textit{Hinode}/EIS.

\subsubsection{\textit{CoMP} inversions and associated uncertainties}
\label{Sect:2.3.1}

Since the \textit{CoMP} data are scans across the line profiles, the intensities of the lines must be derived by fitting the observed points. This task is complicated by seeing conditions and also by pointing inaccuracies. \citet{Tian13} developed a robust analytical inversion, where only the central 3 points of the profile are used to derive the three parameters of a Gaussian, the amplitude $I_0$, Doppler shift $\lambda_\mathrm{D}$, as well as the line width $w$. The corresponding uncertainties of these parameters were derived analytically by \citet{Morton16} using the \textit{CoMP} instrument characteristics and the photon noise in each of the observed points. \citet{Morton16} also showed that the derived uncertainties in the Gaussian parameters are in excellent agreement with the noise as estimated from the variability in time series of \textit{CoMP} data \citep[see Figure 5 of][]{Morton16}. Obviously, this holds only if the observed coronal structures are stable, i.e., do not significantly change over time.

Besides obtaining the robust analytical inversions using the method of \citet{Tian13}, we have separately reduced the raw 7-point per profile \textit{CoMP} measurements, provided by the \textit{CoMP} team, using the original \textit{CoMP} pipeline source code \footnote{\url{ https://github.com/mgalloy/comp-pipeline}} with some custom code changes. The code provides the Stokes $I$ parameter separated from the background light caused by the sky and instrument scattered light, profiting from the special design of the \textit{CoMP} Lyot filter \citep{Tomczyk08}. The code has been adapted by applying a background absorption spectrum subtraction also for the 10797\,\AA~line. Moreover, for Gaussian fitting of both lines the full measured Stokes $I$ profiles of 7 wavelength points has been applied. The spectral behaviour of the background light has been averaged to minimize the effect of atmospheric seeing and/or telescope guiding during the line scans. The noise of the spectral intensities has been estimated statistically and includes both the photon noise and the seeing effects. No other special seeing/guiding correction has been applied to the CoMP data.

Comparisons of the 3-point analytical fits with the fits of the full 7-point line profile are provided in Appendix \ref{Appendix:A}, showing that the differences in the line center intensities $I_0$ between the two datasets is typically about 5\%. Due to some difficulties with the 7-point fits described there, especially in the weaker 10797\,\AA~line, in the remainder of this work we use the 3-point fits \citep{Tian13}. We note that similar data were extensively used in the past to study the coronal dynamics \citep[e.g.,][]{Tomczyk09,Tian13,Liu14,Morton16,French19,Tiwari19,Yang20}. Since only 3 points are used, the derived quantities are less sensitive to both seeing and tracking inaccuracies. 


%
\subsubsection{On the use of line center intensities}
\label{Sect:2.3.2}

Upon cursory inspection, the line intensities $I$ obtained from 3-point fits cannot be used for diagnostics of electron density, as the total Gaussian intensities $I$\,=\,$w I_0(2\pi)^{1/2}$ lead to a ratio $I_{10747} / I_{10797}$ that can be outside of the theoretically calculated values (see Sect. \ref{Sect:3.2}). The reason is that the 3-point fits, unlike the 7-point ones, can lead to errorneous widths $w$ for the stronger \ion{Fe}{13} 10747\,\AA~line. We note however that the line center intensities $I_0$ are not affected by the choice of how many points are included in the fit (see Sect. \ref{Sect:2.3.1} and Appendix \ref{Appendix:A}).

In addition, we consider the influence of stray light, following the description of \citet[][p. 290, chapter 12 therein]{Gray08}. If $I_\mathrm{obs}$ is the observed intensity point at a particular wavelength $\lambda$, and $I_\mathrm{true}$ is the true signal at this position, presence of the stray light in the instrument removes a fraction $s$ from the observed signal, while it adds a scattered contribution originating at other wavelengths. This contribution can be expressed as $s\left<I\right>$, where $\left<I\right>$ is the mean spectral intensity. Therefore,
\begin{equation}
    I_\mathrm{obs}  = I_\mathrm{true} - s I_\mathrm{true} + s \left<I\right>\,,
    \label{Eq:I_obs}
\end{equation}
or
\begin{equation}
    I_\mathrm{true}  = \frac{I_\mathrm{obs} - s \left<I\right>}{1-s}\,.
    \label{Eq:I_true}
\end{equation}
With respect to the local continuum $I_\mathrm{true, \,c}$ one can then write
\begin{equation}
    \frac{I_\mathrm{true}}{I_\mathrm{true, \,c}}  = \frac{I_\mathrm{obs} - s \left<I\right>}{I_\mathrm{obs,\,c} - s\left<I\right>} = \frac{I_\mathrm{obs} / I_\mathrm{obs,\,c} - s\left<I\right> / I_\mathrm{obs,\,c}}{1-s\left<I\right> /I_\mathrm{obs,\,c}}\,.
    \label{Eq:I_true_c}
\end{equation}
In the continuum, where the spectrum is uncrowded by other lines, it holds that $I_\mathrm{obs,\,c} = \left<I\right>$. Therefore, 
\begin{equation}
    \frac{I_\mathrm{true}}{I_\mathrm{true, \,c}}  = \frac{I_\mathrm{obs} / I_\mathrm{obs,\,c} - s}{1-s}\,.
    \label{Eq:I_true_c_approx1}
\end{equation}
as derived by \citet{Gray08}.

Using this result, we note that in the line center, the observed signal with respect to the continuum is much stronger than the fraction of $s$ due to stray light, i.e., $I_\mathrm{obs} / I_\mathrm{obs,\,c} \gg s$. This permits us to neglect the $s$ in the numerator of Equation (\ref{Eq:I_true_c_approx1}). Finally, taking a ratio of  line-center intensities of two lines $\lambda$1 and $\lambda$2 with respect to a close-by continuum results in
\begin{equation}
    \frac{I_\mathrm{true, \,\lambda1} / I_\mathrm{true, \,c}}{I_\mathrm{true, \,\lambda2} / I_\mathrm{true, \,c}} = \frac{I_\mathrm{true, \,\lambda1}}{I_\mathrm{true, \,\lambda2}} =
    \frac{ \left( I_\mathrm{obs,\,\lambda1} / I_\mathrm{obs,\,c} \right) / \left(1-s\right)} {\left(I_\mathrm{obs,\,\lambda2} / I_\mathrm{obs,\,c} \right) / \left(1-s\right) } =
    \frac{I_\mathrm{obs,\,\lambda1}}{I_\mathrm{obs,\,\lambda2}}\,.
    \label{Eq:line_center_intensities}
\end{equation}
Therefore, the ratio of the $I_{0,\,10747} / I_{0,\,10797}$ line center intensities alone can be used to measure the electron densities. We note that the above approximation holds $(i)$ only if the fraction of the stray light $s$ is similar in both lines, which in our case is an assumption justified by the fact that the NIR \ion{Fe}{13} lines are close in wavelength, and $(ii)$ only for areas where the lines are well observed, i.e., their center intensities are much stronger than the local continuum or stray light.

\subsubsection{Coalignment of individual \textit{CoMP} frames}
\label{Sect:2.3.3}

Imperfections in tracking as well as possible image deformations due to atmospheric seeing conditions lead to individual \textit{CoMP} images not being aligned. In our case, the pointing inaccuracies are typically several arc sec, and for the last frame before the gap in the observations, up to about 10$\arcsec$. To remedy this situation, we performed automatic alignment of the \textit{CoMP} datasets (line center intensities) using the \texttt{auto\_align\_images.pro} IDL routine available under SolarSoft. The routine performs cross-correlation of two input images not only by shifting the second image in both solar $X$ and $Y$ directions, but also by exploring mutual rotation and stretching. The routine was applied to a field of view limited to $X$\,$\in$\,$\left<1010\arcsec, 1270\arcsec\right>$ and $Y$\,$\in$\,$\left<-100\arcsec,400\arcsec\right>$, which was chosen to encompass the EIS FOV as well as to avoid the \textit{CoMP} occulter and areas with weak (noisy) signal far off-limb.

We chose the first 10747\,\AA~line center intensity image observed at 22:12:48\,UT as a reference. The following 10747\,\AA~frames were progressively cross-correlated to the previous one; i.e., the $i$-th frame was cross-correlated to the $i-1$ one. We found that the procedure works well, resulting in relative rotations between frames of up to 0.4$^\circ$ and relative stretching of up to 2\%, which we attribute to seeing. The most common occurrence was of almost no rotation and no stretching. This procedure results in very well-aligned 10747\,\AA~data. \textbf{An example is provided in Appendix \ref{Appendix:B}.}

Unfortunately, the procedure does not work so well for the weaker 10797\,\AA~line. However, we found that aligning the respective 10797\,\AA~frame to the preceding 10747\,\AA~frame (typically taken 36\,s earlier) removes most of the difficulties. Satisfactory alignment is achieved by first manually correcting the worst shifts in the 10797\,\AA~images and then running the cross-correlation routine. In this manner, both the 10747\,\AA~and 10797\,\AA~images are aligned with respect to the first 10747\,\AA~image. After this procedure, residual shifts in some frames may still be present, but these are under the \textit{CoMP} pixel size of 4.35$\arcsec$ and are thus difficult to detect with respect to the \textit{CoMP} resolution (see Figure \ref{Fig:Context}f--g).

\subsubsection{\textit{CoMP} coalignment with AIA and EIS}
\label{Sect:2.3.4}

Finally, the \textit{CoMP} observations are coaligned with the \textit{Hinode}/EIS raster using the AIA images as reference. The procedure is somewhat involved.
First, we constructed time-series of data using all six of the coronal AIA filters (94, 131, 171, 193, 211, and 335\,\AA). Then, we calculated the differential emission measure DEM($T$)\,=\,$N_\mathrm{e} N_\mathrm{H} \mathrm{d}l / \mathrm{d}T$ (see Sect. \ref{Sect:3.1}) using the regularized inversion method of \citet{Hannah12,Hannah13}. Subsequently, we synthesized the time-series of the \ion{Fe}{13} 202.0\,\AA~intensities. These synthetic intensities were then used to build a pseudo-raster \citep[see][]{delzanna_etal:2011_aia} to be compared to the observed EIS raster in the \ion{Fe}{13} 202.0\,\AA~line (see Figure \ref{Fig:Context}d--e). In doing so, we explored shifting the EIS coordinates as well as rotating AIA maps manually until a satisfactory alignment was found. As is evident from panels d--e of Figure \ref{Fig:Context}, the two instruments can be aligned quite well. To do that, a rotation of 0.5$^\circ$ of EIS relative to AIA was necessary, with no spatial shift between EIS and AIA.

As the individual \textit{CoMP} frames have already been aligned, only the reference \textit{CoMP} 10747\,\AA~frame needed to be aligned with the AIA observations. To do this, we compared the reference \textit{CoMP} frame with the syntetized AIA \ion{Fe}{13} off-limb coronal emission in the 202.0\,\AA~line. Luckily, we found that the relative rotation of \textit{CoMP} to AIA is the same as for EIS, 0.5$^\circ$. A slight image shift was also found to be present. Finally, the \textit{CoMP} co-alignment with EIS was checked and fine-tuned by producing a pseudo-raster of \textit{CoMP} data (see Figure \ref{Fig:Context}d--g). The pseudo-raster from the time-series of \textit{CoMP} data is built in the same way as for AIA by using the EIS pointing information. We find that the \textit{CoMP} instrument needs to be shifted by [5$\arcsec$, $-26\arcsec$] relative to EIS for the best co-alignment. Finally, the comparisons indicate that the effective spatial resolution of EIS is about 3--4" and COMP about 8".

\section{Synthetic spectra}
\label{Sect:3}
\subsection{Line intensities}
\label{Sect:3.1}

To perform the electron density and emission measure (EM) diagnostics, the intensity $I_{ji}$ of an emission line arising due to electron transition $j \to i$ between levels $j > i$ is calculated as \citep[cf.,][]{Phillips08,DelZanna18}
\begin{equation}
	I_{ji} = \int A_X G_{X,ji}(T,N_\mathrm{e}) N_\mathrm{e} N_\mathrm{H} \mathrm{d}l\,,
	\label{Eq:line_intensity}
\end{equation}
where the integration arises due to the optically thin nature of the solar corona, and is done along the line of sight $l$. In this Equation, $N_\mathrm{e}$ and $N_\mathrm{H}$ are the electron and (total) hydrogen densities, respectively, $A_X$ is the abundance of element $X$, and $G_{X,ji}(T,N_\mathrm{e})$ is the line contribution function, which can be expressed as
\begin{equation}
	G_{X,ji}(T,N_\mathrm{e}) = \frac{hc}{\lambda_{ji}} \frac{A_{ji}}{N_\mathrm{e}} \frac{N(X_j^{+k})}{N(X^{+k})} \frac{N(X^{+k})}{N(X)}\,,
	\label{Eq:G(T)}
\end{equation}
where $hc/\lambda_{ji}$ is the photon energy, $\lambda_{ji}$ is the line wavelength, and $A_{ji}$ is the Einstein coefficient for the spontaneous emission. The ratio $N(X_j^{+k}) / N(X^{+k})$ denotes the relative population of the upper level $j$ within the ion $X^{+k}$, while the ratio $N(X^{+k})/N(X)$ is the relative ion abundance of this ion.

To calculate the contribution functions and the line intensities, we use the CHIANTI database and software, version 9 \citep{Dere97,Dere19}. The excitation cross-sections and $A_{ji}$ values for \ion{Fe}{12} and \ion{Fe}{13} are from \citet{DelZanna12c} and \citet{DelZanna12a}, respectively.

%
\begin{figure}
   \centering
   \includegraphics[width=8.8cm,viewport= 0 50 495 335,clip]{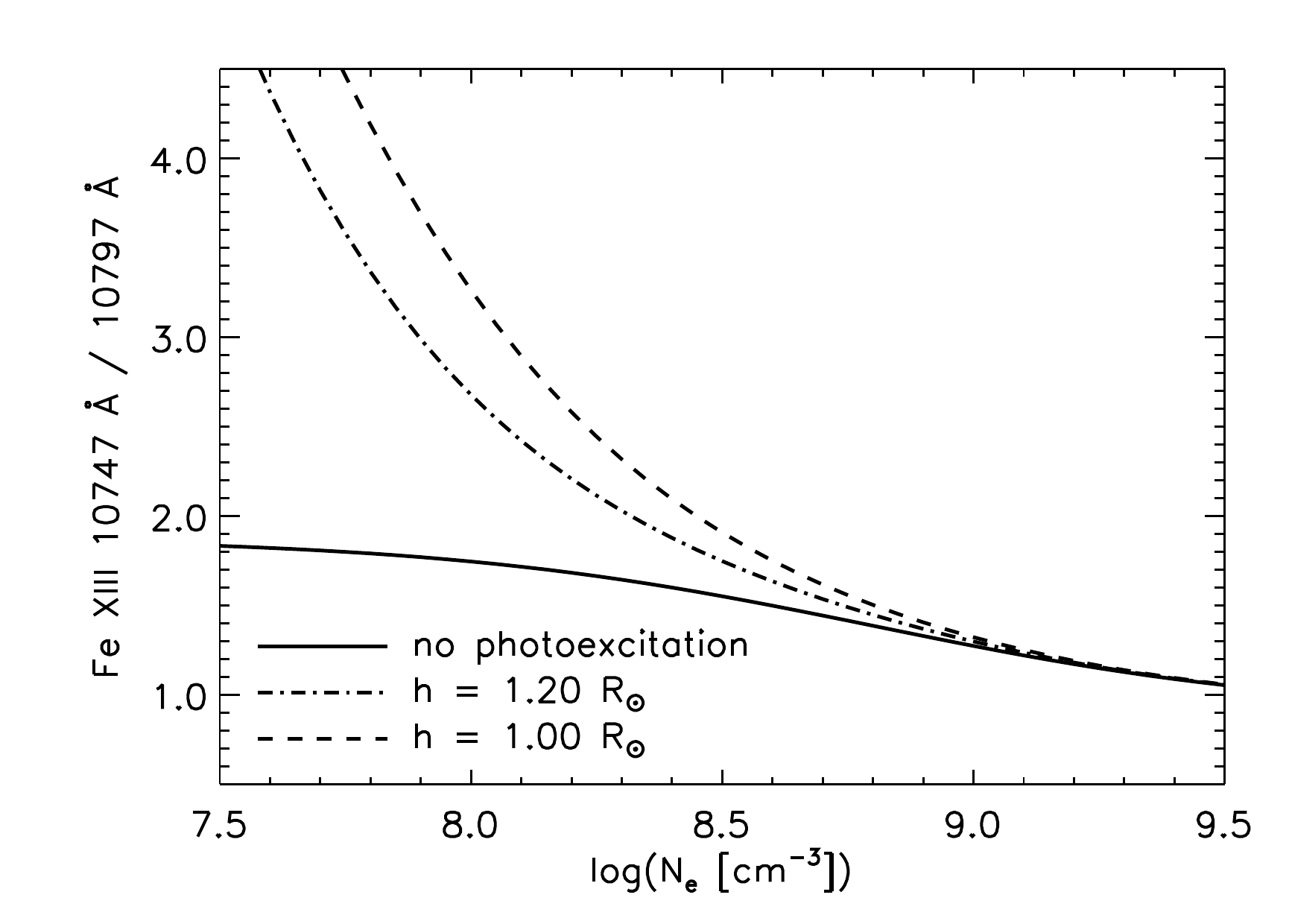}
   \includegraphics[width=8.8cm,viewport= 0 50 495 335,clip]{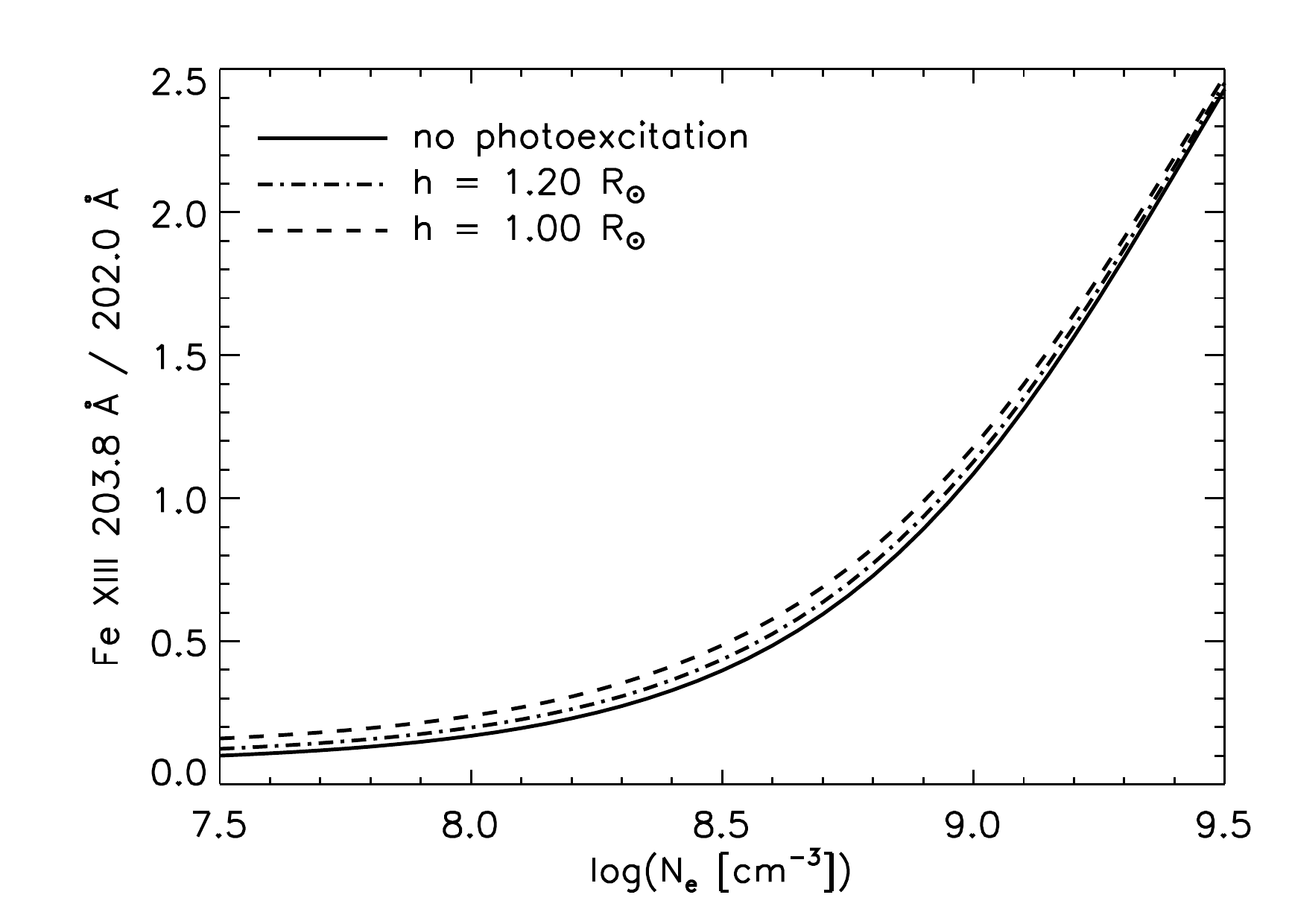}
   \includegraphics[width=8.8cm,viewport= 0  0 495 335,clip]{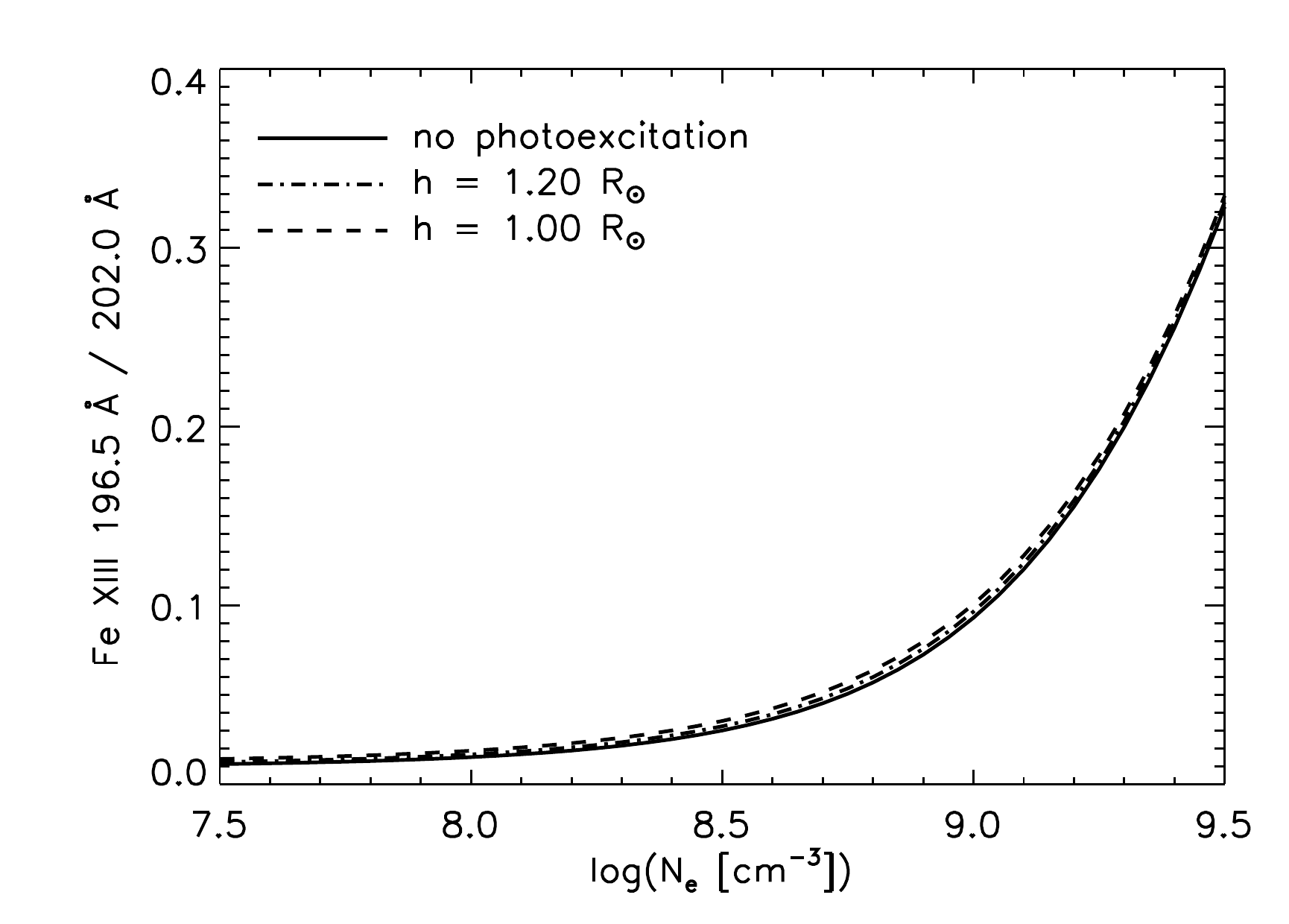}
   \caption{\ion{Fe}{13} line intensity ratios for diagnostics of electron density. The ratios are shown with and without inclusion of photoexcitation.}
   \label{Fig:photoexcit}
\end{figure}
%
%
\subsection{Photoexcitation}
\label{Sect:3.2}

In the low-density outer corona, the photoexcitation (PE) due to the disk radiation (mostly photospheric) becomes a non-negligible excitation mechanism compared to electron-ion collisions.
The magnitude of this effect is dependent on the height above photosphere, due to dilution of the photospheric radiation, as well as on the ambient electron densities. At constant electron density, the PE contribution decreases with height due to the dilution of the radiation, while at constant height, the contribution of photoexcitation increases with decreasing electron densities. The effects can be subtle, depending on the atom considered. Resonant PE occurs when the same transition is pumped from the disk radiation. This is an important process for the 
visible and NIR lines, considering the large number of photons emitted by the disk. But indirect PE can also occur, i.e. PE can increase the populations of some states which then affects other transitions.

As \ion{Fe}{13} has two strong NIR transitions within the ground configuration, PE must be considered when calculating the synthetic line intensities \citep[see, e.g.][]{chevalier_lambert:69,Pineau73,Young94}. We include it by considering that the photospheric radiation can be approximated by a black-body with radiation temperature of 6000\,K, and perform a grid of line intensity calculations with the height $h$ above the photosphere ranging from $h$ = 1.0\,R$_\odot$ to 1.3\,R$_\odot$ with a step of 0.02\,R$_\odot$. Interpolating this grid to a given value of $h$ determined from observations leads to line intensities within 0.5\% of their values if calculated at the same $h$.

It is well known that PE affects significantly the intensities of the NIR lines. In turn, the ratio of the \ion{Fe}{13} NIR lines is also strongly affected (top panel of Figure \ref{Fig:photoexcit}). However, as PE affects the population of the ground configuration states, it also leads to a small (but non-negligible) effect for some of the EUV line intensities \citep[as noted by][]{Young09}. The \ion{Fe}{13} levels affected by PE via the two NIR transitions are level 2 (3s$^2$\,3p$^2$ $^3$P$_1$), which shows significant enhancement at \logne\,$\lesssim$\,9 and level 3 (3s$^2$\,3p$^2$ $^3$P$_2$), which shows significant enhancement at \logne\,$\lesssim$\,7. The changes in the population of these levels in turn affect the ground state $^3$P$_0$ as well as level 4 (3s$^2$\,3p$^2$ $^1$D$_2$). Table \ref{Table:fe13} lists the strong \ion{Fe}{13} lines we use for diagnostics, together with their selfblends and the corresponding levels between which the transitions occur.  It is clear that PE affects our diagnostic lines.

To illustrate this effect, we show in Figure \ref{Fig:photoexcit} the \ion{Fe}{13} density-sensitive line intensity ratios. The most conspicuous changes are in the 203.83\,\AA~/\,202.04,\AA~ratio, which is shifted towards lower \logne than if no photoexcitation is included \citep[cf., Figure 2 in][]{Young09}. At densities below \logne =\,8.5, this effect can be noticeable, being larger than 0.1 dex (Figure \ref{Fig:photoexcit}, middle panel). This effect is mostly due to the behaviour of the \ion{Fe}{13} 203.83\,\AA~selfblend, whose $G(T)$ increases by up to $\approx$20\% with photoexcitation at \logne\,=\,8.0. In contrast to that, the $G(T)$ of the strong resonance 202.04\,\AA~line (a decay to the ground state) is \textit{decreased} by up to 15\%. The 196.53\,\AA~/\,202.04\,\AA~line ratio is less affected (Figure \ref{Fig:photoexcit}, bottom panel), since the 196.53\,\AA~line has its $G(T)$ increased by only $\lesssim$5\%.

Finally, as the black-body is an actual approximation to the solar spectrum, we verified that the density-sensitive line ratios are not strongly sensitive to the choice of the black-body temperature. At \logne\,=\,8.0, changing the radiation temperature by $\pm$100\,K results in changes to  the intensities of the 10747\,\AA~and 10797\,\AA~IR lines by less than about 2\% and 1\%, respectively. The resulting changes to the density-sensitive line intensity ratios are negligible, being within than about 0.05 dex in electron density.

%
\begin{figure}
   \centering
   \includegraphics[width=4.18cm,viewport=  0 0 225 575,clip]{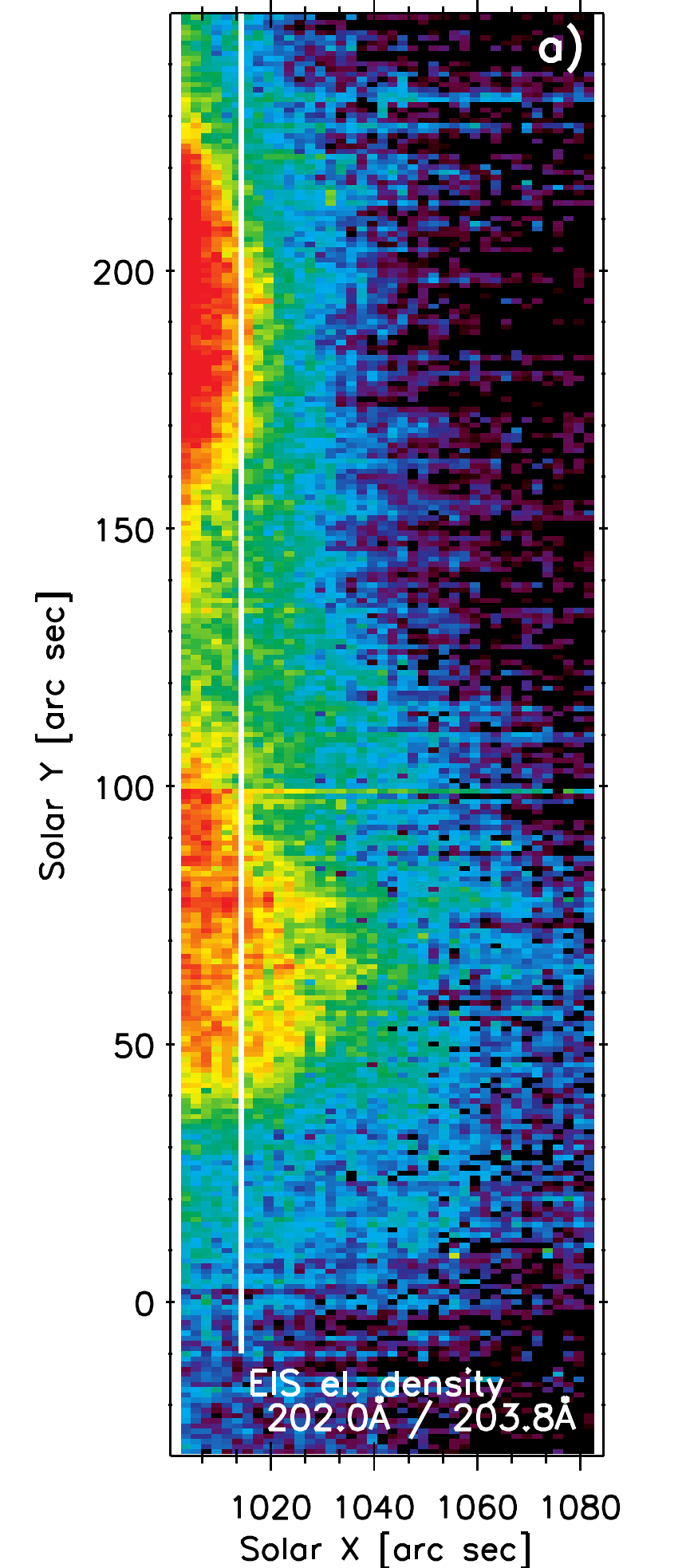}
   \includegraphics[width=3.07cm,viewport= 60 0 225 575,clip]{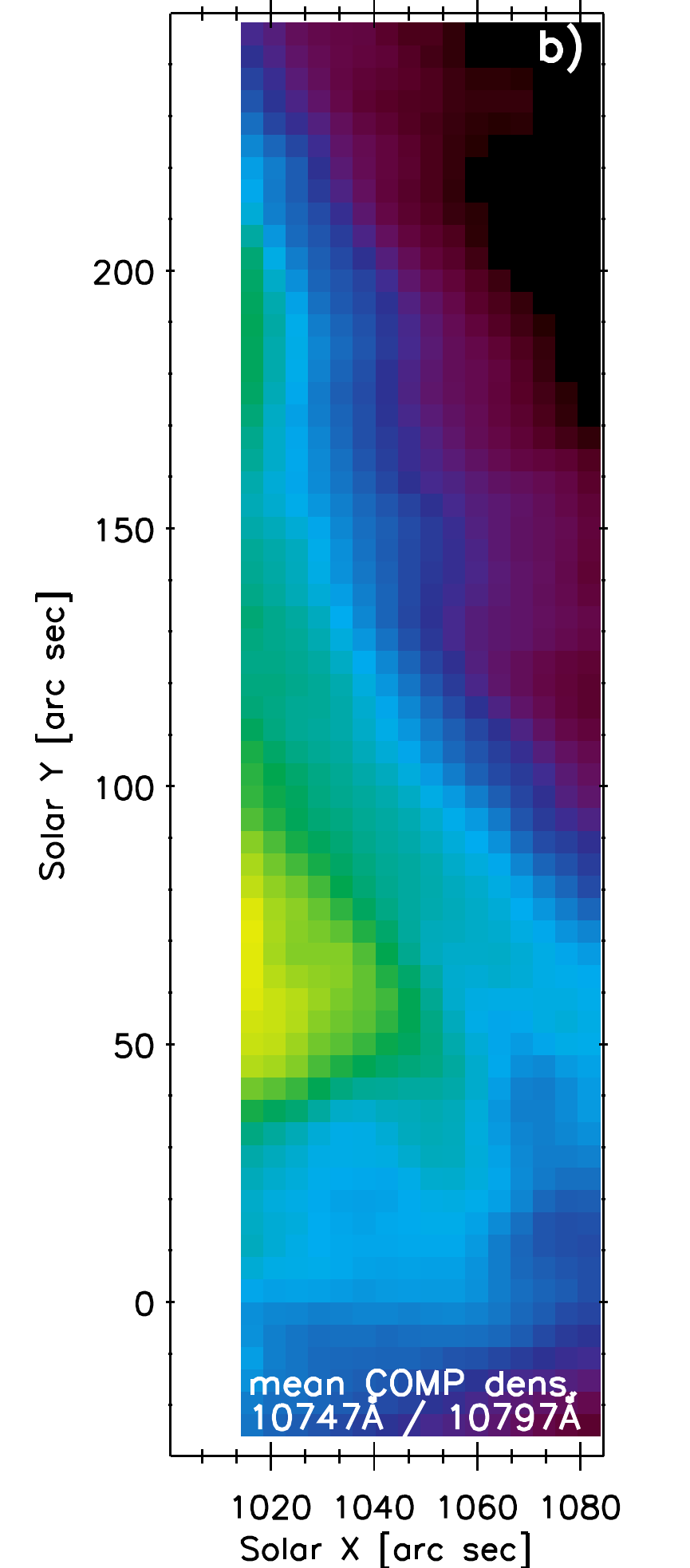}
   \includegraphics[width=1.05cm,viewport=155 0 215 575,clip]{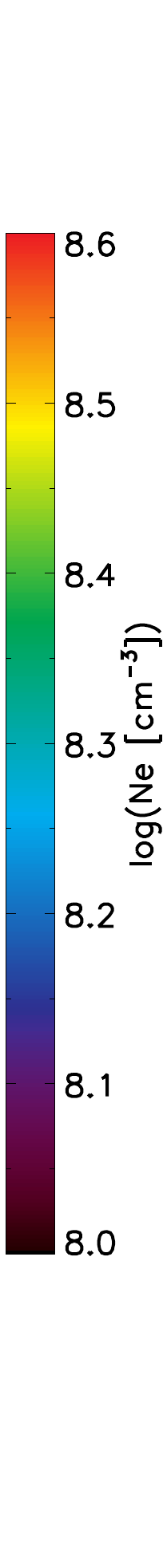}
   \caption{Maps of the electron density derived from (a) the EIS \ion{Fe}{13} 202.0\,/\,203.8\,\AA~and (b) \textit{CoMP} \ion{Fe}{13} 10747\,/\,10797\,\AA~ratios. Note that the COMP densities are averaged over the duration of the EIS raster. The vertical white line in panel (a) denotes the edge of \textit{CoMP} observations.}
   \label{Fig:Density_maps}
\end{figure}

%
\begin{figure*}
\centering
\includegraphics[width=8.00cm,viewport=30 10 480 360,clip]{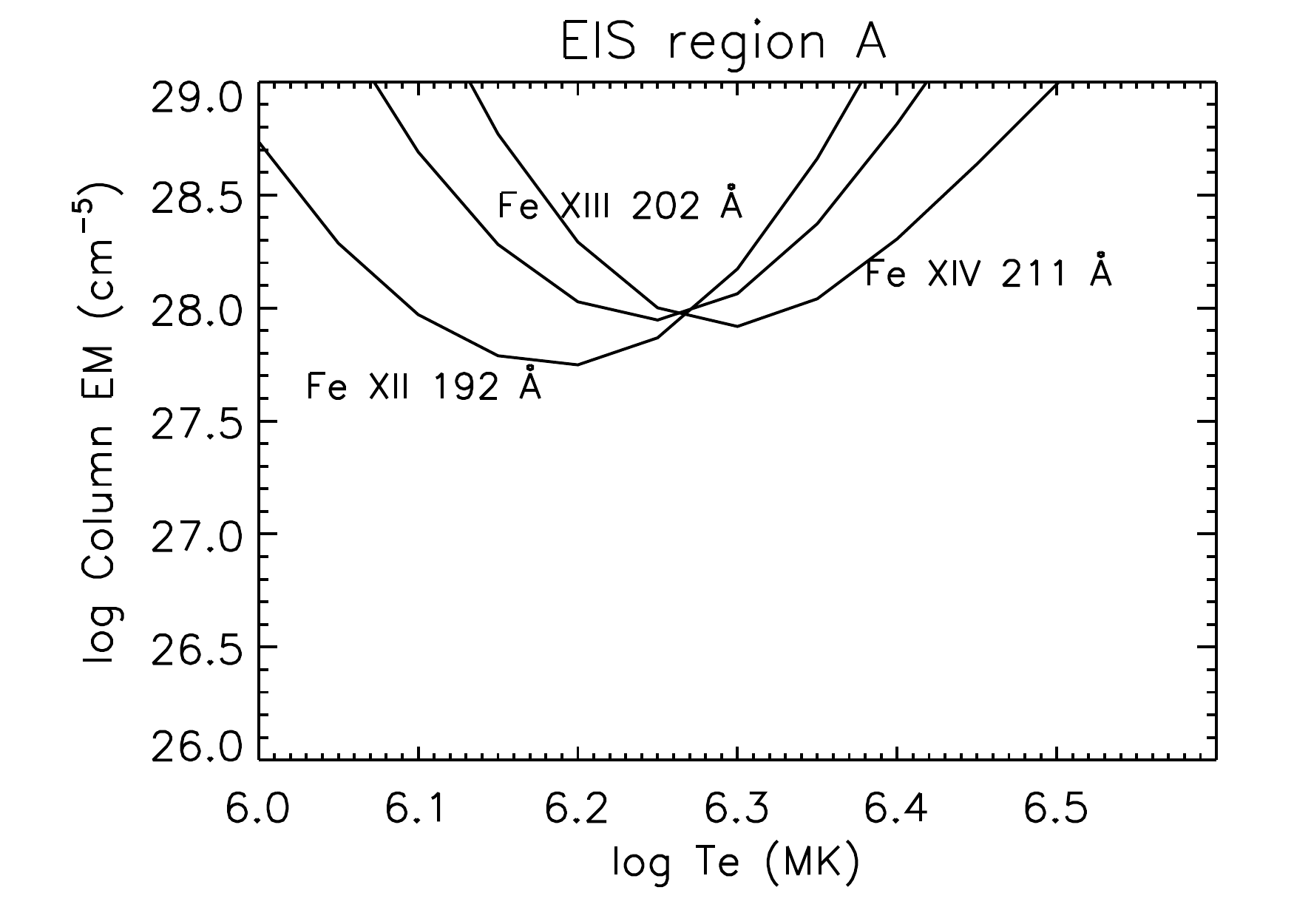}
\includegraphics[width=8.00cm,viewport=30 10 480 360,clip]{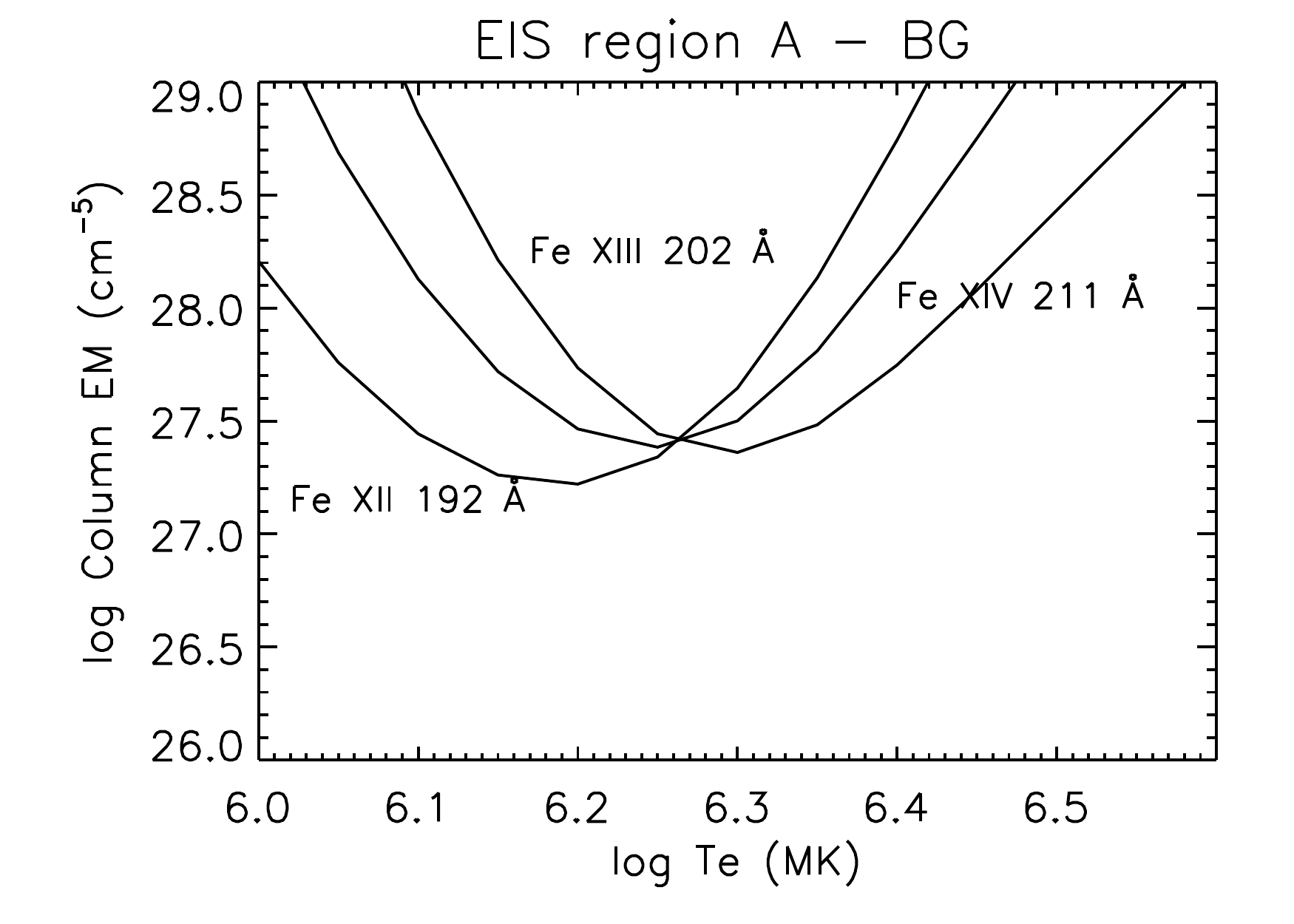}
\includegraphics[height=8.00cm,angle=-90,viewport= 0 40 500 720,clip]{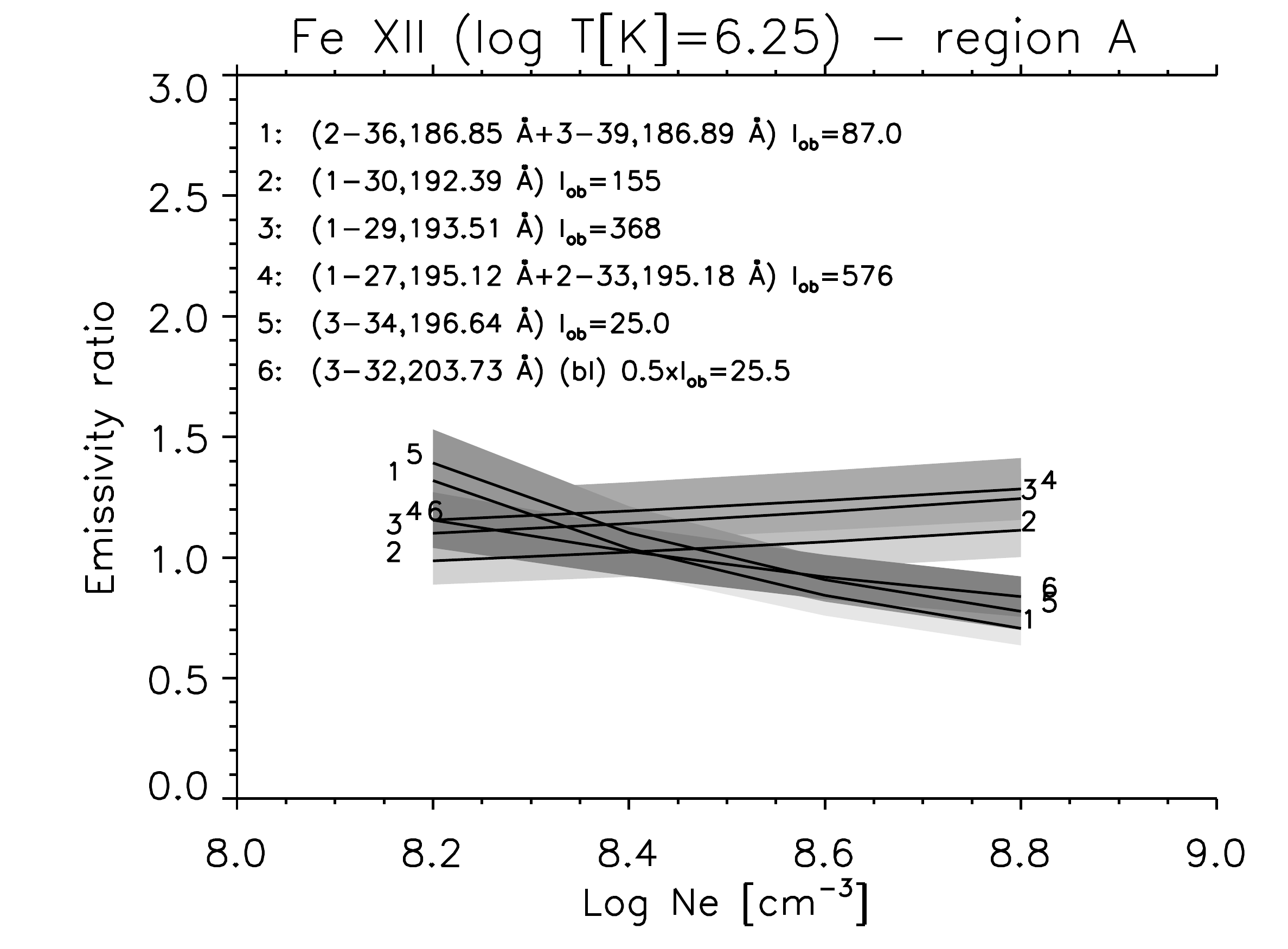}
\includegraphics[height=8.00cm,angle=-90,viewport= 0 40 500 720,clip]{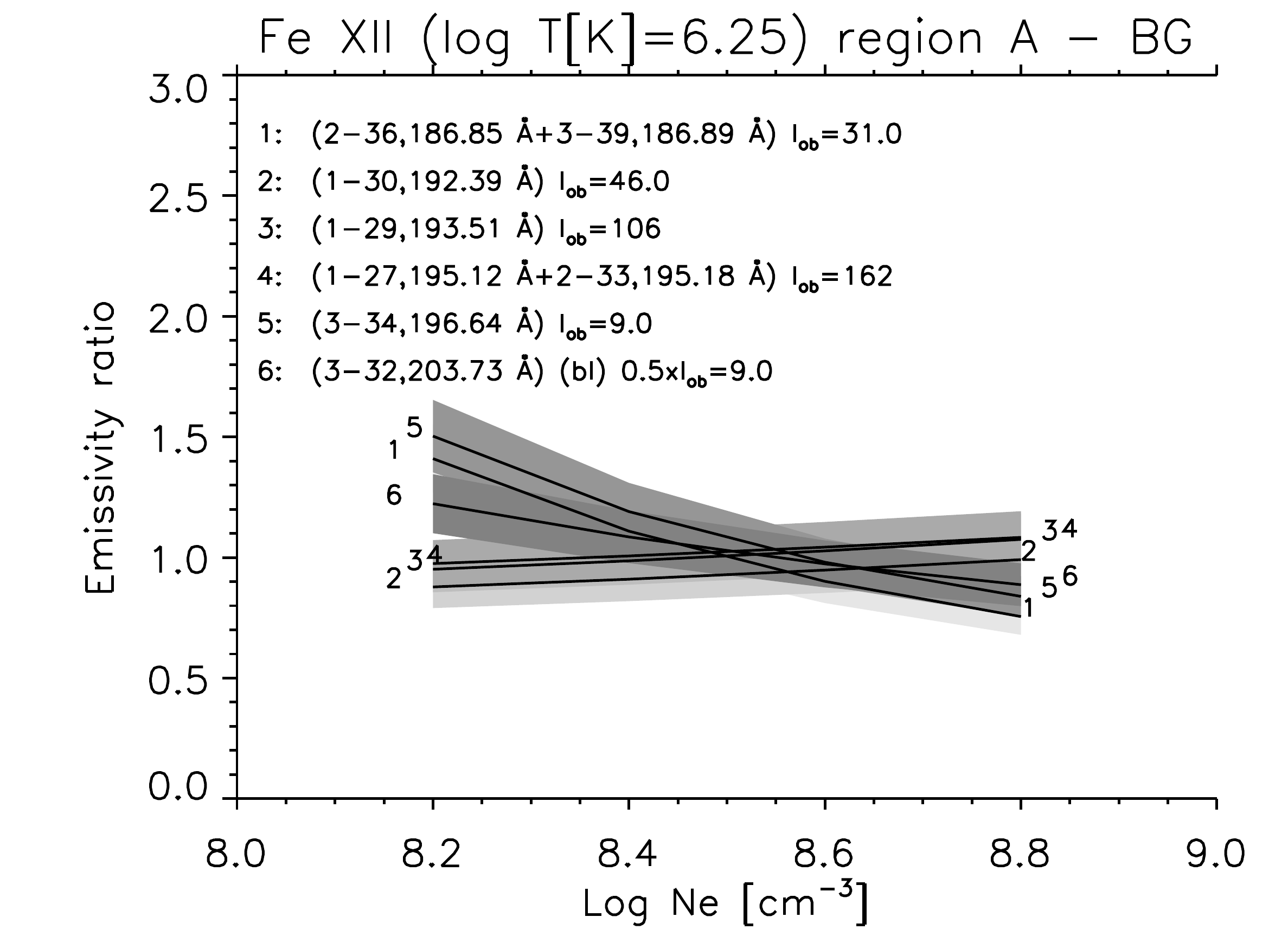}
\includegraphics[height=8.00cm,angle=-90,viewport= 0 40 430 720,clip]{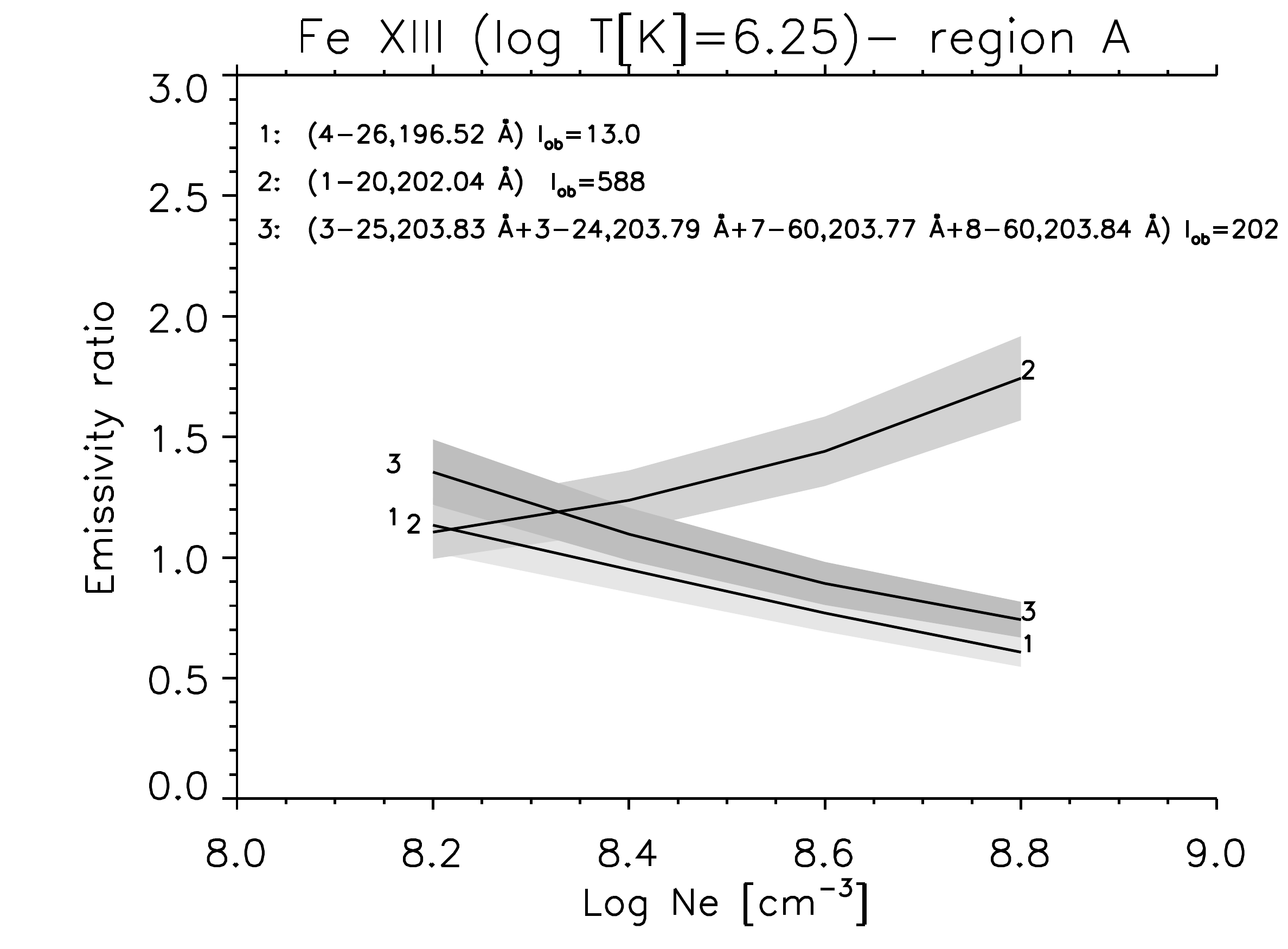}
\includegraphics[height=8.00cm,angle=-90,viewport= 0 40 430 720,clip]{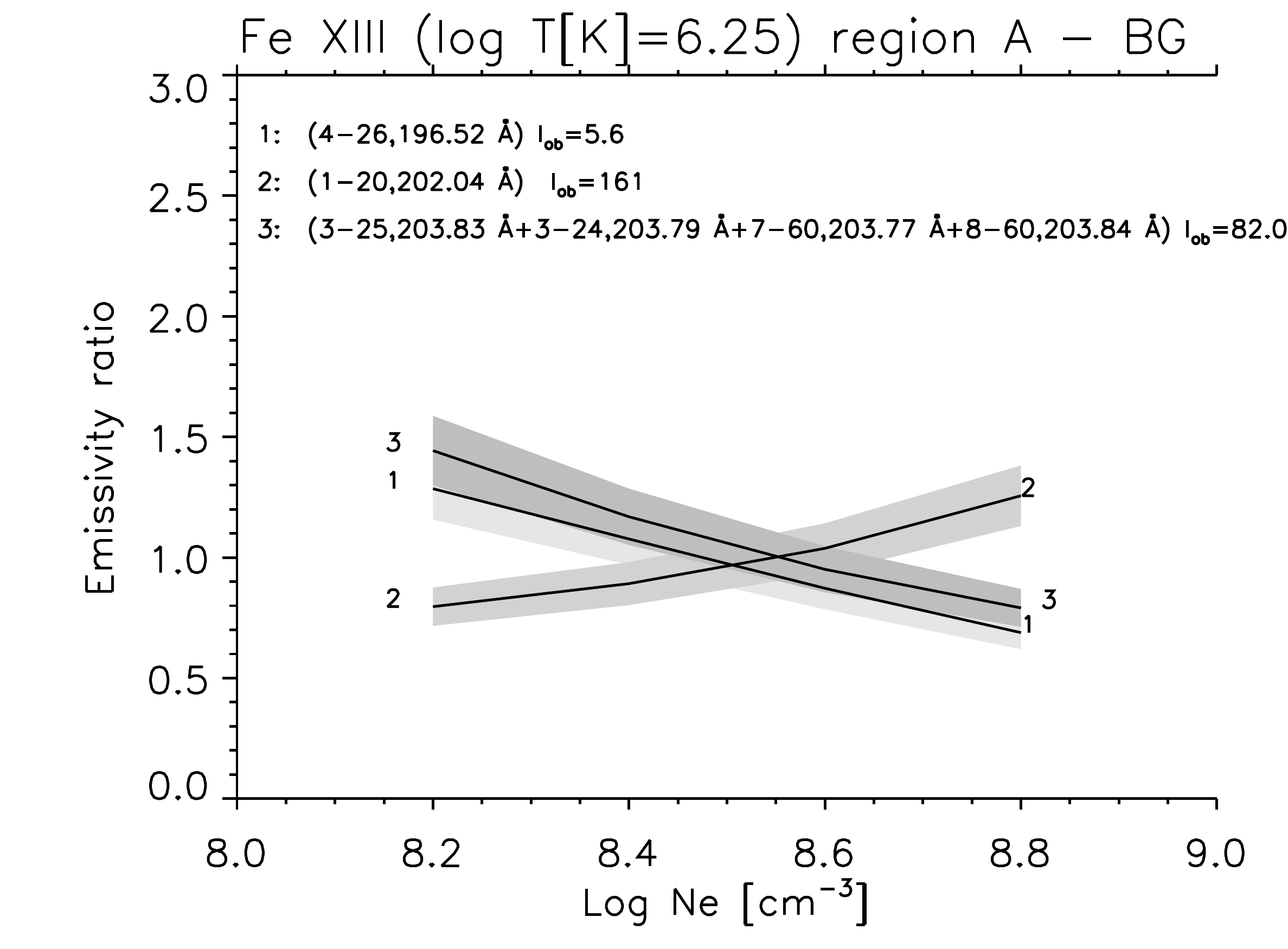}
\includegraphics[width=8.00cm,viewport=40  0 720 468,clip]{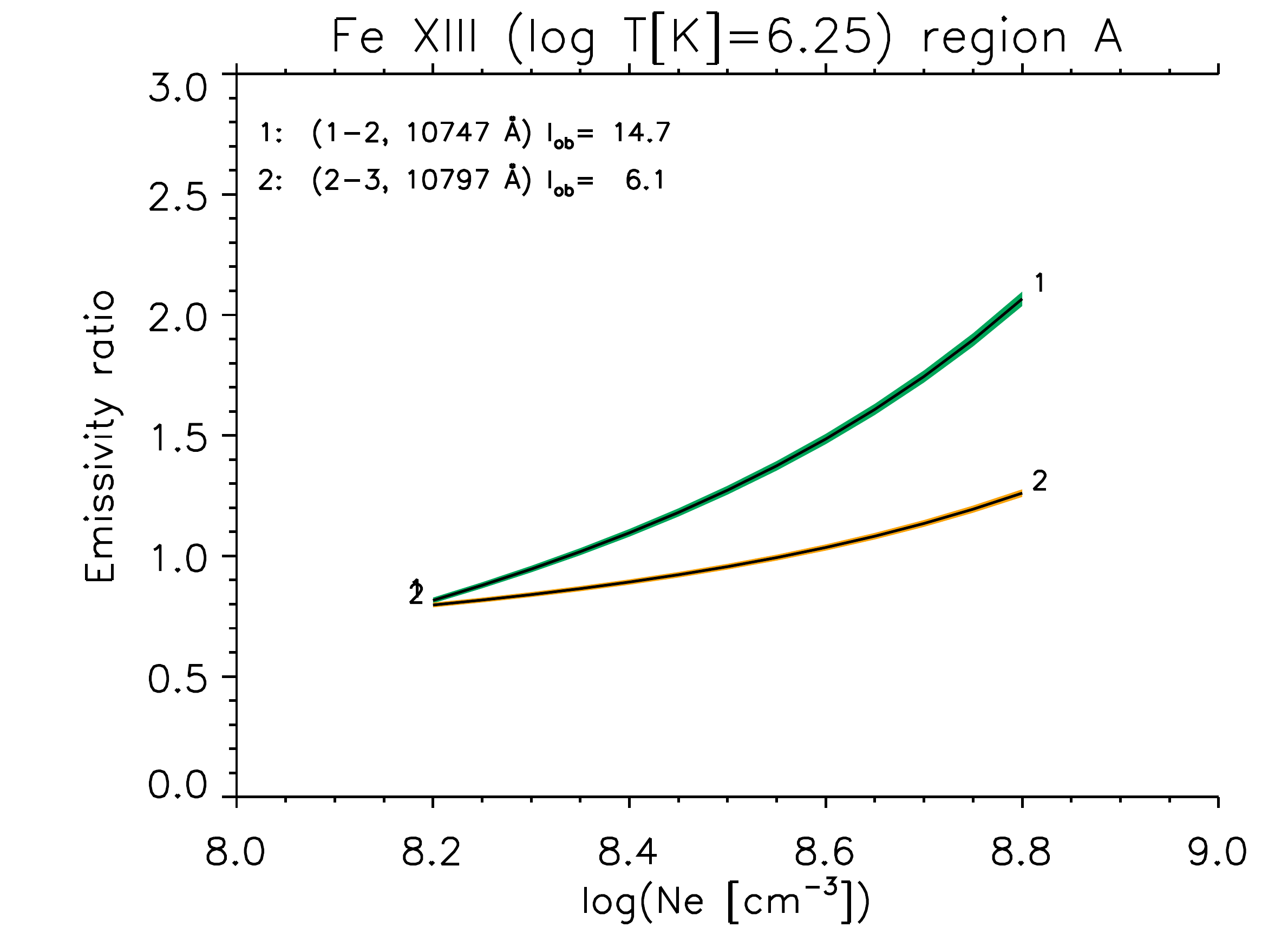}
\includegraphics[width=8.00cm,viewport=40  0 720 468,clip]{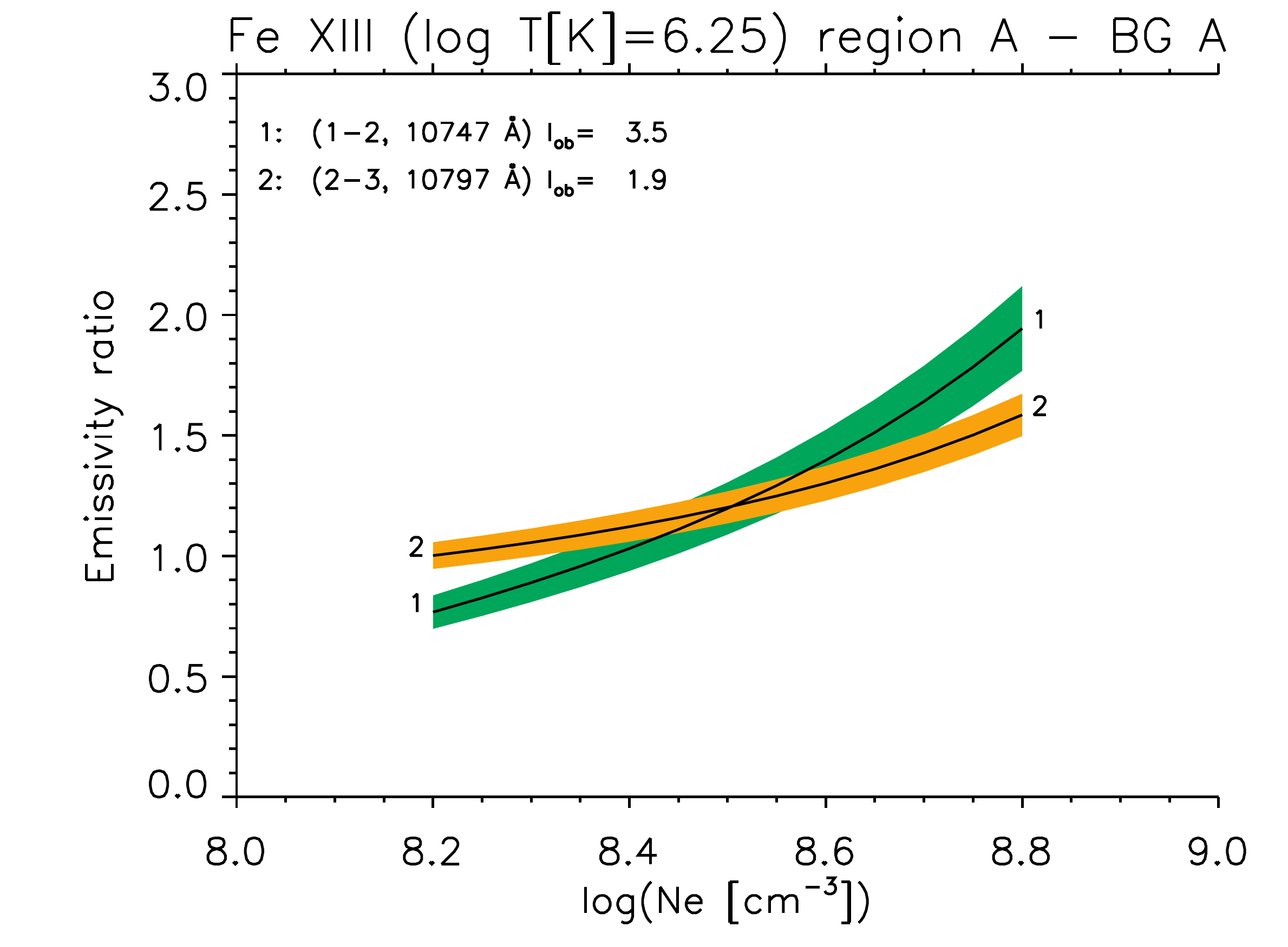}
\caption{Plasma diagnostics in region A. \textit{Left} and \textit{right} show the results without and with background subtracted. The first row shows the emission measure loci plots, while rows 2--4 show the emissivity ratio plots for \ion{Fe}{12} (row 2) and \ion{Fe}{13} (rows 3--4) from EIS and \textit{CoMP} instruments. Emission lines and their observed intensities are indicated in the legend of each panel. The units are ergs\,cm$^{-2}$\,s$^{-1}$\,sr$^{-1}$ for EIS and in milionths of solar disk intensity for \textit{CoMP}. }
\label{Fig:region_A}
\end{figure*}
%
%
\begin{figure}
\centering
\includegraphics[height=8.00cm,angle=-90,viewport= 0 40 500 720,clip]{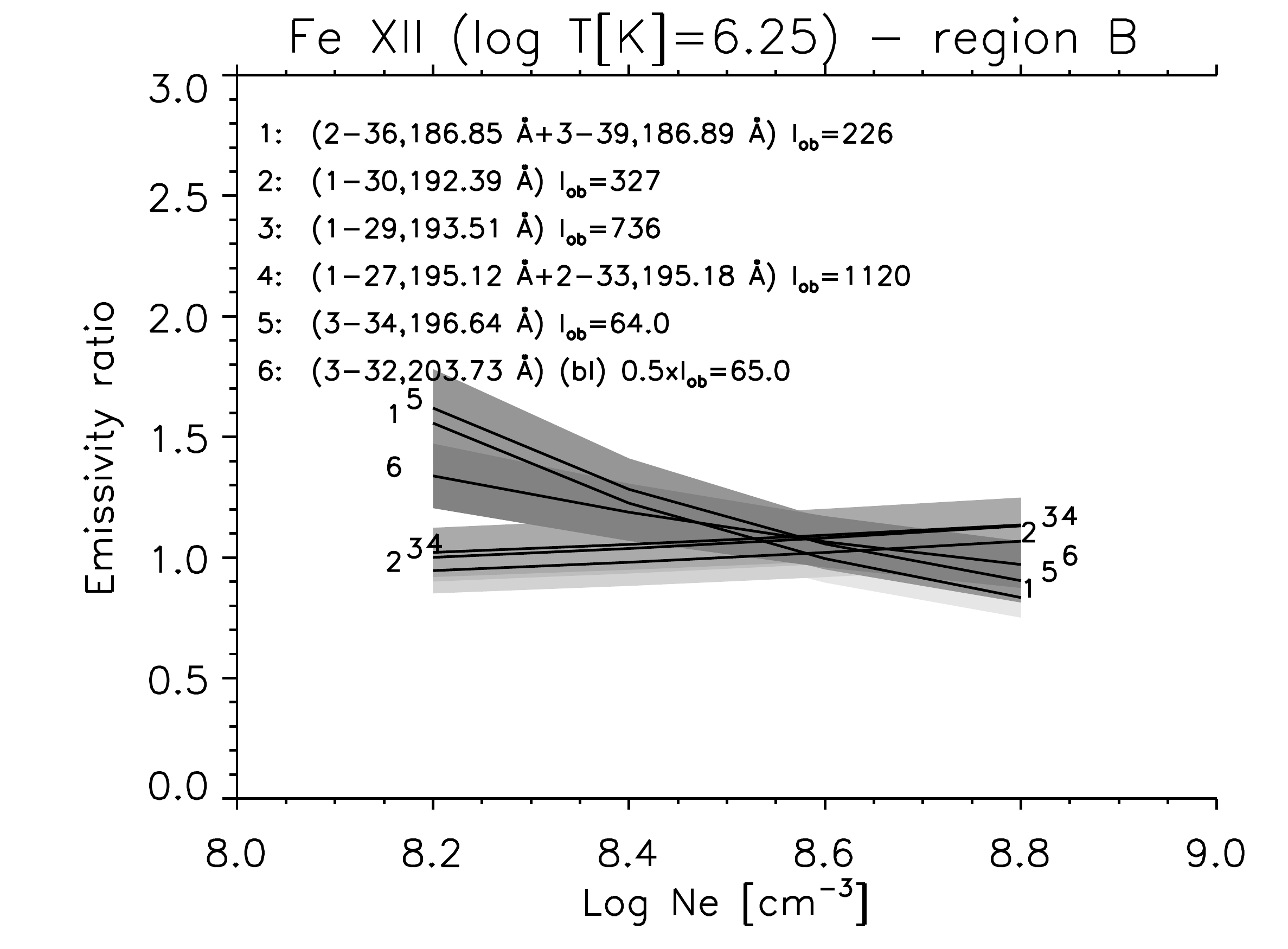}
\includegraphics[height=8.00cm,angle=-90,viewport= 0 40 430 720,clip]{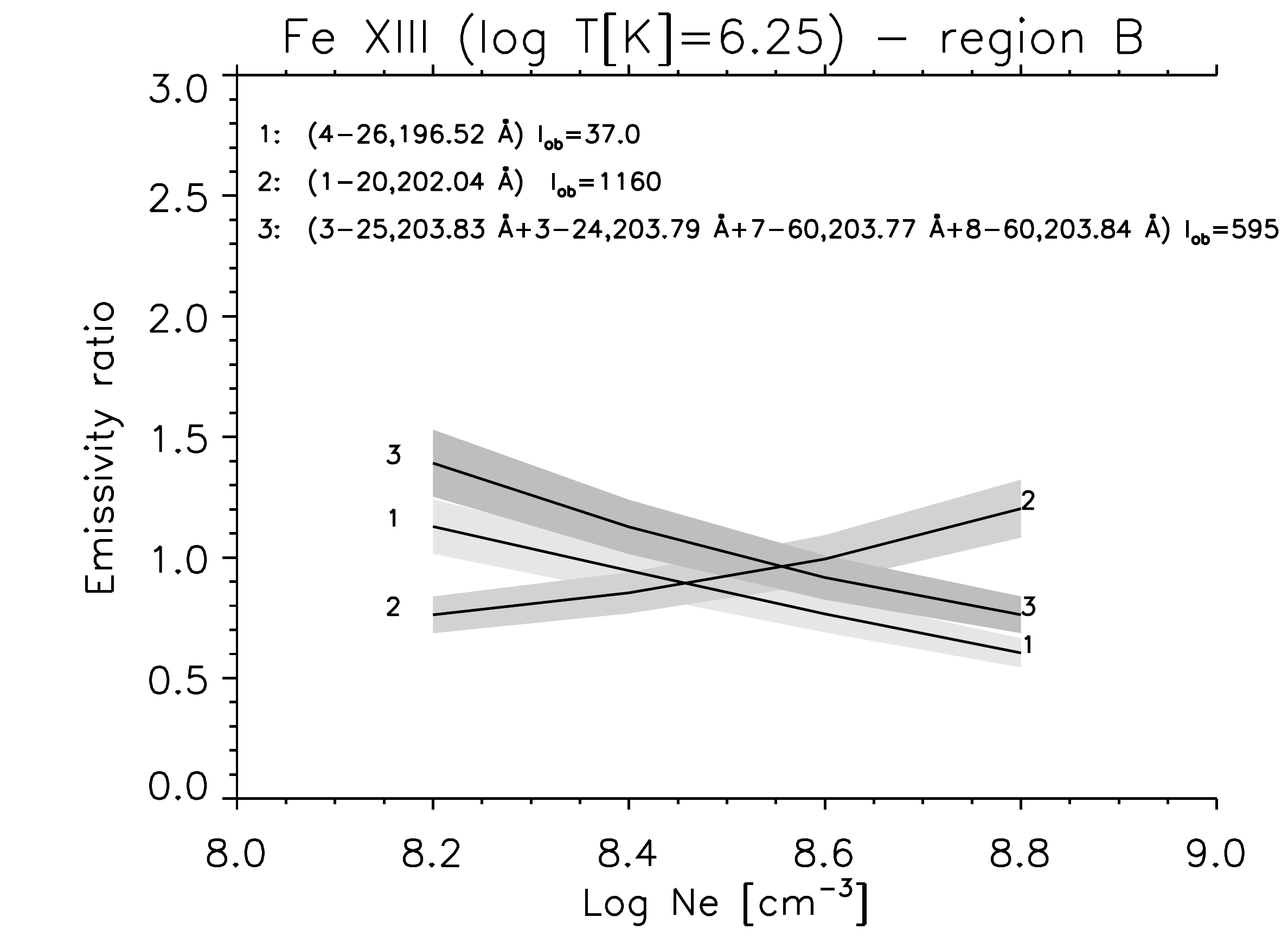}
\includegraphics[width=8.00cm,viewport=40  0 720 468,clip]{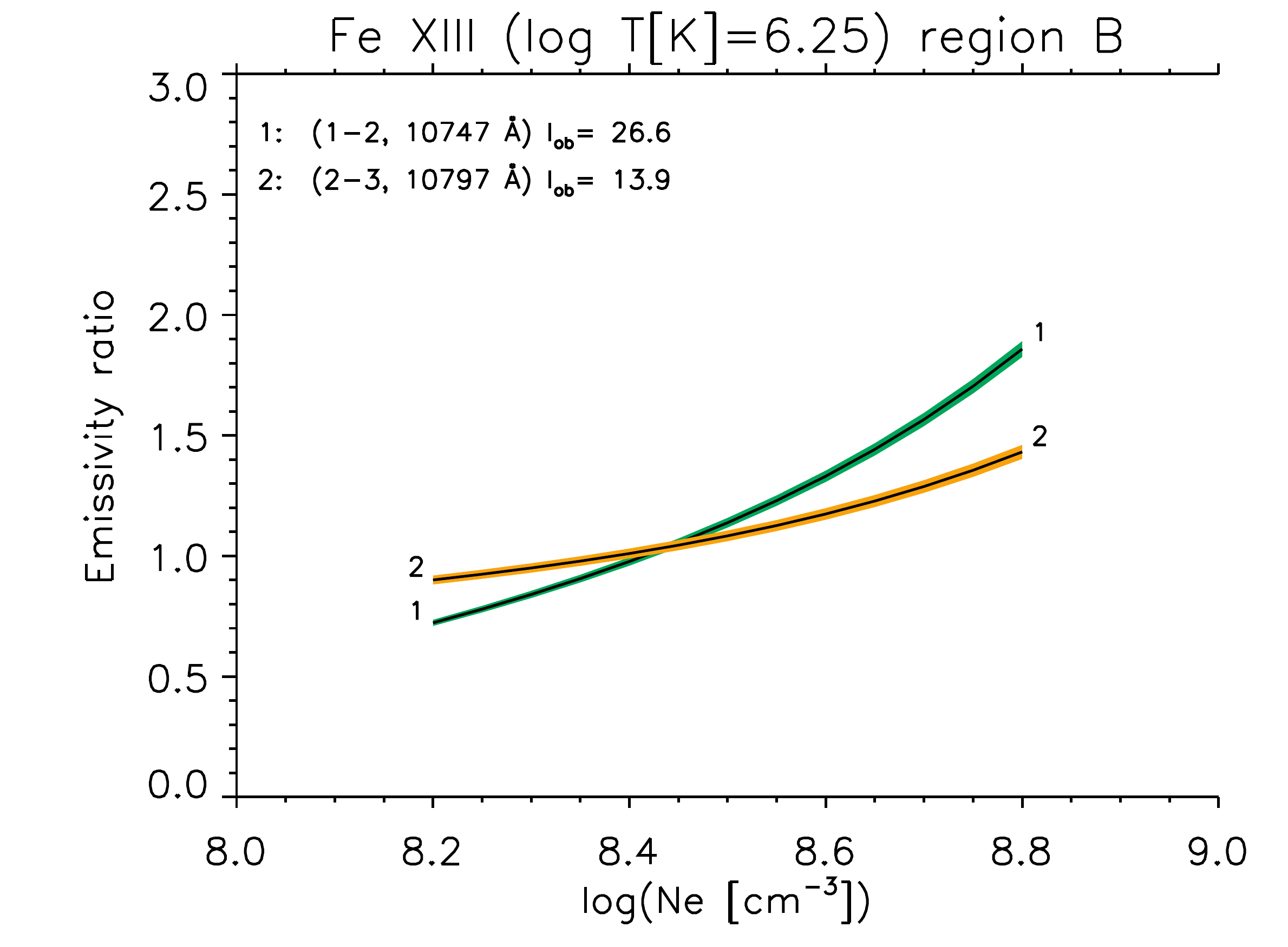}
\caption{Density diagnostics using the emissivity ratio method in Region B. Emission lines observed by EIS and \textit{CoMP} are indicated in the legend of each panel together with their observed intensities.}
\label{Fig:region_B}
\end{figure}
%
%
\begin{figure}
\centering
\includegraphics[height=8.00cm,angle=-90,viewport= 0 40 500 720,clip]{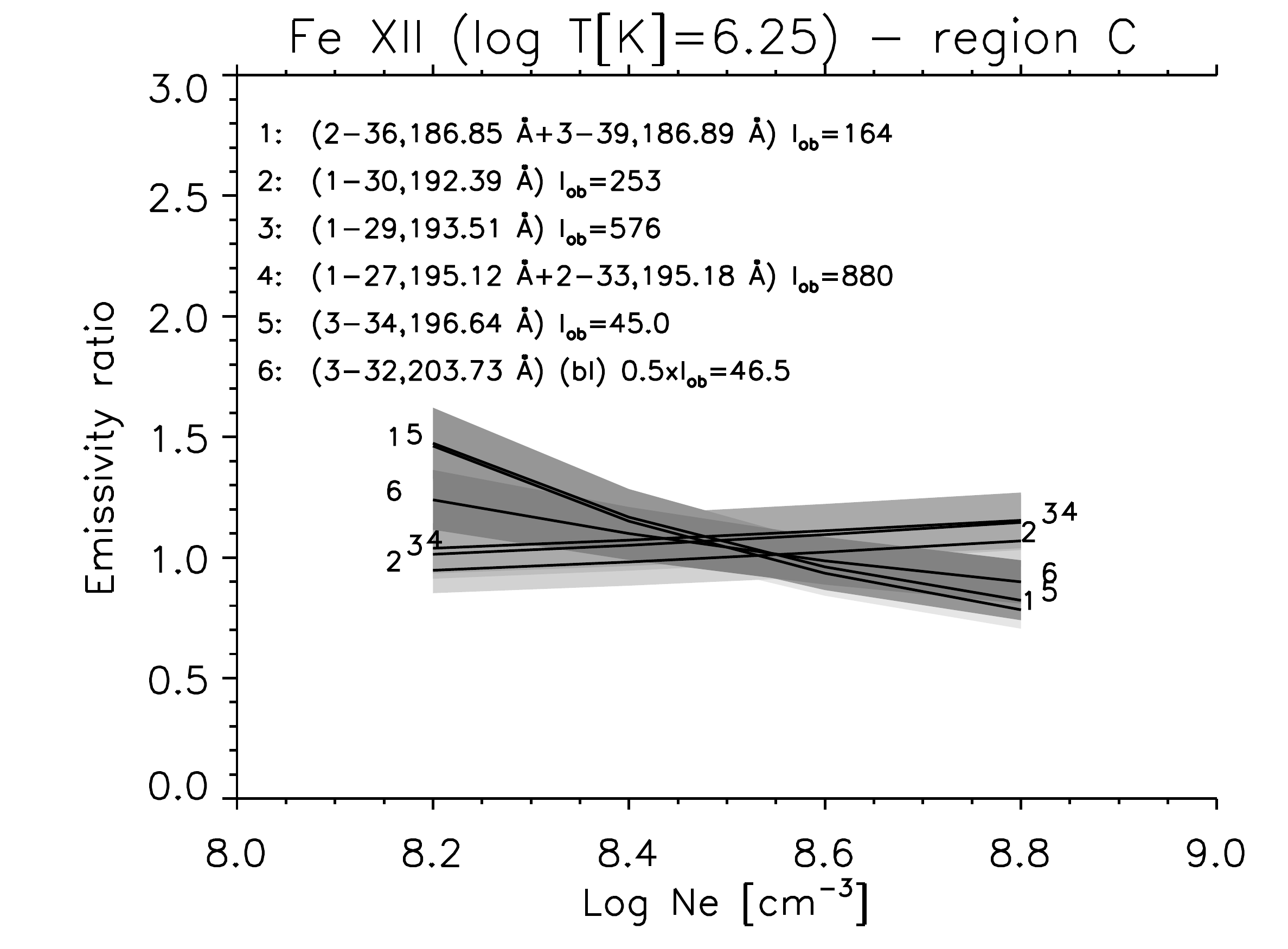}
\includegraphics[height=8.00cm,angle=-90,viewport= 0 40 430 720,clip]{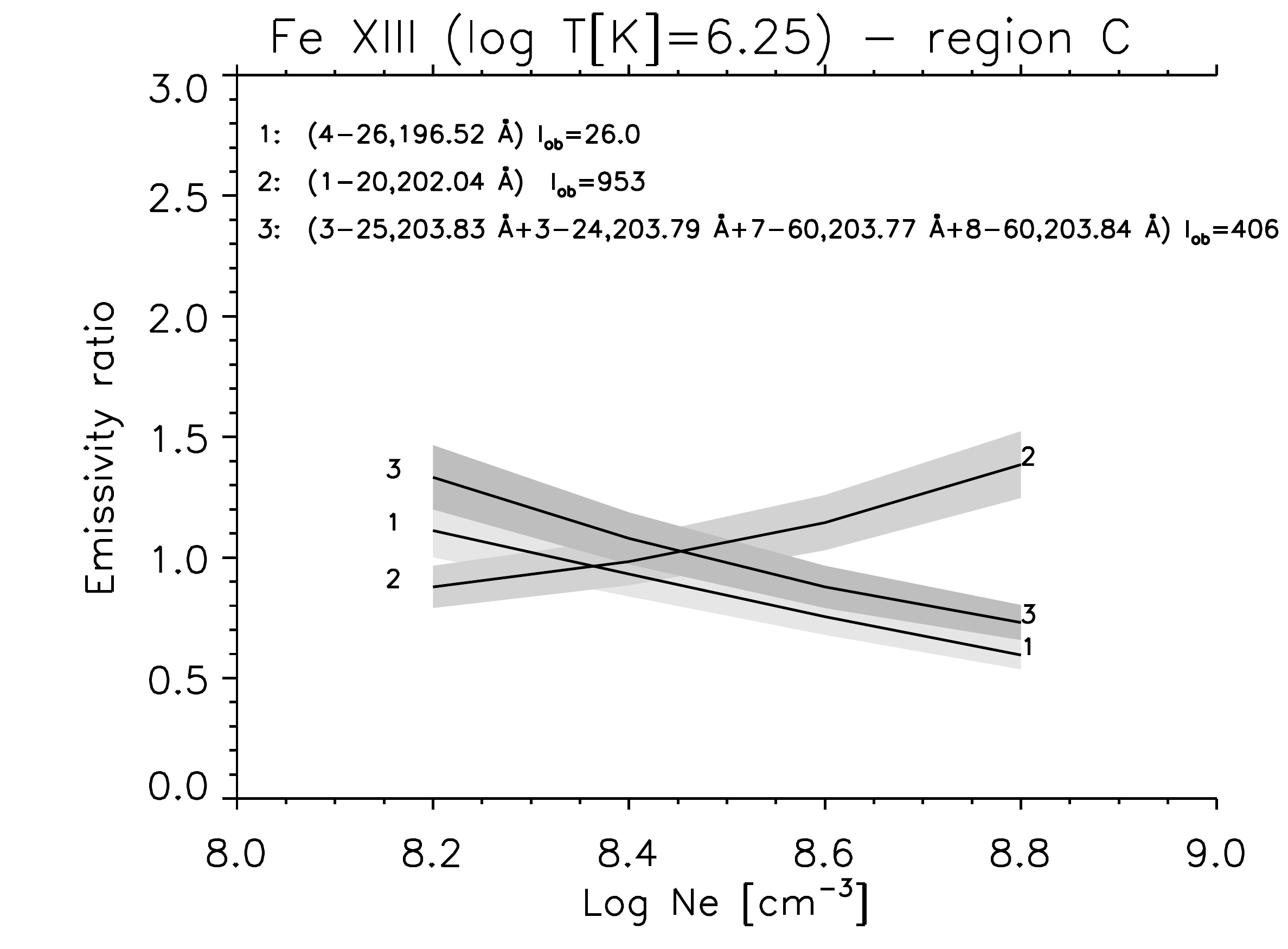}
\includegraphics[width=8.00cm,viewport=40  0 720 468,clip]{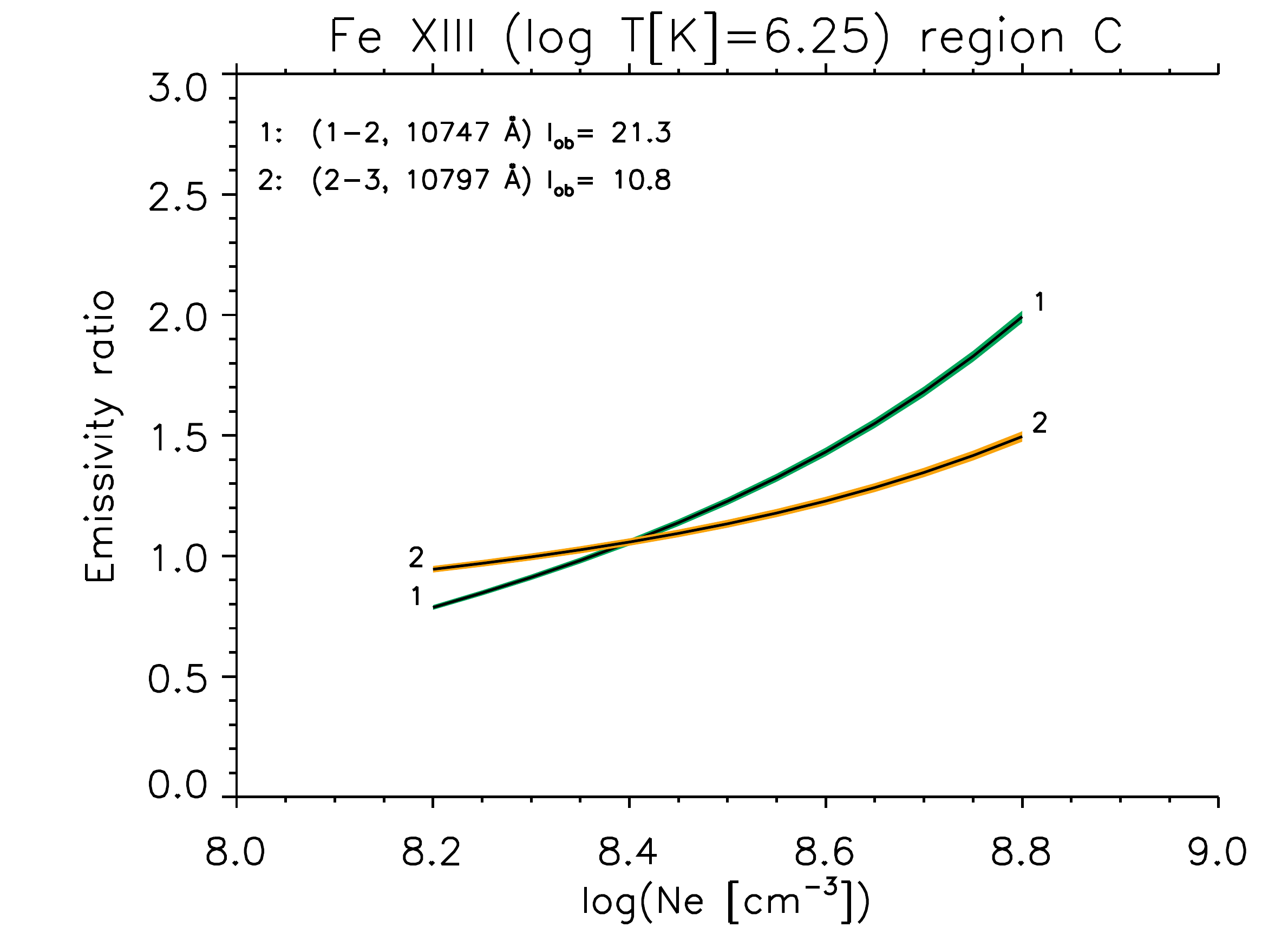}
\caption{Same as in Figure \ref{Fig:region_B}, but for Region C.}
\label{Fig:region_C}
\end{figure}
%
%
\begin{figure}
\centering
\includegraphics[height=8.00cm,angle=-90,viewport= 0 40 500 720,clip]{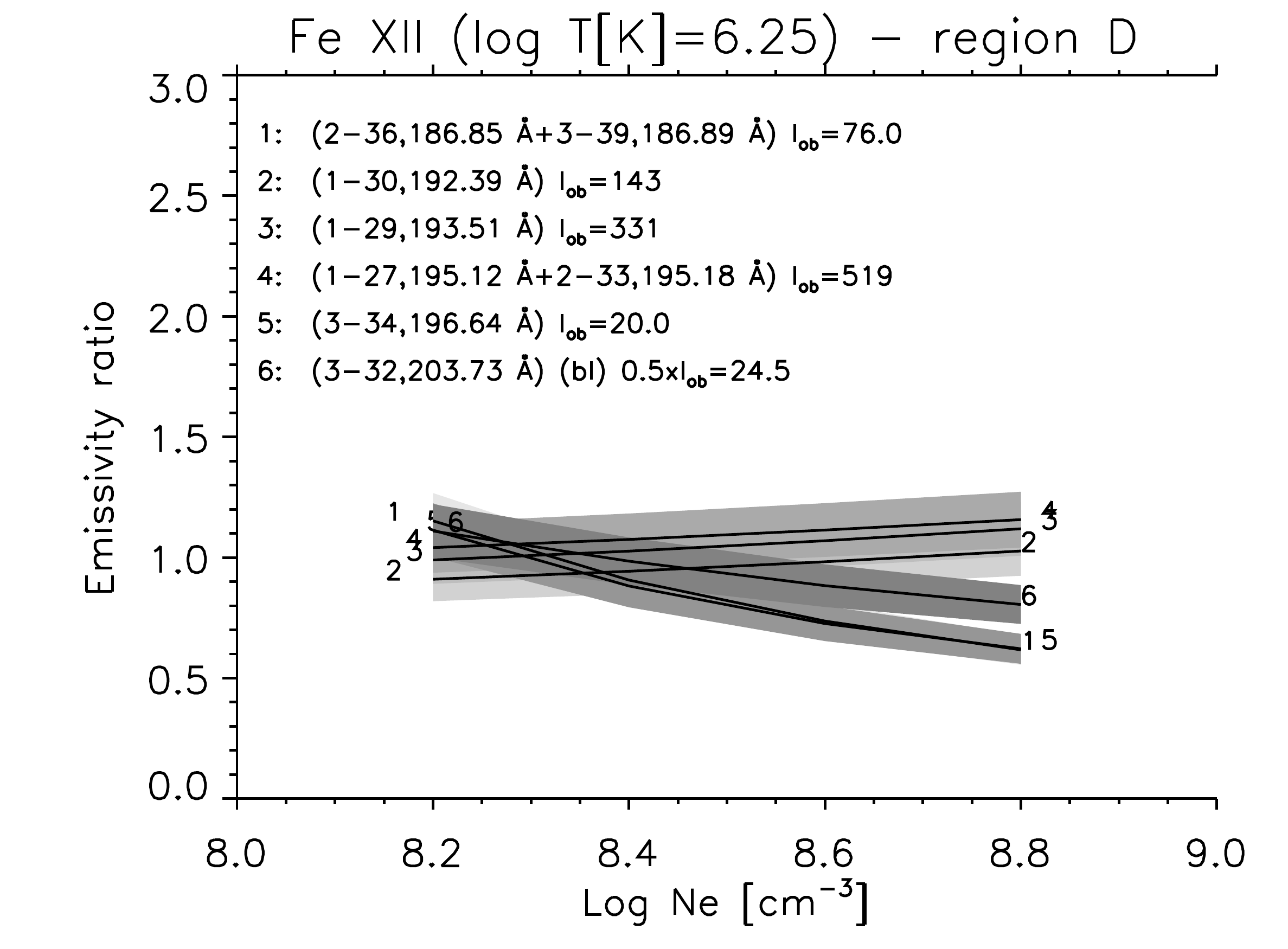}
\includegraphics[height=8.00cm,angle=-90,viewport= 0 40 430 720,clip]{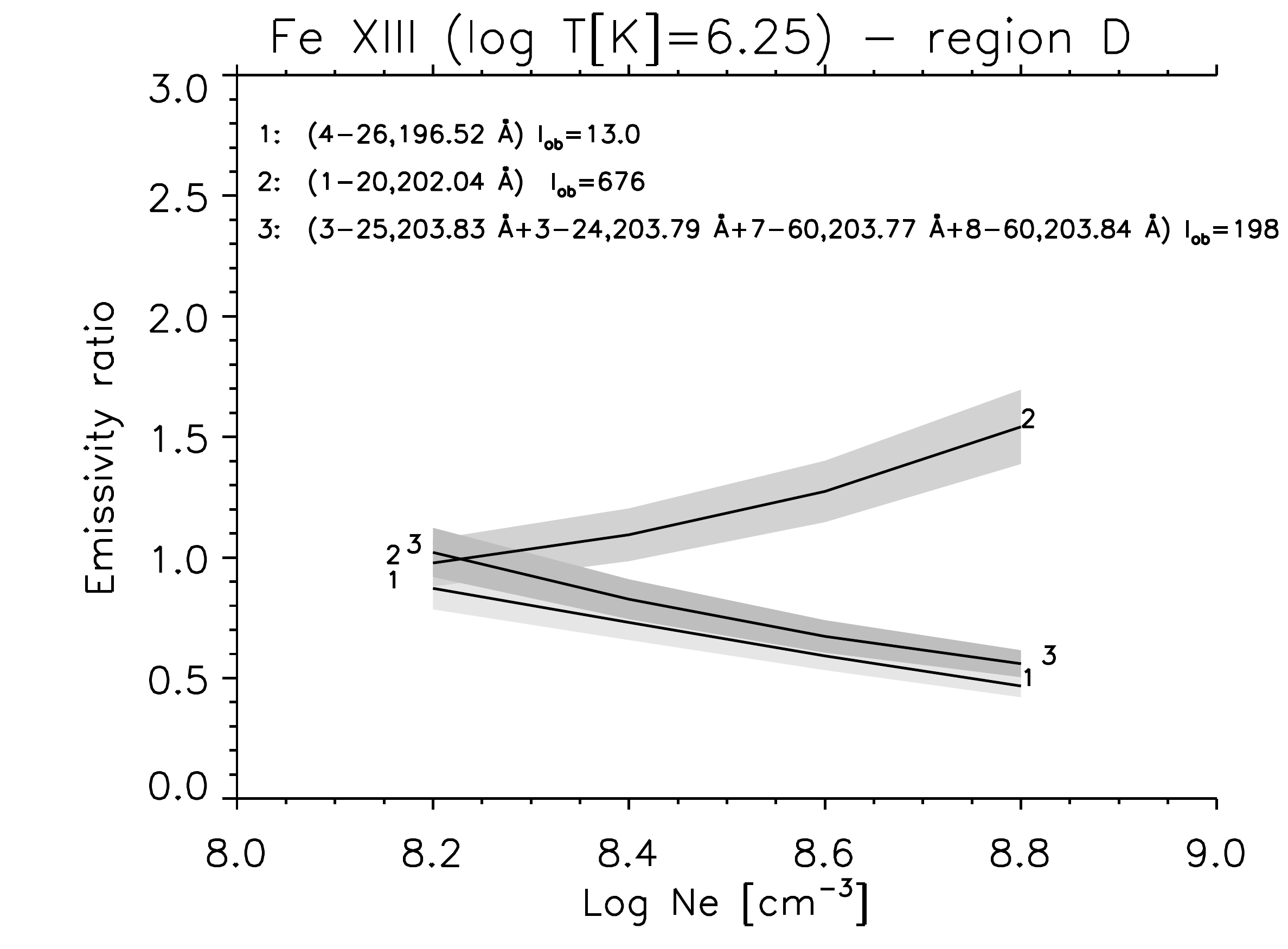}
\includegraphics[width=8.00cm,viewport=40  0 720 468,clip]{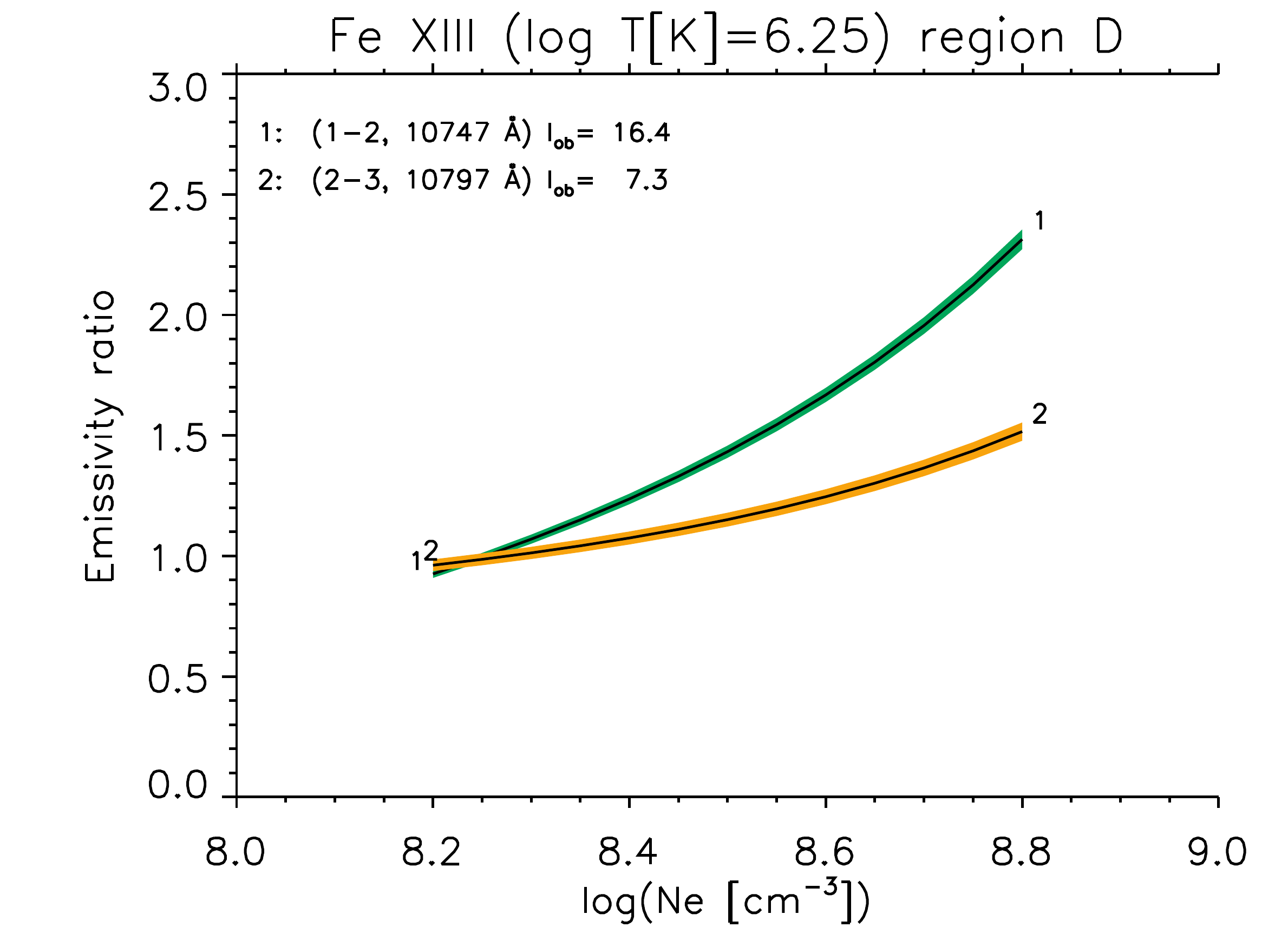}
\caption{Same as in Figure \ref{Fig:region_B}, but for Region D.}
\label{Fig:region_D}
\end{figure}
%
%
\begin{figure}
\centering
\includegraphics[height=8.00cm,angle=-90,viewport= 0 40 500 720,clip]{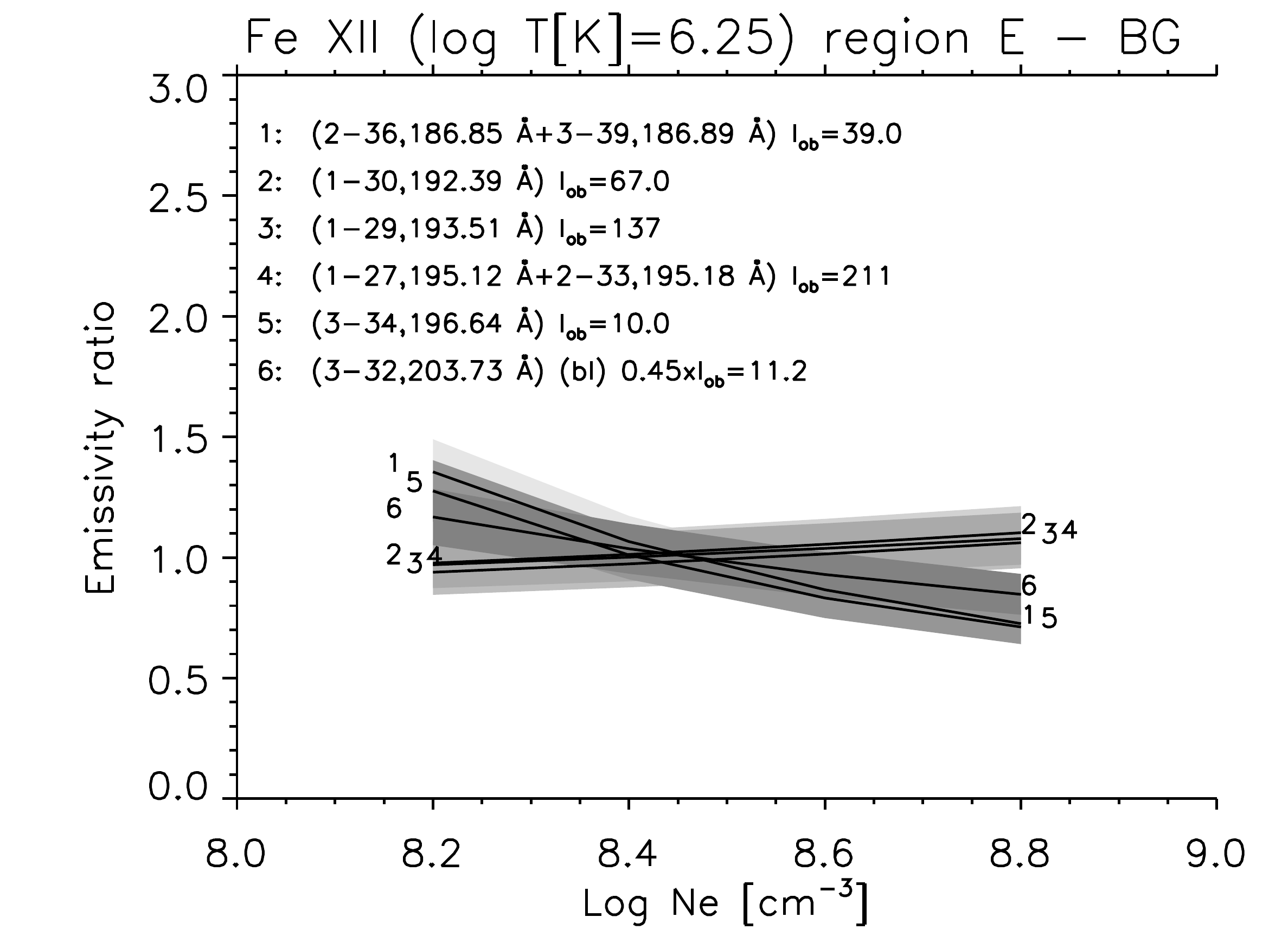}
\includegraphics[height=8.00cm,angle=-90,viewport= 0 40 430 720,clip]{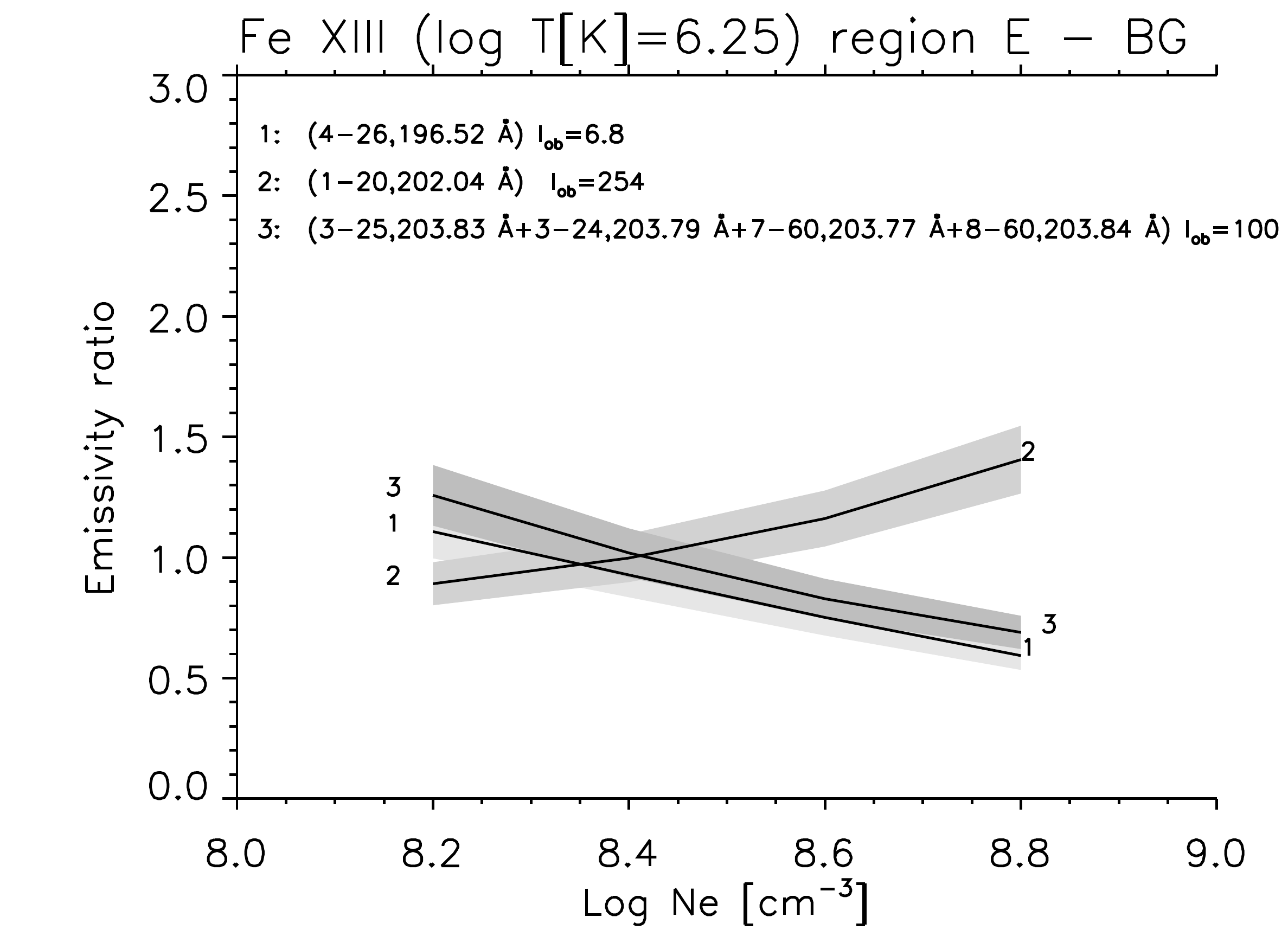}
\includegraphics[width=8.00cm,viewport=40  0 720 468,clip]{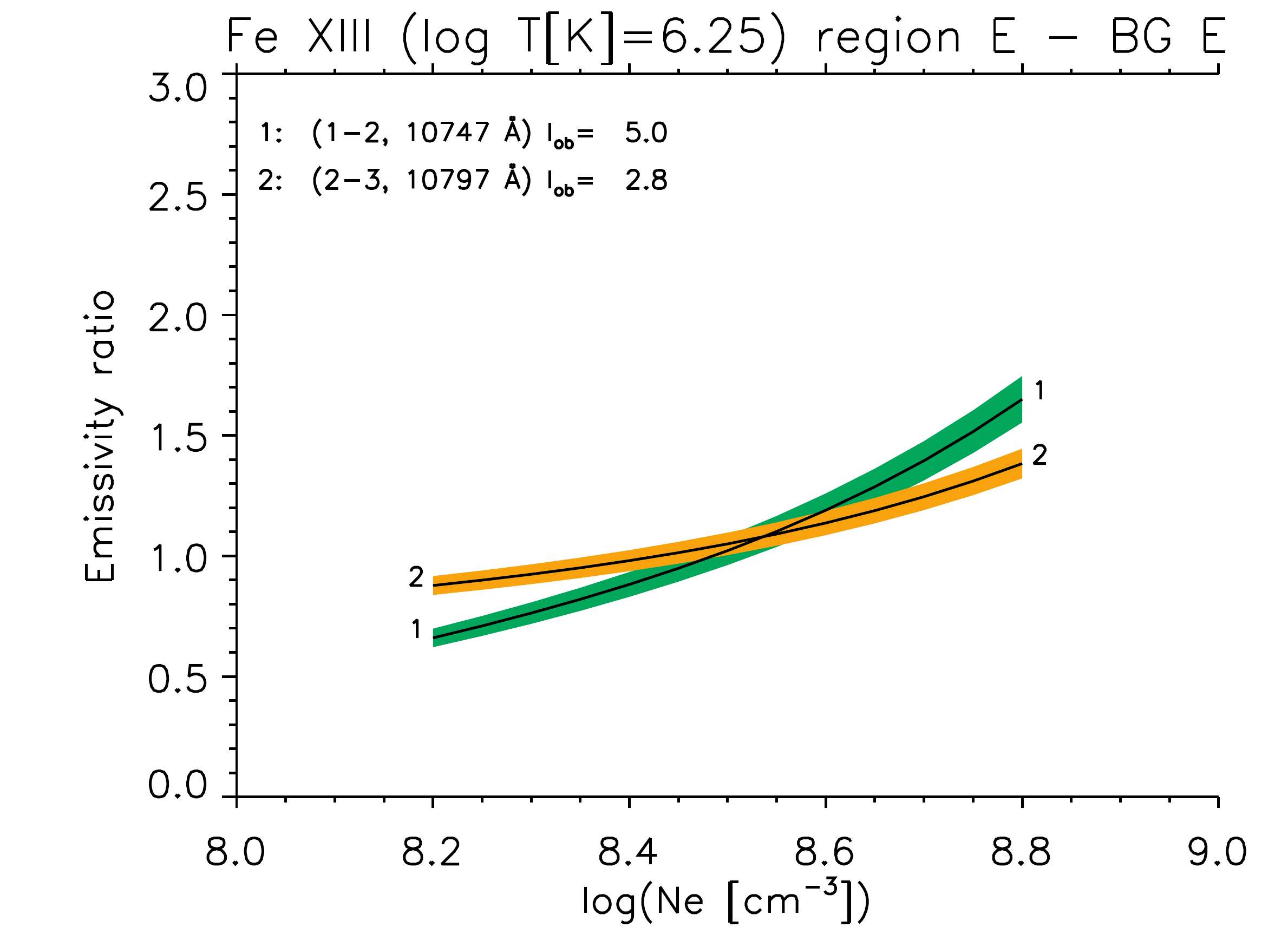}
\caption{Same as in Figure \ref{Fig:region_B}, but for Region E. Note that unlike Figures \ref{Fig:region_B}--\ref{Fig:region_D}, the observed intensities represent background-subtracted values.}
\label{Fig:region_E}
\end{figure}
%
%
%
\section{Electron Density Diagnostics}
\label{Sect:4}

\subsection{Electron density maps}
\label{Sect:4.1}

To give an overview of the electron densities diagnosed by both EIS and \textit{CoMP} instruments, we produce electron density maps (Figure \ref{Fig:Density_maps}) using the plane-of-sky approximation. This is of course only an approximation. A  better method is to assume spherical symmetry and a radial variation of density and temperature, integrate along the line of sight and then compare to observations \citep[see][]{delzanna_etal:2018_cosie,delzanna_deluca:2018}.
However, this method, the same routinely used to infer densities from polarized brightness of the visible continuum, is only applicable to quiet Sun regions, which show a degree of homogeneity.
It does not work when observing active regions, which have loop structures on top of the diffuse, long-range\textbf{, and} more quiet emission \textbf{\citep[see also][]{Subramanian14,Brooks19}}. The issue of "background" contribution to the observed intensities is dealt with further in Sect. \ref{Sect:4.2}.

Having obtained the electron density maps (Figure \ref{Fig:Density_maps}), we do not however proceed to compare the results point-by-point. There are several reasons for this. First, there are large differences in pixel size and spatial resolution between the two instruments. Second, in areas with weaker signal, the fitting of the EIS spectra can be difficult, resulting in a grainy appearance of the electron density map (see Figure \ref{Fig:Density_maps}a). Third, small fitting uncertainties can result in changes in the electron densities determined from \textit{CoMP}. The steep dependence of the density on the observed ratio (see Figure \ref{Fig:photoexcit}), results in non-negligible differences in the diagnosed electron densities even if the observed intensities do not change by more than their $\approx$5\% uncertainties. Therefore, tracking difficulties and the associated changes in occulter position relative to the solar photosphere lead to changes in contributions of the continuum or stray light, which could impact the analysis, even though the line center intensities measured by \textit{CoMP} do not change by more than its uncertainty among the subsequent frames (Section \ref{Sect:2.3.1}). Therefore, to obtain a more reliable density map from \textit{CoMP}, in Figure \ref{Fig:Density_maps}b we show the electron densities diagnosed from the \textit{mean} intensities observed by the \textit{CoMP} instrument, averaged over the entire duration of the EIS raster. Note that the spatial distribution of density is broadly similar between the two instruments, with the \textit{CoMP} instrument being capable of measuring weaker densities approaching log($N_\mathrm{e}$\,[cm$^{-3}$])\,=\,8.0 in regions further off-limb than EIS (Figure \ref{Fig:Density_maps}) at a similar cadence.

\begin{table}[!h]
\begin{center}
\caption{Electron density diagnostics in selected regions. Values listed represent \logne based on the crossings only. See \ref{Sect:4.2} for discussion on uncertainties.}
\label{Table:Results}
\begin{tabular}{lccc}
\tableline\tableline
Region    & EIS \ion{Fe}{12} & EIS \ion{Fe}{13}    &  \textit{CoMP} \ion{Fe}{13}	 \\
\tableline
 A -- BG A  & 8.50 -- 8.65  & 8.50 -- 8.55         	& 8.50	\\
 B          & 8.40 -- 8.55  & 8.40 -- 8.50       	& 8.45	\\
 C          & 8.50 -- 8.65  & 8.45 -- 8.55          & 8.40  \\
 D          & 8.25 -- 8.40  & $<$ 8.25				& 8.20  \\
 E -- BG E   & 8.40 -- 8.50  & 8.35 -- 8.45          & 8.55  \\
\noalign{\smallskip}
\tableline\tableline
\end{tabular}
\end{center}
\end{table}
%
%
%
%
\subsection{Electron densities in selected areas}
\label{Sect:4.2}

We now proceed to diagnose the plasma parameters in detail in five selected regions, labeled A to E, and shown in Figure \ref{Fig:Context}d. These correspond to locally brighter coronal features (loops) clearly discernible in AIA, but still visible with both EIS and \textit{CoMP}. 
The spatial size of these regions is chosen to be $8\arcsec\times8\arcsec$ so that averaging over at least several, typically $2\times2$ \textit{CoMP} pixels is performed to reduce the noise 
and take into account the spatial resolution.
The EIS spectra are first averaged and then fitted with Gaussian components including known blends. For EIS, the averaging significantly improves the signal, as each EIS pixel is $2\arcsec\times1\arcsec$. We calculated the resulting photon noise, taking into account the pseudo-continuum and the detector noise, and obtained relative values below 1\% for the strong lines, and about 2--3\% for the weaker lines. 

Unlike for the electron density maps (Figure \ref{Fig:Density_maps}) produced from one line ratio at a time, we take advantage of multiple available \ion{Fe}{12} and \ion{Fe}{13} lines for diagnostics of electron density in the selected areas. The electron densities are determined using the emissivity ratio method (\citeauthor{DelZanna04} \citeyear{DelZanna04}, see also Section 9.2 of \citeauthor{DelZanna18} \citeyear{DelZanna18}), where the ratios of the observed to the theoretical intensities are plotted as a function of electron density $N_\mathrm{e}$
\begin{equation}
R_{ji} = C \frac{I_\mathrm{obs} \lambda_{ji} N_\mathrm{e}}{N_{j}(T,N_\mathrm{e}) A_{ji}}\,,
\label{Eq:emissivity_ratio}
\end{equation}
where $N_j$\,=\,$N(X^{+m}_j)/N(X^{+m})$ and the constant $C$ is chosen so that the intersections of the curves for each line $\lambda_{ji}$ occur near unity. If the plasma were isodensity, the plotted ratios should intersect at a single point.  If the curves do not intersect at a single point, the scatter of the intersection points provides a measure of the relative differences that could originate either due to uncertainties in the measurements, atomic data, instrument calibration,  or how the plasma is multi-density, given the different sensitivity of the lines to the electron density distribution along the line of sight. Note also that the emissivity ratio method is in principle slightly dependent on the electron temperature $T$, which ought to be determined using other means. 

We note that when plotting the emissivity ratios, uncertainties should be included. For \textit{CoMP}, we use the 5\% uncertainty (Appendix \ref{Appendix:A}) and propagate it through the averaging over selected regions. The averaged intensities together with their uncertainties are then plotted in Figures \ref{Fig:region_A}--\ref{Fig:region_E} as stripes along the emissivity ratio curves. For EIS, we use a constant $\pm$10\% uncertainty on the $I_\mathrm{obs}$ averaged over the selected regions. This estimate  reflects the uncertainty in the EIS relative calibration of these lines close in wavelength \citep[see, e.g.,][Figure 8 therein, and discussion below]{DelZanna13a}.

\subsubsection{Region A and the effect of background subtraction}
\label{Sect:4.2.1}

Region A comprises of a loop in the northern part of the AR complex. At the EIS resolution, the loop does not show substructure (Figure \ref{Fig:Context}d), although some individual sub-structures are present at AIA resolution (Figure \ref{Fig:Context}a--b), especially in the 171\,\AA, whose emission occurs at a lower temperature than \ion{Fe}{13}. The region A has the advantage that a background (which would also include foreground emission) 
region is available nearby, denoted as "BG A" in Figure \ref{Fig:Context}d. The center-to-center distance of the BG A region to region A is 16$\arcsec$, or 0.014 R$_\odot$.

As expected \citep[e.g.,][]{DelZanna03},
the background subtraction modifies the observed line intensities significantly. This holds for both the EUV and NIR \ion{Fe}{13} lines. The total and background-subtracted intensities are listed in the legend of Figure \ref{Fig:region_A}, where the left column gives the values measured in the region A, and the right column gives the background-subtracted intensities. The most affected by the background emission are the 10747 and 10797\,\AA~NIR lines, which retain only 23\% and 21\% of the total intensities measured in region A. The EUV lines are less reduced. While the 202.0\,\AA~line retains only 27\% of the value measured in region A, the 196.5\,\AA~and the 203.8\,\AA~lines retain 43\% and 41\%. For the \ion{Fe}{12}, the reductions range from 28 to 35\%.

Before proceeding with the density diagnostics using the emissivity ratio method, we constructed the emission measure loci plots, where the quantity $I_\mathrm{obs}/G(T_i)$ is plotted as a function of temperature \citep{Strong79,Veck84}, see the first row of Figure \ref{Fig:region_A}. A common crossing of the emission measure loci curves indicates an isothermal temperature \citep[e.g.,][]{DelZanna03,DelZanna18,Schmelz09}. Using the strong EUV lines observed in the SW channel of EIS, we find that the \ion{Fe}{12} 192\,\AA, \ion{Fe}{13} 202\,\AA, and the \ion{Fe}{14} 211\,\AA~indicate an isothermal plasma at log($T$\,[K])\,=\,6.25 both before and after background subtraction (see the top panels of Figure \ref{Fig:region_A}).  We use this temperature for the emissivity ratio plots.

Note that we do not use the \ion{Fe}{14}--\ion{Fe}{16} lines observed in the LW channel of EIS \citep[see, e.g., Table 1 of][]{DelZanna12b}, as the in-flight degradation correction (for data before 2012) by \cite{DelZanna13a} does not apply to the data presented here. The intensities of the LW lines would have to be increased by a factor of up to 10 (compared to the ground calibration) to bring their emission measure loci curves into a common crossing with the lines from the SW channel. Similarly, the cooler \ion{Fe}{9}--\ion{Fe}{11} lines are not used, since these are weak and could be contaminated by other coronal loops, such as the open fan loops overlying the northern AR, seen in AIA 171\,\AA~(Figure 1a).

The emissivity ratio plots (Figure \ref{Fig:region_A}) also indicate a strong influence of the background subtraction on the resulting densities. If the background is not subtracted (left column), the densities diagnosed range from \logne\,=\,8.2 from the \textit{CoMP} \ion{Fe}{13} lines to 8.2--8.35 from EIS \ion{Fe}{13} lines, while the \ion{Fe}{12} crossings are spread over 8.2--8.45. The background subtraction improves the results considerably, with the EIS \ion{Fe}{12} lines indicating crossings over \logne\,=\,8.50--8.65 and EIS \ion{Fe}{13} lines \logne\,=\,8.50--8.55. These values are also listed in Table \ref{Table:Results}. Note however that when accounting for the 10\% uncertainty (gray bars in Figure \ref{Fig:region_A}), the range of possible densities increases. The \textit{CoMP} results indicate a crossing at \logne\,=\,8.50, consistent with EIS results. For \textit{CoMP}, the background subtraction increases the propagated uncertainty, indicating a $\pm$0.15 uncertainty in the resulting \logne. Nevertheless, the crossing point itself is in excellent agreement with the EIS results.  

In summary, as we  expected, the background subtraction for both 
EIS and \textit{CoMP} increases the measured density, brings the results of the two into closer agreement and indicates a plasma close to being isodensity.

\subsubsection{Regions B, C, and D}
\label{Sect:4.2.2}

Next we chose several bright loop regions at different altitudes above the solar limb, where no neighboring background is available (Figure \ref{Fig:Context}d--g). We note that for active regions observed off-limb, this non-ideal situation, possibly with many overlapping structures, is rather typical. The densities diagnosed are then expected to only yield average values.

Region B comprises the brightest loop-tops observed. Although not possible to discern at \textit{CoMP} resolution, there are many loops located inside the $8\arcsec\times8\arcsec$ region. Here, the \ion{Fe}{12} emissivity ratio plots indicate electron densities of \logne\,$\approx$\,8.55--8.60 (Figure \ref{Fig:region_B}) although the range within uncertainties is wider. The EIS \ion{Fe}{13} emissivity ratios indicate a somewhat lower \logne\,=\,8.5, with the 203.8\,\AA~line  indicating higher densities than the 196.5\,\AA~line. In contrast to that,  \textit{CoMP} yields \logne\,$\approx$\,8.45 with relatively small photon noise uncertainties. Considering the various uncertainties, and the fact that the background/foreground would affect somewhat these results (to a lesser extent than in region A), the EIS and \textit{CoMP} results are consistent. 

Region C contains a bright loop-top, but is possibly also overlapped by a portion of the loop E (Figure \ref{Fig:Context}d). In AIA data (Figure \ref{Fig:Context}e), another thin loop is seen to pass through this region. This loop leaves the FOV at $[X,Y]=[1010\arcsec,120\arcsec]$ and it is not readily discernible in EIS, while it is completely lost at the \textit{CoMP} resolution. The density diagnostics from EIS indicates electron densities in the range 8.4--8.5, with \ion{Fe}{12} again yielding slightly larger values than \ion{Fe}{13}. The \textit{CoMP} ratio indicates \logne\,=\,8.4, again consistent with the results from EUV when taking into account the uncertainties (Figure \ref{Fig:region_C}).

In contrast to the other regions, region D is located inside an outer and longer loop (Figure \ref{Fig:Context}d--g) that extends outside of the EIS FOV (cf., Figure \ref{Fig:Context}a--c and Figure \ref{Fig:COMP_fits}). This loop appears as an almost monolithic, relatively-wide structure. Both EIS and \textit{CoMP} indicate densities of \logne\,=\,8.2--8.3 (Figure \ref{Fig:region_D}), with \ion{Fe}{12} as measured by EIS again leading to densities higher than the \ion{Fe}{13}.

\subsubsection{Region E}
\label{Sect:4.2.3}

The last region investigated is Region E (Figure \ref{Fig:Context}d--g), located near the top of a bright loop, with a suitable nearby background. 
We found that the background comprises a significant portion of the signal, as in region A. For the NIR \ion{Fe}{13} lines observed by \textit{CoMP}, the background-subtracted 10747\,\AA~and 10797\,\AA~intensities are only 31\% and 37\% of the original ones, respectively. The EIS \ion{Fe}{13} lines at 196.5\,\AA, 202.0\,\AA, and 203.8\,\AA~retain 41\%, 33\%, and 44\% of the total intensities in region E, respectively. The contribution of the EIS background for the region E is similar to the case of region A (Sect. \ref{Sect:4.2.1}); however, for the \textit{CoMP} lines, the background is weaker. Since the background can vary with spatial location, this highlights the importance of background subtraction.

The emissivity ratios from EIS (Figure \ref{Fig:region_E}) indicate tightly-clustered \ion{Fe}{12} lines at \logne\,$\approx$\,8.45. The \ion{Fe}{13} indicates slightly lower densities, \logne\,= 8.40--8.45, with the 196.5\,\AA~line again giving somewhat lower density than the 203.8\,\AA. In contrast to that, \textit{CoMP} indicates \logne\,$\approx$\,8.5, but with an uncertainty of about 0.15 dex arising from photon noise alone. Thus, the electron densities measured by both EIS and \textit{CoMP} are again consistent within their respective uncertainties.

%
\section{Discussion}
\label{Sect:5}

\subsection{\textit{Hinode}/EIS and \textit{CoMP} results}
\label{Sect:5.1}

Having obtained the density diagnostics from both EIS and \textit{CoMP}, we summarize the main results in Table \ref{Table:Results}. This table lists the (approximate) crossing positions from the emissivity ratio plots, without considering the uncertainties. Even then, the agreement is surprisingly good, often within 0.1 in \logne. Considering the uncertainties, the agreement is excellent; in all cases within the 10\% uncertainty as estimated from \textit{Hinode}/EIS. This is in part due to the emissivity ratio curves crossing at shallow angles; meaning that even a small uncertainty broadens the resulting range of possible electron densities by a non-negligible amount. This is true especially when considering the background subtraction, which increases the uncertainty considerably also for \textit{CoMP} (last row of Figure \ref{Fig:region_A}).

We recall that we performed our diagnostics using the plane-of-sky assumption \citep[as e.g. in][]{Yang20} rather than assuming a radial dependence of density and temperature distributions. The reasons for adopting the plane-of-sky approximation here were twofold. First, subtracting the background for density diagnostics in a particular structure, as in cases of regions A and E, already accounts for additional contributions along the line of sight. Second, the radial temperature and density distributions are typically valid for the quiet Sun, rather than for an active region, whose size and densities can vary considerably. In addition, the active regions can be highly structured. Finally, we note that even in cases where we did not perform the background subtraction (regions B--D), we still obtain consistent electron densities from both EIS and \textit{CoMP}. This is interesting, given that the EUV allowed and NIR forbidden lines have different sensitivities to electron density.

We further note that the electron densities obtained here are by no means unique. At temperatures where \ion{Fe}{13} is formed, the electron densities of the order of \logne\,$\approx$\,8.5 are often seen in apex regions of coronal loops (e.g., Figure 7 of \citet{Tripathi09}; Figure 8 of \citet{Gupta15}) as well as lower down along the loops (see, e.g., Figures 11-15 of \citet{Young09}; Figure 10 of \citet{DelZanna19}; Figure 10 of \citet{Ghosh20}). These densities are also typical in off-limb observations at similar heights above the limb, where some features cannot be distinguished (e.g., Figure 2 of \citet{Penn94}; Figures 1 and 5 of \citet{Warren09}; Figure 6 of \citet{ODwyer11}; Figure 7 of \citet{Madsen19}). Densities of a similar magnitude have also been reported for the star Procyon \citep{Young94}. In the solar corona, higher electron densities are seen at lower heights \citep{ODwyer11}, where the coronal loops are shorter, and presumably located in regions with stronger magnetic field and thus stronger heating \citep[e.g.,][]{Ugarte-Urra19}.

Furthermore, electron densities of similar order of magnitude are observed not only in active region loops, but can be present in quiescent regions \citep[e.g.,][]{Banerjee09,Brooks09,Schmit11,Young14,Landi14,Lorincik20,Yang20}. This highlights the possibility of using the NIR measurements for reliable density diagnostics also outside active regions. Finally, erupting structures during CMEs viewed off limb can in some cases also have electron densities of similar order of magnitude \citep{Landi10,Landi13} and thus present another interesting target for ground-based NIR observations.

\subsection{Atomic data uncertainties}
\label{Sect:5.2}

The atomic data used in the spectral synthesis (Sect. \ref{Sect:3}) are another possible source of uncertainties in the measurements and we now discuss their magnitude. The atomic data for coronal iron ions as available in CHIANTI were obtained with large-scale calculations. Generally, agreement with observations to within a few percent is achieved when comparing intensities of the strongest transitions \citep[see e.g. the benchmark papers:][]{delzanna_mason:05_fe_12,DelZanna11}. This indicates that this is the level of accuracy in the line emissivities of the strongest transitions. Another way to estimate the uncertainty in the theoretical line intensities, developed by one of us (GDZ), is to run Monte Carlo simulations of the level populations of one ion, by varying all the 
atomic rates within some prescribed bounds. One way to estimate the bounds is to compare each rate as obtained by different calculations. As several accurate calculations for the coronal iron ions exist, this is now possible. The uncertainties in the emissivities of the 
strongest EUV lines in \ion{Fe}{13} are at most of the order of 10\%, as in the case of the main diagnostic ratio used here \citep[see][]{Yu18}. A similar procedure was applied to estimate the uncertainty in the \ion{Fe}{13} NIR ratio, which resulted in 2--5\%, for the typical density values obtained here \citep{Yang20}, although we note that the uncertainty increases to about 10\% for higher densities.

As the atomic data uncertainties are comparable to the uncertainties in the observations, we have not considered them when deriving the electron densities. Adding them in quadrature to the measurement uncertainties would lead to total uncertainty being only about $\sqrt{2}$-times larger. Finally, we have already demonstrated that the densities diagnosed from both EIS and \textit{CoMP} are already in excellent agreement, when considering the 10\% EIS calibration uncertainty for lines close in wavelength.

\subsection{Non-Maxwellian effects}
\label{Sect:5.3}

Although the synthetic line intensity calculations in this work were obtained under the standard assumption of a Maxwellian distribution, portions of the solar corona in active regions might in fact be non-Maxwellian. \citet{Dudik15} and \citet{Lorincik20} found that the coronal loops and moss can be characterized by the non-Maxwellian $\kappa$-distributions with extremely low values of $\kappa$\,$\approx$\,2, meaning presence of significant high-energy tails. The diagnostics of the $\kappa$-distributions is a challenging task, requiring analysis of weak lines observed in the LW channel of EIS with respect to relatively-stronger lines observed in the SW channel. Although the IHOP 316 observations were originally designed for non-Maxwellian diagnostics, we did not perform these, since it would first require an update on the in-flight calibration of both EIS channels, which is not currently available. 

Nevertheless, we note that the $\kappa$-distributions do not have a strong effect on density diagnostics \citep{Dzifcakova10}. Typically, the individual density-sensitive ratios of EUV line intensities do not change by more than 0.1--0.2 in \logne \citep[see, e.g., Figure 7 in][]{Dudik14b}. Since the calculations of the synthetic spectra in \citet{Dudik14b} did not include the photoexcitation, we verified that the NIR ratio behaves in a similar way than the EUV ones once photoexcitation is included; that is, strongly non-Maxwellian $\kappa$\,=\,2 distributions result in about a 0.1 dex decrease in the diagnosed densities similarly as for the EUV lines. Therefore, even if the observed regions were non-Maxwellian, the consistency in electron densities diagnosed from both EIS and \textit{CoMP} would still be maintained.

Finally, we note that the non-Maxwellian $\kappa$-distributions lead to an increase of the forbidden line intensities relatively to the EUV ones \citep[see Figure 3 in][]{Dudik14b}. In some cases, for example for \ion{Fe}{10}, the increase can be by a significant amount. In the case of \ion{Fe}{13}, the increase is of only several tens of per cent. Therefore, diagnostics of $\kappa$ from combined EIS and \textit{CoMP} observations remains a challenging task, since the absolute cross-calibration of the two instruments would need to be known to a significant accuracy.

%
\section{Summary}
\label{Sect:6}

We performed diagnostics of electron density in an off-limb active region complex using EUV and NIR emission lines of \ion{Fe}{13}. The EUV lines were observed by the \textit{Hinode}/EIS spectrometer, while the NIR lines were observed by the ground-based \textit{CoMP} polarimeter. The EIS analysis was further supplemented by the EUV lines of \ion{Fe}{12}, which are more numerous, and occur at similar wavelengths. We found excellent agreement, within about 10\%, between the densities diagnosed from EUV and NIR. 

The densities were diagnosed in five selected regions, all comprising portions of relatively-bright coronal loops seen off-limb. The densities found were between about \logne\,=\,8.2--8.6, depending on the region. For two regions, neighbouring background could be subtracted. The background was found to be a dominant contributor to the total observed signal. After subtraction, the loop signal constitutes only about 20--45\%, and the contribution of background strongly depends on the line used. For the NIR lines, the loop signal can be at the lower end of the range mentioned.

The density diagnostics itself was performed using the emissivity-ratio method \citep{DelZanna04,DelZanna18}, where the ratios of the observed to the synthetic intensities are plotted together using all available lines. Our spectral synthesis accounted for photoexcitation, and in line with \citet{Young09} we found that photoexcitation impacts the intensities of the EUV \ion{Fe}{13} lines via changes in relative level population. To reduce the calibration uncertainty in EIS measurements, we used only EIS lines located close in wavelength in the same short-wavelength channel. A 10\% uncertainty accounted reasonably for the calibration uncertainty, while the photon noise uncertainty was considerably smaller, only reaching a few per cent for the weaker lines.

The \textit{CoMP} observations were fitted using the 3-point analytical fits of \citet{Tian13} as well as 7-point Gaussian fits to the full profile plus continuum. We found that the fits give comparable results, and estimated the differences to be within 5\% (half-width at half-maximum of the histograms). The 3-point fits are less sensitive to seeing and tracking inaccuracies, since the data are acquired over a shorter period of time compared to 7 points per profile. They yield reliable central intensities, which alone are recommended for density diagnostics, since they also do not depend on the stray light in the instrument. 

Finally, the agreement between EIS and \textit{CoMP} in the densities diagnosed in this active region complex highlights the usefulness of the ground-based NIR observations for future diagnostics of the plasma in the solar corona alongside the polarimetric measurements of the coronal magnetic field, for example with \textit{DKIST}.

\acknowledgements
J.D., J.L., and E.Dz. acknowledge Grant No. 18-09072S of the Grant Agency of the Czech Republic, as well as institutional support RVO:67985815 from the Czech Academy of Sciences. J.R. acknowledges supported by the Science Grant Agency project VEGA 2/0048/20 (Slovakia). Help of the Bilateral Mobility Project SAV-18-01 of the SAS and CAS is acknowledged as well. J.L. acknowledges the Charles University, project GA UK 1130218. J.D., J.L., G.D.Z., and H.E.M. acknowledge support from the Royal Society via the Newton International Alumni Programme. G.D.Z. and H.E.M. also acknowledge STFC funding via the consolidated grants to the atomic astrophysics group (AAG) at DAMTP, University of Cambridge (ST/P000665/1. and ST/T000481/1).
The National Center for Atmospheric Research is a major facility sponsored by the National Science Foundation under Cooperative Agreement No. 1852977.
Hinode is a Japanese mission developed and launched by ISAS/JAXA, with NAOJ as domestic partner and NASA and STFC (UK) as international partners. It is operated by these agencies in cooperation with ESA and NSC (Norway). 
CHIANTI is a collaborative project involving in the UK the University of Cambridge,
and in the USA the George Mason University, GSFC, and the University of Michigan.


\bibliographystyle{aasjournal}      
\bibliography{2020_EIS_COMP}        

\appendix

%
\begin{figure}
   \centering
   \includegraphics[width=4.76cm,viewport=  0 0 245 415,clip]{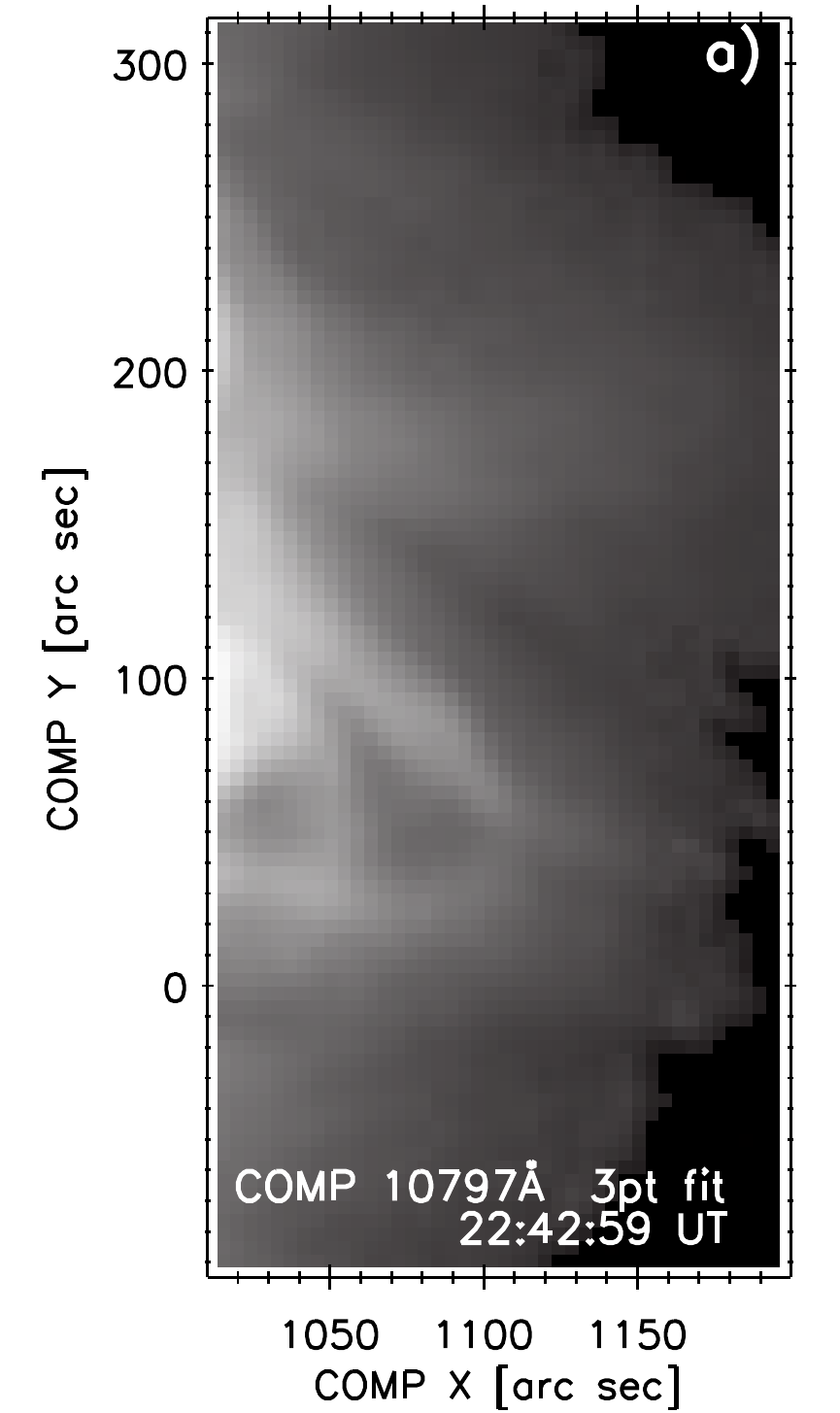}
   \includegraphics[width=3.56cm,viewport= 60 0 245 415,clip]{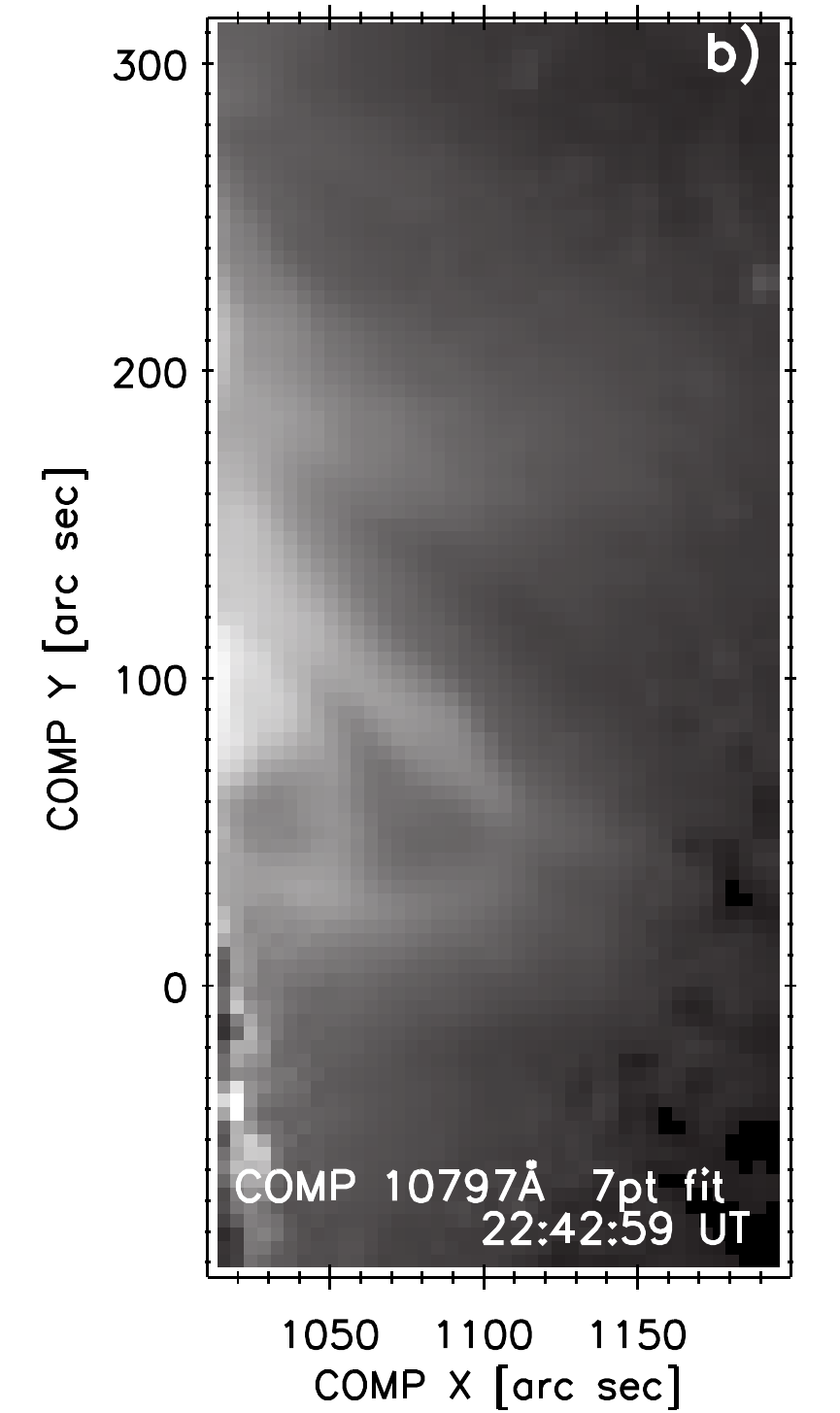}
   \caption{Comparison of two \textit{CoMP} inversions in the 10797\,\AA~line: the 3-point (a) and 7-point (b). The area where the 7-point fits fail are evident by their checquered appearance. }
   \label{Fig:COMP_fits}
\end{figure}
%
\begin{figure}
   \centering
   \includegraphics[width=8.8cm,viewport=  0 0 495 270,clip]{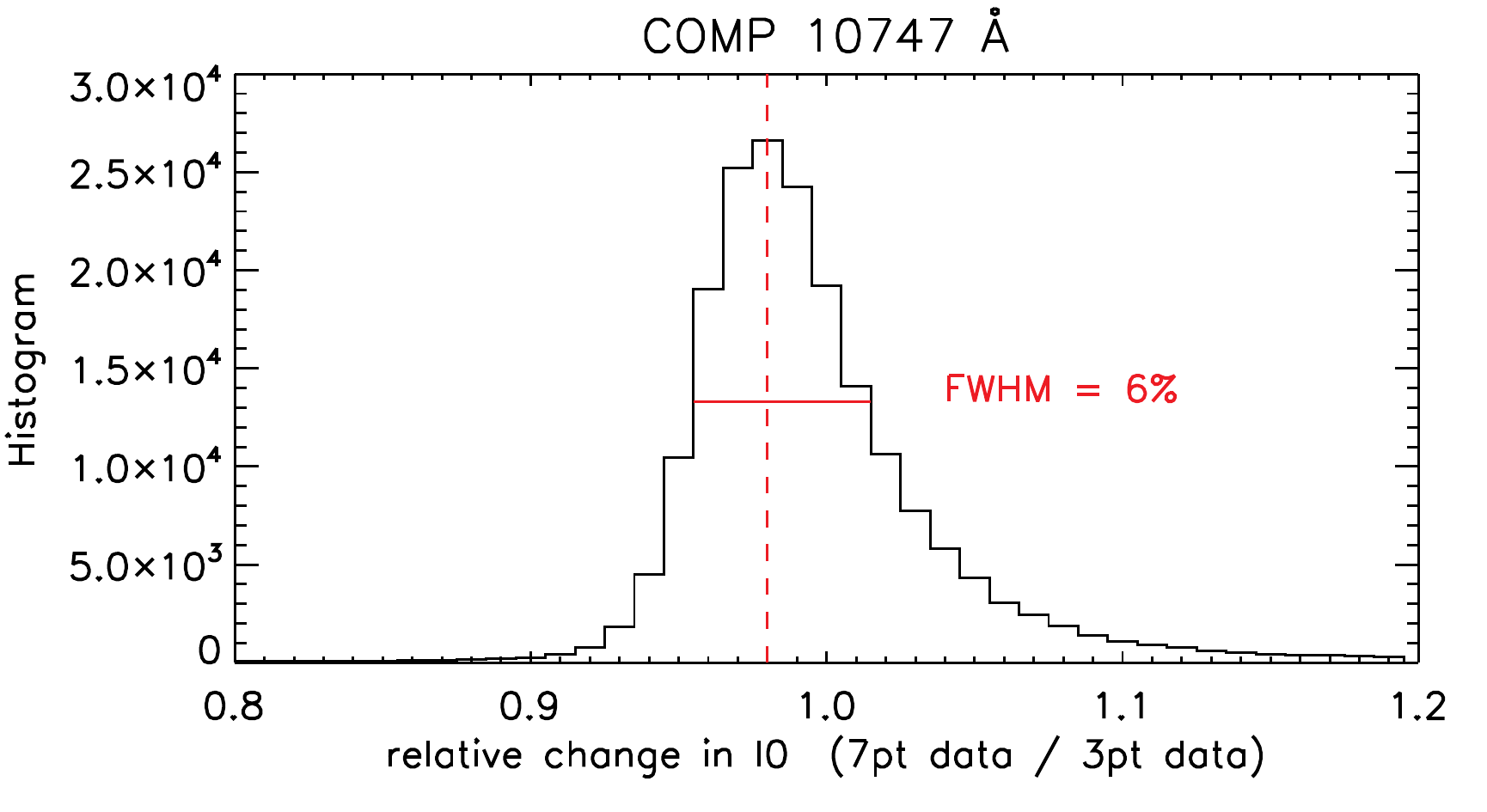}
   \includegraphics[width=8.8cm,viewport=  0 0 495 270,clip]{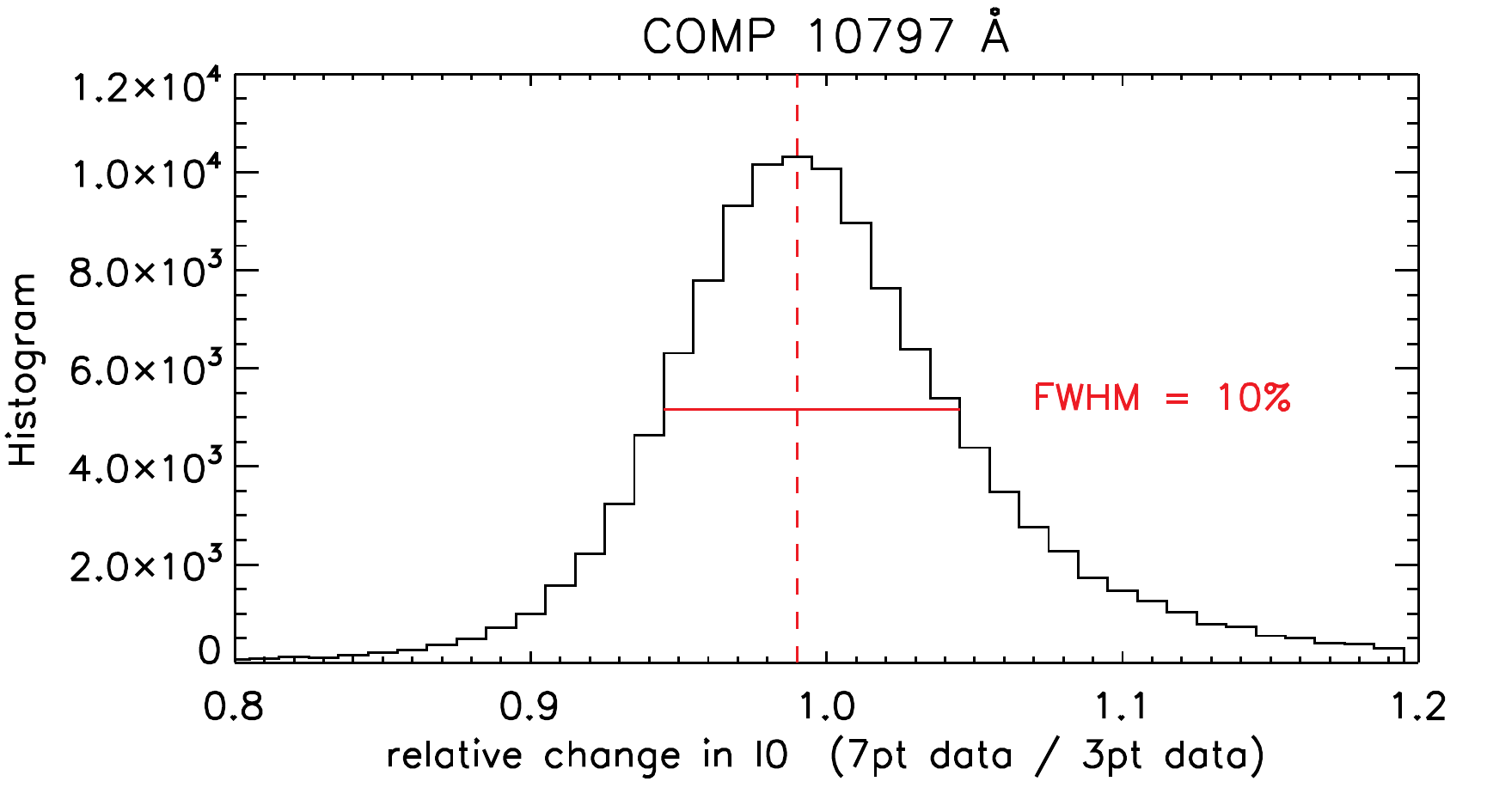}
   \caption{Histograms of the ratios of \textit{CoMP} central intensities derived from 7pt fits to 3pt fits, spanning the whole duration of EIS raster. The full-width at half maximum is indicated by full red lines. The location of the histogram peak is indicated by the red vertical dashed line.}
   \label{Fig:COMP_fits_histograms}
\end{figure}
%
%
%
\section{Comparison of \textit{CoMP} inversions}
\label{Appendix:A}

Having obtained both the 3-point analytical fits of the \textit{CoMP} data, as well as fits of all the 7-points per profile (see Sect. \ref{Sect:2.3.1}), we now proceed to compare the line center intensities $I_0$ from these two datasets.

A review of the co-aligned images showed that there are locations where the Gaussian fits of the 7 points fail. The Figure \ref{Fig:COMP_fits} provides an example of the $I_0$ image derived from \textit{CoMP} 10797\,\AA~data observed at 22:42:59\,UT. The fits fail at both $X$\,$\approx$\,1010$\arcsec$, $Y$\,$<$\,0 close to the limb (panel b) and at larger heights, where $X$\,$\gtrapprox$\,1150$\arcsec$. These locations stand out with their "checquered" and not a smooth appearance as the rest of the image.

Since we obtained the $I_0$ by two inversion methods, we use these to estimate the uncertainty of this parameter. We consider the \textit{CoMP} data observed during the entire duration of the EIS 60\,s raster, as well as shortly before; that is, during 22:12 -- 23:00\,UT. The histograms of the $I_0$ ratios (7-point fits to the 3-point fits) are shown in Figure \ref{Fig:COMP_fits_histograms}. The histogram for the 10747\,\AA~line is peaked, with the peak being located at 0.98 (i.e., a median 2\% difference). The half-width at half-maximum of about 3\%, indicating good correspondence between the two inversions. We note that this value is comparable to the uncertainty derived from the scatter of \textit{CoMP} observations, which is also about 3\% \citep[cf.,][]{Morton16}. Nevertheless, the histogram is skewed to the right, and differences in $I_0$ of up to 10\% also occur. Inspection of the spatial distribution shows that this happens especially in the weaker areas, including the BG E region. The histogram for the 10797\,\AA~line is broader, with half-width at half-maximum of 5\%, and tails extending up to 20\% and even further, again mostly in areas of lower signal. 

Since the 7-point fits require averaging of the pseudo-continuum at all 7 points per profile, and since the Gaussian fits still fail in some locations, we consider them to be less reliable for obtaining the $I_0$ parameter than the 3-point inversions of \citet{Tian13}. Nevertheless, the 7-point fits yield consistent line widths for both NIR lines. This is unlike the 3-point fits, where the width can vary between the lines as it is unconstrained by the outlying points of the profile. Since we do not require the width for density diagnostics (see Sect. \ref{Sect:2.3.2}), we do not investigate this issue further. Finally, as stated in Sect. \ref{Sect:2.3.1}, we adopt the central intensities $I_0$ from the 3-point inversions and consider the 5\% uncertainties to be a reasonable estimate.

%
\begin{figure}
   \centering
   \includegraphics[width=4.37cm,viewport=  0 35 247 245,clip]{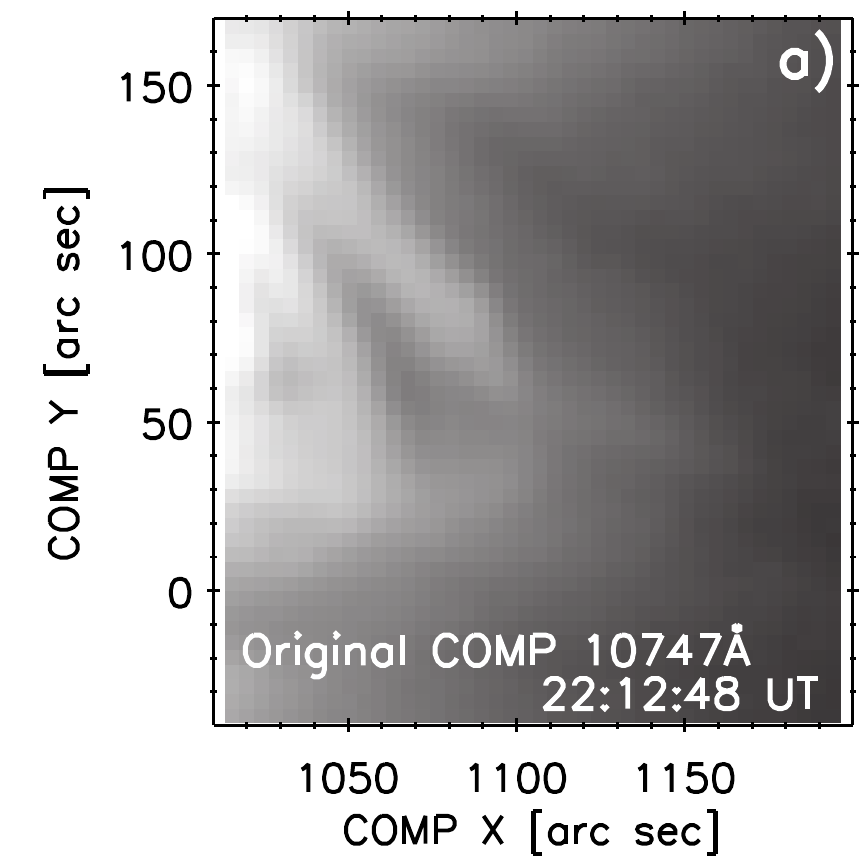}
   \includegraphics[width=3.30cm,viewport= 60 35 247 245,clip]{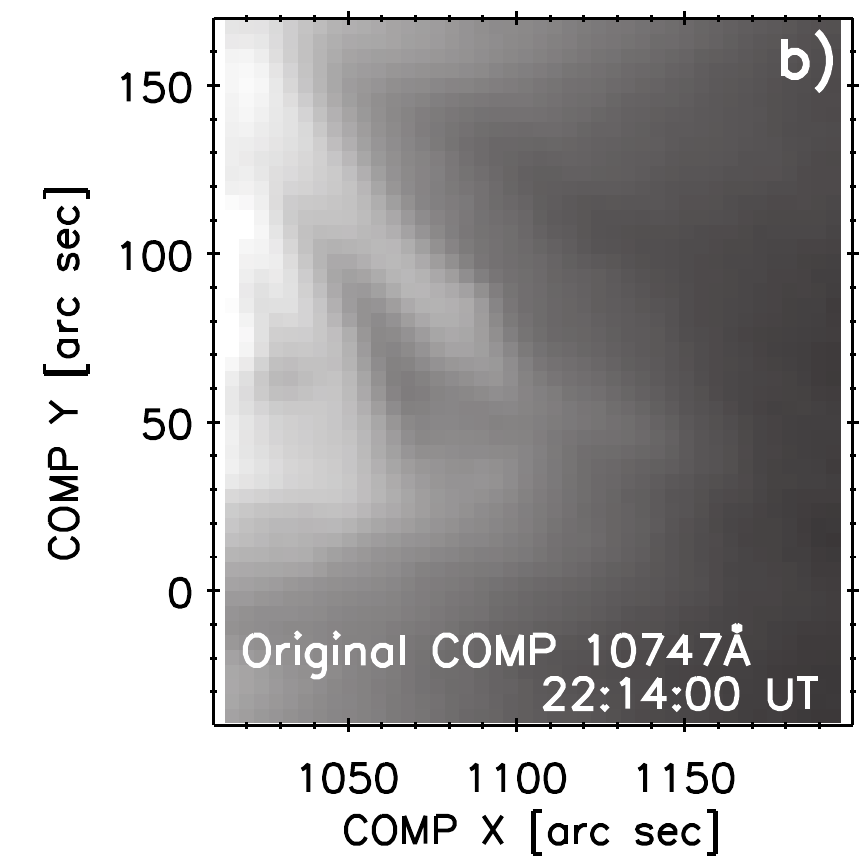}
   \includegraphics[width=3.30cm,viewport= 60 35 247 245,clip]{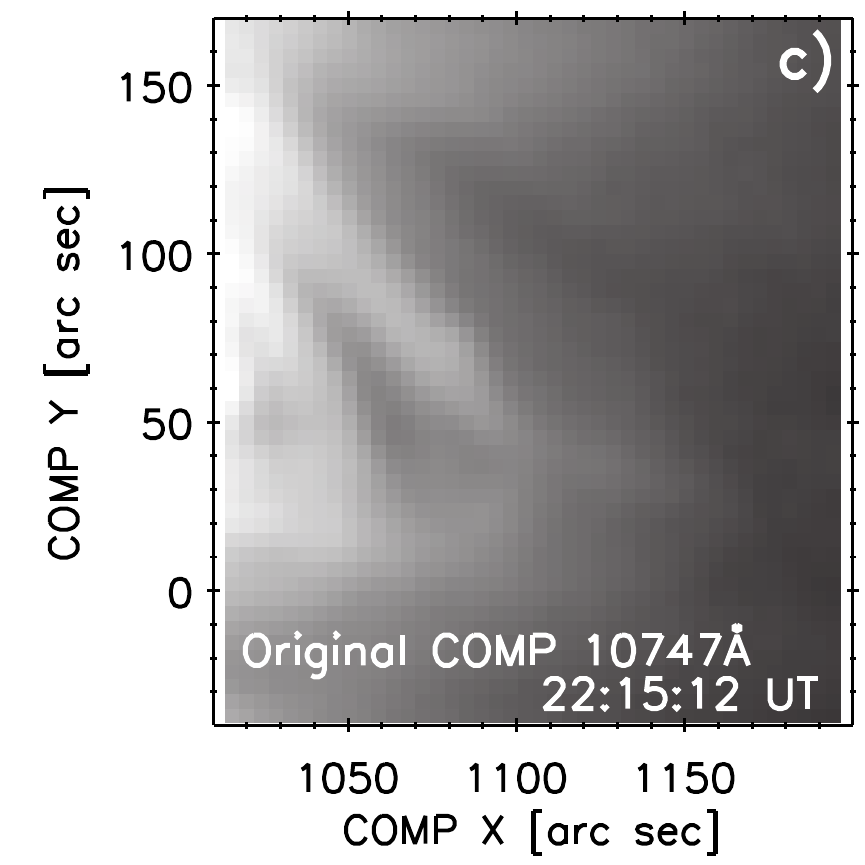}
   \includegraphics[width=3.30cm,viewport= 60 35 247 245,clip]{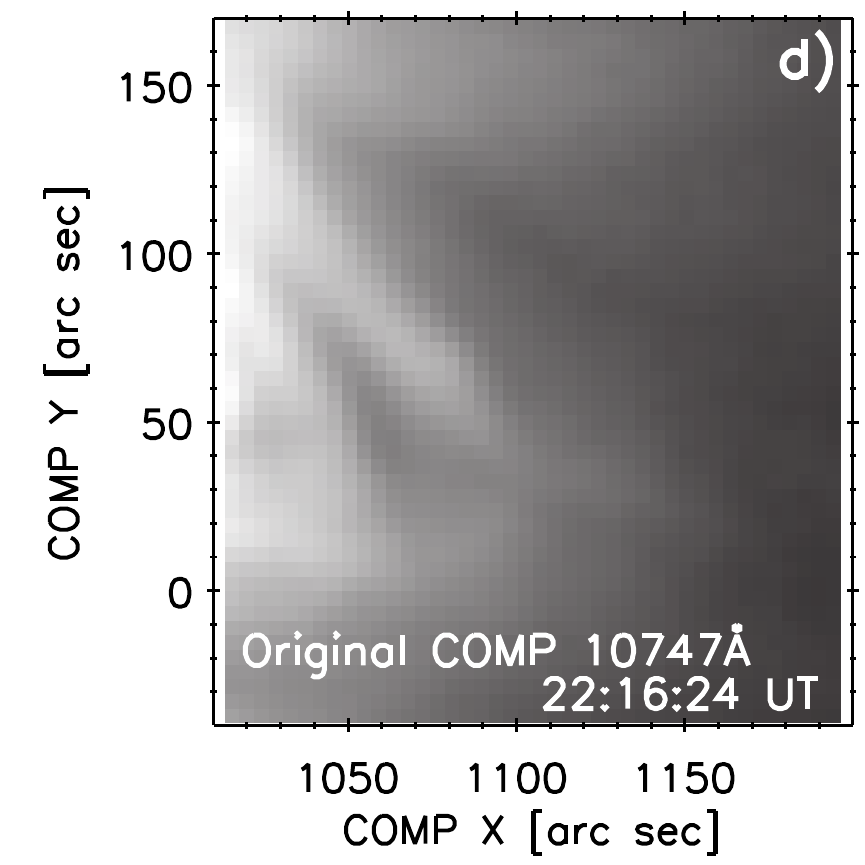}
   \includegraphics[width=3.30cm,viewport= 60 35 247 245,clip]{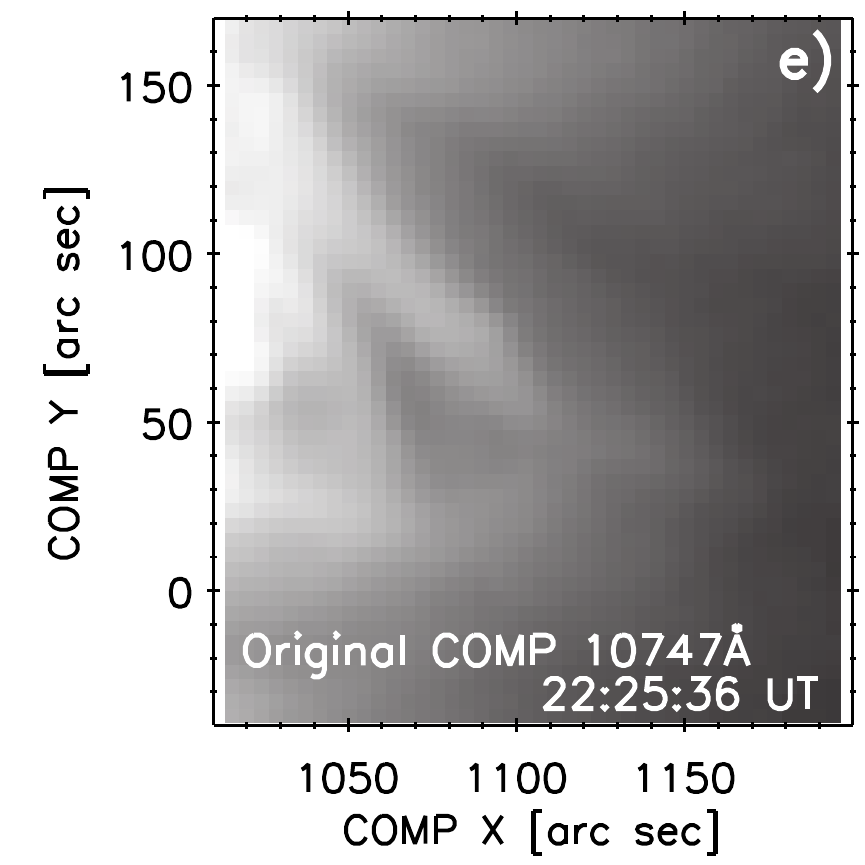}
   \includegraphics[width=4.37cm,viewport=  0  0 247 245,clip]{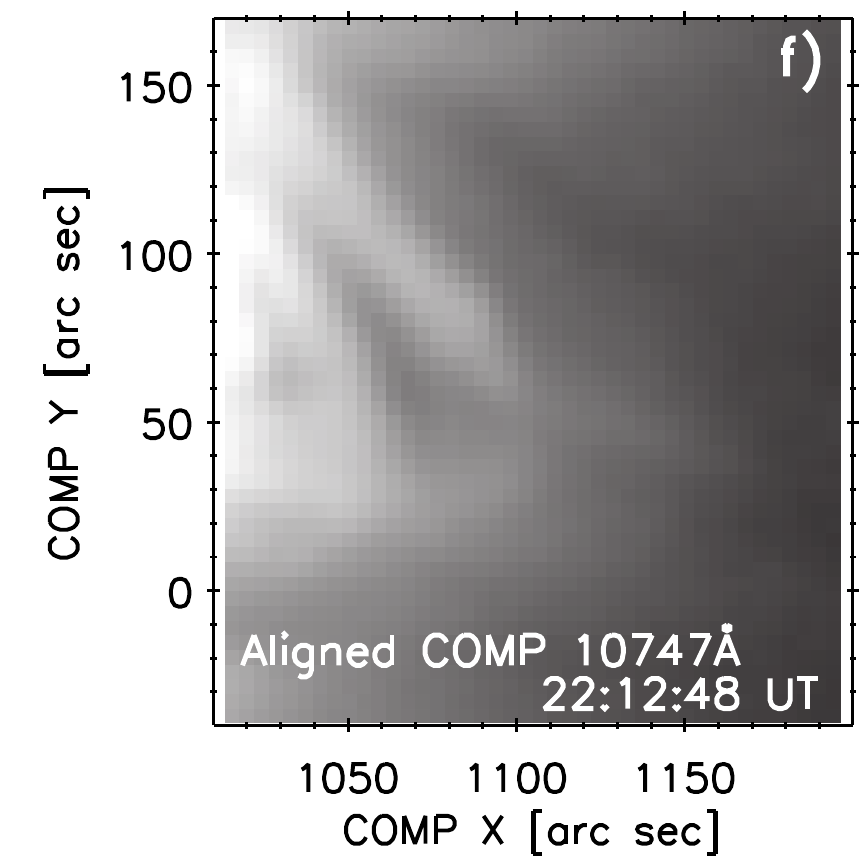}
   \includegraphics[width=3.30cm,viewport= 60  0 247 245,clip]{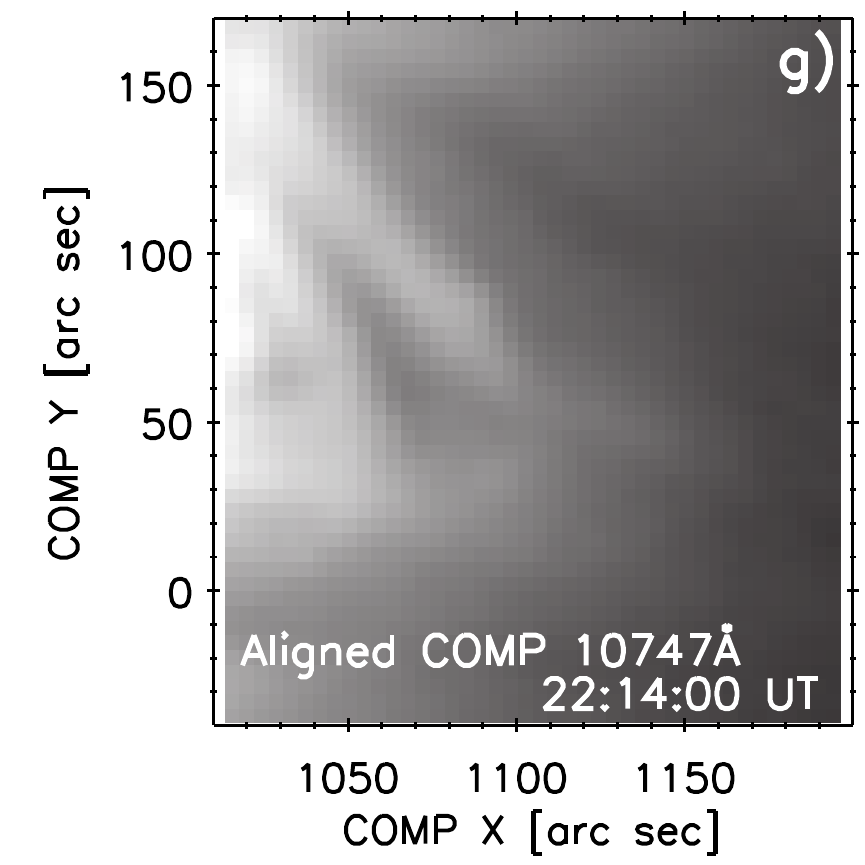}
   \includegraphics[width=3.30cm,viewport= 60  0 247 245,clip]{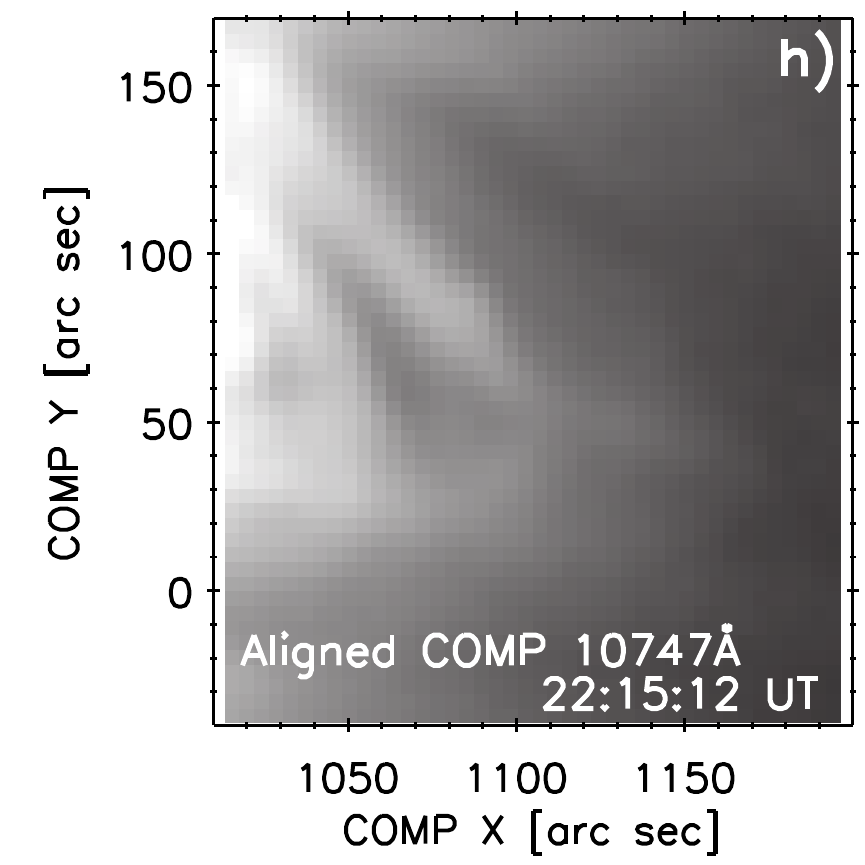}
   \includegraphics[width=3.30cm,viewport= 60  0 247 245,clip]{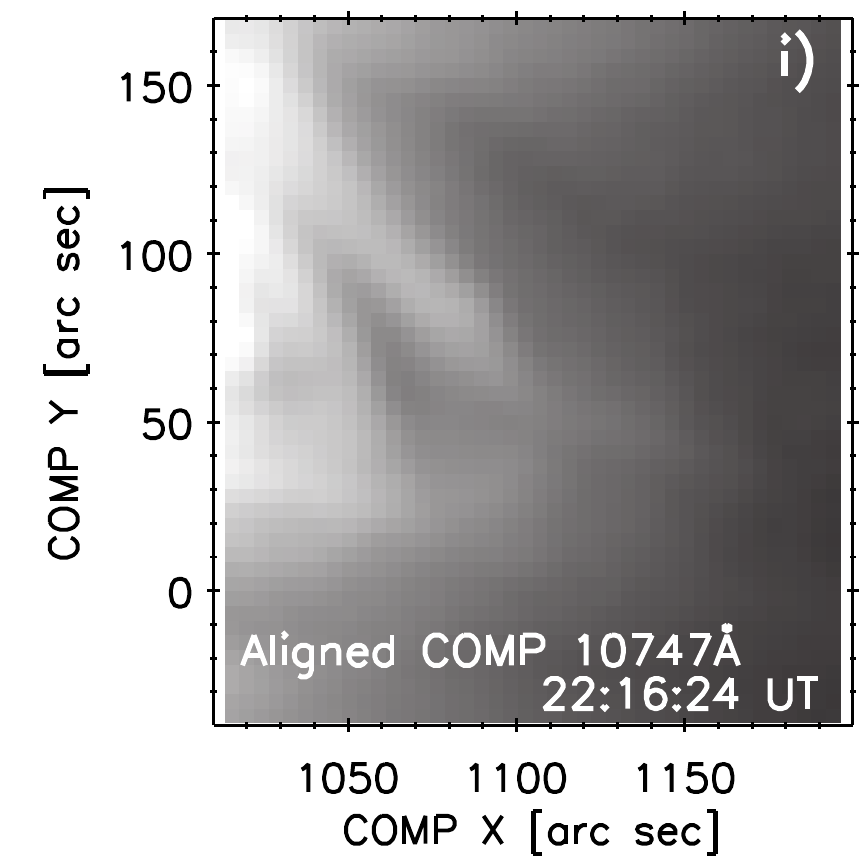}
   \includegraphics[width=3.30cm,viewport= 60  0 247 245,clip]{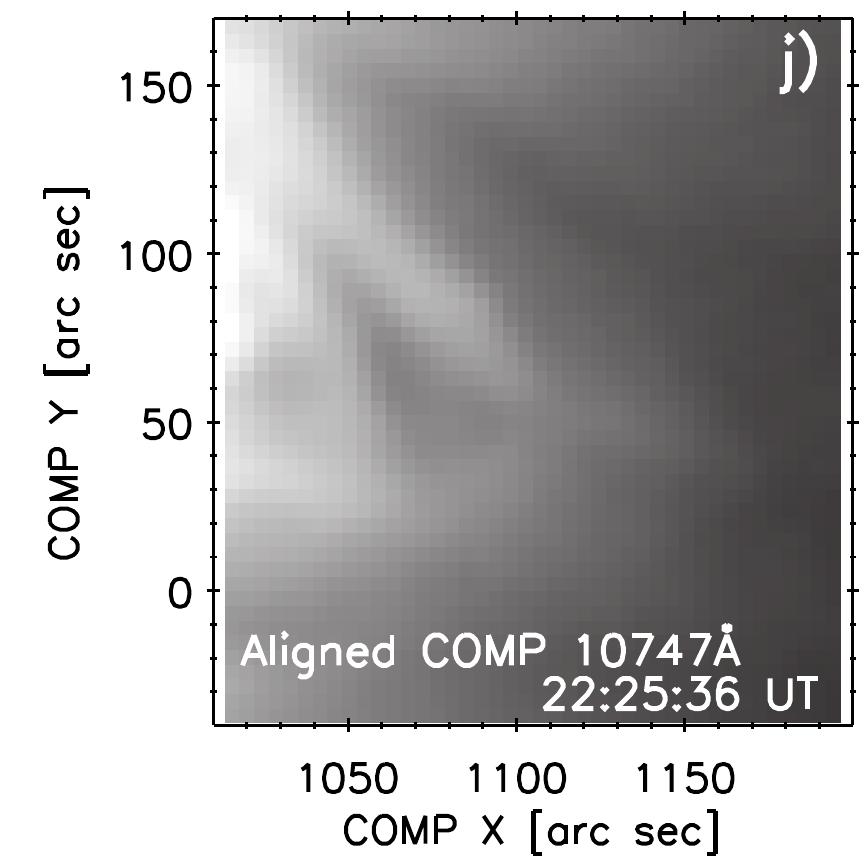}
   \caption{\textbf{Examples of the original \textit{COMP} 3-point line center intensity data in the 10747\,\AA~line (panels a--e) and the corresponding data aligned to the first frame (panels f--j). See Section \ref{Sect:2.3.3} and Appendix \ref{Appendix:B} for details.}}
   \label{Fig:COMP_align}
\end{figure}
%

%
%
{\bf
\section{Example Coalignment of individual \textit{COMP} frames}
\label{Appendix:B}

Individual \textit{COMP} frames are not perfectly spatially aligned (Section \ref{Sect:2.3.3}) as a result of tracking inaccuracies and possibly seeing-induced deformations. Figure \ref{Fig:COMP_align} provides an example of the 10747\,\AA~line center intensities observed at five different times (panels a--e). The times chosen correspond to the first \textit{COMP} frame, which is chosen as a reference for coalignment (panel a), and subsequently to frames 3, 5, 7, and 9 (top row of Figure \ref{Fig:COMP_align}) out of 31 frames acquired during the EIS raster.

The data aligned using the \texttt{auto\_align\_images.pro} SSW routine are shown in panels (f)--(j) of Figure \ref{Fig:COMP_align}. As noted in Section \ref{Sect:2.3.3}, this routine progressively aligns an $i$-th frame to the $i$--1 one by using cross-correlation. After application of this routine, the resulting data are well-aligned.

}

\end{document}